\documentclass[aps,prx,reprint,superscriptaddress]{revtex4-2}

\usepackage{amsmath,amssymb,amsfonts}
\usepackage{dsfont}  %
\usepackage{graphicx}
\usepackage{hyperref}
\usepackage{xcolor}

\graphicspath{{Figures/}{../outputs/}}

\begin{document}
\title{Poincar\'e Vision Transformer, a foundational model to analyze Poincar\'e maps}

\author{Vittorio Peano}
\affiliation{Max Planck Institute for the Science of Light, Staudtstr.~2, 91058 Erlangen, Germany}

\begin{abstract}
Much of the progress in physics rests on a scientist's ability to extract
understanding from a diagnostic plot, yet scientific visualizations remain
largely outside the reach of modern vision models.
We introduce the Poincar\'e Vision Transformer, a foundation model trained to
extract physical insight from Poincar\'e maps.
These are among the most information-dense images in physics: a single picture
captures the dynamics of a Hamiltonian system across an extended region of phase
space.
We recast the expert visual analysis of such maps as a set of image-to-image
tasks, and introduce a physics-inspired pretraining for them: from a map built by
following a set of initial conditions for a short time, the network predicts the
map that the same initial conditions produce after a much longer evolution.
We show that this pretraining yields better performance on the downstream tasks
than either random initialization or the state-of-the-art masked-autoencoder
pretraining used for vision models. Although trained entirely on randomly generated discrete symplectic maps, the
model transfers to Poincar\'e sections of real Hamiltonian systems, from periodically
driven one-dimensional systems to two-degree-of-freedom flows such as the
H\'enon--Heiles system.
\end{abstract}

\maketitle

\begin{figure}
  \centering
  \includegraphics[width=\columnwidth]{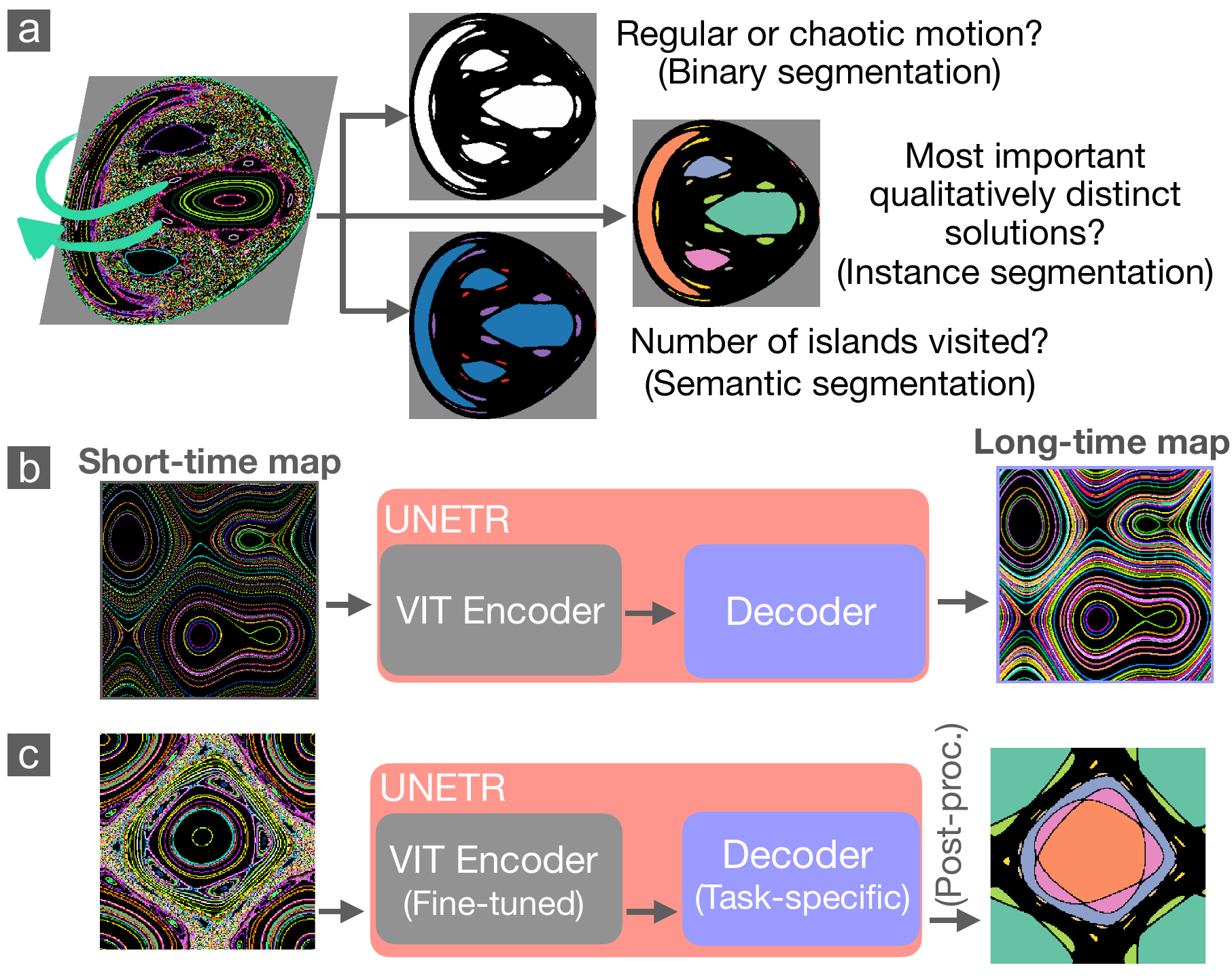}
  \caption{
    Overview of the Poincar\'e Vision Transformer.
    (a)~(left) Sketch of the Poincar\'e construction: a continuous trajectory in the higher-dimensional phase space repeatedly pierces a fixed lower-dimensional section, and the recorded crossings form the two-dimensional Poincar\'e map. (right) Possible AI scientist tasks. (top) Distinguishing regular and chaotic phase-space regions. (middle) Grouping the regular regions visited in turn by a common orbit into a single object---the islands of one chain, or a band whose orbits circulate rather than librate. This is an instance segmentation task. (bottom) Mark each region by the number of distinct regions an orbit visits before returning to the one it started in---for a chain of islands, the number of islands. This is the \emph{first-return period} of Sec.~\ref{sec:tasks}, and labelling it is a semantic segmentation task.
    (b)~Physics-inspired pretraining: long-time map completion. The network is shown a Poincar\'e map painted from short trajectory samples and must predict how the same map will look after a much longer evolution. Encoder and decoder are pretrained together as a UNETR, and the target requires no labelling, only further iteration of the map. (c) Fine-tuning and Inference. The ViT encoder is combined with a task-specific decoder (overall creating a UNETR architecture), and is fine-tuned to perform a specific downstream task.  For instance segmentation, the final output involves a cheap post-processing step. 
    Throughout the paper, Poincar\'e maps are displayed by colouring each pixel
    according to the identity of the orbit that painted it, with unvisited
    pixels in black. These display colours are arbitrary and are not
    the $30$-channel encoding the network receives
    (Sec.~\ref{sec:architecture}). In the predictions, we use black for the chaotic pixels, white for the regular pixels (binary segmentation), and a different colour for each class (semantic segmentation) or instance (instance segmentation).}
  \label{fig:overview}
\end{figure}

\section{Introduction}

Much of the progress in physics comes from a scientist's ability to
extract understanding from visual representations---phase
portraits, spectra, field maps, and other diagnostic plots.
Since most physical models are too complex to solve analytically,
their study relies on numerical exploration, in which visual analysis
plays a central role: a physicist runs simulations for a
batch of parameters, inspects the resulting plots, and uses the
acquired insight to choose the next set of parameters to explore.
Recent AI-driven workflows have begun to close this loop for
specific problems, autonomously generating simulations and steering
parameter exploration~\cite{Naegele2026,Vishwasrao2026,Somasekharan2026}.
However, these approaches predominantly rely on numerical analysis of the output rather than on the rich visual
representations that human physicists so effectively exploit, reflecting the stronger programming compared to vision capabilities of current large language models.

Computer vision has made striking advances on natural
images~\cite{LeCun2015,He2016,Ronneberger2015,Hatamizadeh2022,Dosovitskiy2021,Radford2021,Kirillov2023}, but scientific visualizations
remain largely outside the scope of current foundation models.
Scientific plots differ fundamentally from photographs: they
feature sharp, precise structures rather than natural textures,
their semantics are abstract and domain-specific, and visual
elements such as color, line style, or symbol shape encode
quantities that have no analogue in photographic data.
Building vision models that can parse scientific plots with the
fluency of a domain expert is an open challenge, one that would
bring the power of modern computer vision to bear on problems
across physics, engineering, and applied mathematics.

In this work, we take a first step toward this goal by targeting
Poincar\'e maps---a cornerstone of nonlinear dynamics since
Poincar\'e's study of the three-body
problem~\cite{Poincare1890}, and one of the most information-dense
visualizations in physics: a single image captures the complex
dynamics of a system across an extended region of phase space. We introduce the Poincar\'e Vision Transformer (PVT), a foundational model trained to extract physical insight from Poincar\'e maps.
Such insight, which allows a physicist for example to distinguish
regular from chaotic motion, identify resonance structures,
and delineate continuous invariant sets from a finite number of discrete
intersections, can be naturally formulated as a set of image-to-image
mapping tasks (Fig.~\ref{fig:overview}a).
Following the standard foundation model
paradigm~\cite{Radford2018,Devlin2019,Brown2020,Hoffmann2022,Bao2022,HeMae2022,Dosovitskiy2021,Radford2021,Kirillov2023},
recently extended to the natural
sciences~\cite{Jumper2021,Lam2023,Merchant2023,Bodnar2025,Parker2024,Sheng2025,Khakzad2025,Jakobsen2025},
we first pretrain our model on a large corpus of unlabeled data in a
self-supervised fashion.
Rather than relying solely on the generic masked-autoencoding pretext task
inherited from computer vision~\cite{Bao2022,HeMae2022}, in which the network
learns to restore randomly hidden portions of an image, we introduce a
physics-inspired pretraining task: the network is shown a
Poincar\'e map generated from short trajectories and must predict how the
same map will look after a much longer evolution (Fig.~\ref{fig:overview}b).
Followed long enough, each orbit visually fills out the invariant set that
contains it, so this task is precisely the extrapolation from finitely many
discrete intersections to the underlying invariant sets that a physicist
imagines when looking at such a plot.
Unlike the downstream tasks, its target requires no labelling, only further
iteration of the map.
The pretrained model is then fine-tuned separately for each
downstream task (Fig.~\ref{fig:overview}c).
This pretraining yields better downstream performance than both training
from scratch and standard MAE pretraining.  It is also useful in its own right,
since predicting the long-time map from short trajectories replaces a long
numerical integration with a single forward pass.
To our knowledge, this is the first foundation model trained to
extract physical understanding directly from scientific
visualizations.
By contrast, existing domain-specific foundation models in the sciences---such as
AstroCLIP~\cite{Parker2024} and AION-1~\cite{Jakobsen2025} for
galaxy survey images, self-supervised models for electron
microscopy~\cite{Khakzad2025}, and the seismic foundation
model~\cite{Sheng2025}---operate on raw instrument data
(photometric observations, micrographs, acoustic reflectivity
profiles) rather than on the human-constructed diagnostic
visualizations that scientists use to reason about underlying
physics.

Deriving the pretext task from the physics of the data is not in itself new.
Denoising perturbed equilibrium molecular geometries amounts to inferring the
effective molecular force field, and pretrains state-of-the-art predictors of
molecular properties~\cite{Zaidi2023}; autoregressive prediction of the next
state of a physical system pretrains surrogate models that transfer across
fluid-dynamical and PDE families~\cite{McCabe2023,Herde2024}, in the same
spirit as the forecasting objectives of modern weather and Earth-system
models~\cite{Lam2023,Bodnar2025}.
In all of these the network operates on the physical state itself---atomic
coordinates, or a field on a grid---and its target is a local increment of that
state: one step forward in time, or one step downhill toward equilibrium.
Our pretext task instead operates on a visualization and predicts its long-time
limit: what has to be inferred is not the next state of a single trajectory, but
the geometry that a whole ensemble of trajectories will eventually reveal.

\section{Poincar\'e Maps and Analysis Tasks}
\label{sec:tasks}
Poincar\'e maps were introduced by Henri Poincar\'e in his study of the
restricted three-body problem~\cite{Poincare1890}, where a test particle
moves in the gravitational field of two massive bodies on circular
orbits.
For planar motion, the system has two degrees of freedom and a
four-dimensional phase space.
In the co-rotating frame the equations of motion become autonomous and
admit a conserved quantity (the Jacobi integral) that confines each
orbit to a three-dimensional energy surface; recording the successive
intersections of a trajectory with a two-dimensional surface
(the \emph{Poincar\'e section}) within this energy surface yields a
two-dimensional discrete map.
By analyzing the intersections of stable and unstable manifolds of a
hyperbolic fixed point in this map, Poincar\'e discovered that a
perturbed separatrix splits into a homoclinic tangle of infinite
complexity, the first glimpse of deterministic
chaos~\cite{LichtenbergLieberman1992}.
More than a century later, Poincar\'e maps remain the primary visual
diagnostic for Hamiltonian systems, compressing the full complexity
of phase space into a single two-dimensional image~\cite{LichtenbergLieberman1992}.

The construction is general: given a Hamiltonian flow, one records
the successive intersections of a trajectory with a fixed
lower-dimensional surface (the \emph{Poincar\'e section}), reducing
the continuous dynamics to a discrete map
(Fig.~\ref{fig:overview}a, left).
For an autonomous system with $N$ degrees of freedom, the section is
$(2N{-}2)$-dimensional; the case $N{=}2$ yields a two-dimensional
map that can be visualized directly as an image~\cite{HenonHeiles1964}, which is the
setting we consider here.
The Hamiltonian nature of the underlying flow endows the resulting
map with important structural properties: it is area-preserving
(symplectic), deterministic---each initial condition generates a
unique orbit---and admits neither attractors nor repellers, so that
regular and chaotic structures coexist indefinitely.
Two-dimensional Poincar\'e sections arise naturally in a wide range of physical
contexts beyond autonomous Hamiltonian flows: periodically driven one-dimensional systems~\cite{LichtenbergLieberman1992},
magnetic field-line tracing in
tokamak plasmas~\cite{Boozer2004},
ray propagation in optical
resonators~\cite{NoeckelStone1997},
and, most directly, discrete area-preserving maps such
as the Chirikov standard map~\cite{Chirikov1979} and the
H\'enon map~\cite{Henon1969}.

A typical Poincar\'e map reveals a rich coexistence of qualitatively
distinct dynamical structures (Fig.~\ref{fig:overview}a).
Since a regular orbit lies on an invariant torus~\cite{Kolmogorov1954,Arnold1963,Moser1962},
its repeated crossings of the Poincar\'e section delineate one or more
closed curves---the intersections of the torus with the
two-dimensional section (Fig.~\ref{fig:overview}a). A group of
topologically equivalent orbits typically appears as a series of nested closed
curves, forming a single \emph{island} or a \emph{chain of islands}
around a stable periodic orbit. 
Because we plot the points of each orbit in a distinct colour, the
islands belonging to the same chain---all visited in turn by the same
orbits---are easy to spot for a trained human. 
Regular regions displaying qualitatively different motion are
separated by chaotic regions. Such regions initially appear as thin
layers around the \emph{separatrices} of the integrable limit and, as
the perturbation grows, develop into a large-scale \emph{chaotic
sea}~\cite{Chirikov1979}.
In a typical map, many chaotic trajectories each wander irregularly
and densely cover the whole sea; their overlapping colours interleave
into a noisy, speckled pattern, in sharp contrast to the ordered
nested curves of the regular regions.  The number of distinct regular regions
visited by a single trajectory before returning to the one it started in is a
key topological invariant; we call it the \emph{first-return period} of the
region.  For an island chain it is the number of islands in the chain, and for
a region whose orbits circulate rather than librate it is one.  Note that it is
a property of the region, defined by the permutation the map induces on the
regular regions, and not of any periodic orbit: a circulating region contains
no elliptic periodic orbit at all.

As described above, a trained physicist can gain rich physical insight
about a large ensemble of trajectories by inspecting a single
image. We recast the expert visual analysis as a set of computer vision
tasks (Fig.~\ref{fig:overview}a, right):
\begin{itemize}
  \item \textbf{Binary segmentation}: classifying each pixel as
        regular or chaotic (Fig.~\ref{fig:overview}a, top);
  \item \textbf{Instance segmentation}: distinguishing distinct regular regions. Each regular instance (e.g., a chain of islands) is assigned a unique label (Fig.~\ref{fig:overview}a, middle);
  \item \textbf{Semantic segmentation}: labelling each regular region by its
        first-return period and identifying the chaotic sea as a
        single class (Fig.~\ref{fig:overview}a, bottom);
  \item \textbf{Map completion}: predicting the long-time map---the phase
        portrait as it appears once every orbit has densely traced the structure
        it explores---from a map drawn with short trajectory samples
        (Fig.~\ref{fig:overview}b).
\end{itemize}
The last of these stands apart from the first three.  Its target is not a
labelling but another map, and it is the only one of the four that a physicist
can already carry out by brute force, simply by integrating for longer; its
value is therefore computational rather than interpretive.  Solving it also
requires inferring which invariant set each orbit belongs to,
which makes it an effective pretraining task for the other three.  We develop
both roles in Sec.~\ref{sec:completion}.

Posing the visual analysis as a series of pixel-level segmentation tasks
presupposes that a map contains a definite, finite set of structures to
label. However, the phase-space structure of a Poincar\'e map is
self-similar: each island harbours its own sub-islands, chaotic
layers, and separatrices, repeating at progressively finer scales and,
thus, defining a fractal structure~\cite{Meiss1992}.
This scale invariance is at once an asset and a challenge for our approach.
On the one hand, visual patterns learned at one resolution remain
relevant at all others.
On the other hand, it makes an exhaustive segmentation ill-posed---a
genuinely fractal image carries structure below any finite resolution,
so no pixel-level labelling can ever be complete.
We resolve this tension by segmenting only the structures that are
\emph{resolved at the scale of the image}: the chains of islands born
from the dominant nonlinear resonances visible at that scale, together
with the thin chaotic layers that separate them.
Even at a fixed scale this remains demanding, since the layers
bordering an island chain can be exponentially thin in the
perturbation strength and must still be detected to delineate the
chains correctly.

So defined, the tasks are not merely convenient benchmarks.
Automating them would enable systematic surveys of high-dimensional
parameter spaces---for example, mapping out stability boundaries in
particle accelerator lattices~\cite{Papaphilippou2014} or identifying the
onset of chaos in celestial mechanics~\cite{Laskar1989}---at a scale that
is infeasible by manual inspection.  In such surveys the cost is dominated by
integrating every map to convergence, which is what makes the fourth task
valuable in its own right; the same holds wherever individual trajectories are
expensive to obtain, as in field-line tracing in a tokamak equilibrium or a
Poincar\'e section reconstructed from experimental data.

Several machine-learning approaches have addressed related problems,
though none operates on the full Poincar\'e map as an image.
\citet{Celletti2022} trained a one-dimensional CNN to classify
individual orbits as chaotic, librational, or rotational from
coordinate time series.
\citet{Lee2020} used a CNN to distinguish chaotic from regular
trajectories in the standard map.
\citet{Kamath2022} classified orbit topologies in tokamak
Poincar\'e maps from point coordinates.
\citet{Stephens2023} rendered return maps as images to estimate
dynamical-system parameters via a CNN.
All of these are supervised and task-specific.
Our approach differs in treating the entire Poincar\'e map as an
image, segmenting all structures simultaneously, and learning
general-purpose representations through self-supervised pretraining.

\begin{figure*}
  \centering
  \includegraphics[width=\textwidth]{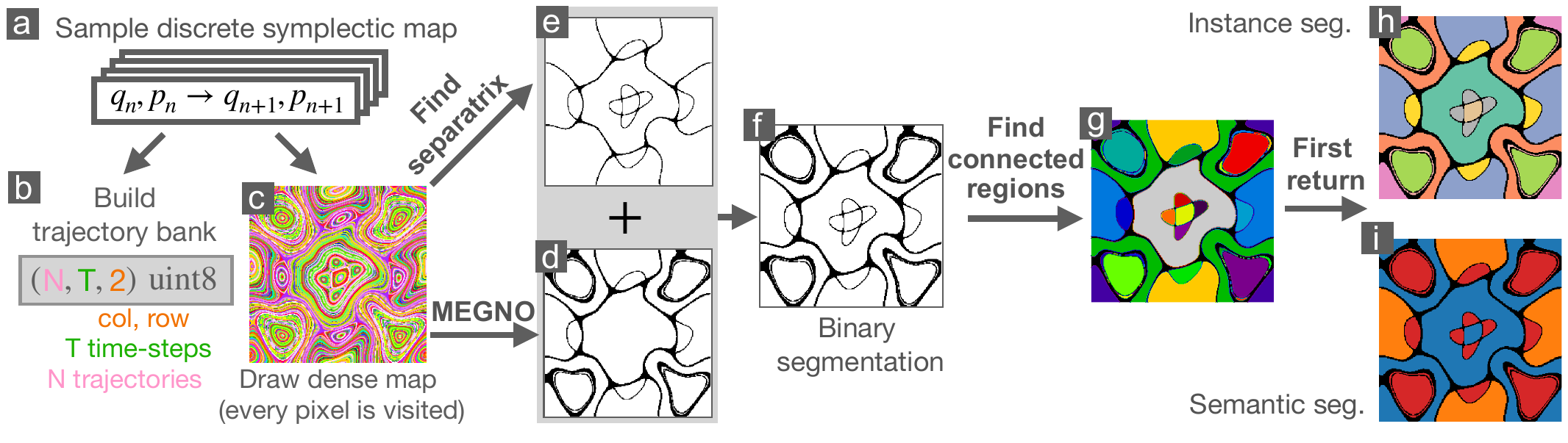}
  \caption{
    Data generation pipeline, illustrated on a single map.
    \textbf{(a)}~A discrete symplectic map $(q_n,p_n) \to (q_{n+1},p_{n+1})$ is
    sampled from one of the constructions of Appendix~\ref{app:map_families}.
    \textbf{(b)}~A fixed bank of $N$ orbits from random initial conditions is
    integrated once for $T$ iterations and stored as pre-binned pixel indices
    (\texttt{uint8} column and row); the network inputs are painted from it.
    \textbf{(c)}~A second, independent integration adds initial conditions
    until every pixel has been visited, giving the dense map on which the
    labels are computed.
    \textbf{(d)}~Each of these covering orbits is classified as regular or
    chaotic by the time-averaged MEGNO indicator (Appendix~\ref{app:megno}); the chaotic
    pixels are the union of the chaotic orbits' footprints.
    \textbf{(e)}~Separatrices are located independently, by finding the
    hyperbolic periodic orbits that generate them and tracing their invariant
    manifolds (Appendix~\ref{app:separatrix_heuristic}).
    \textbf{(f)}~The union of (d) and (e) is the binary segmentation ground
    truth; its complement is the regular foreground.
    \textbf{(g)}~Connected components of the regular foreground are the
    individual islands.
    \textbf{(h)}~Grouping the components visited in turn by a common orbit
    yields the instance segmentation ground truth, one label per island chain.
    \textbf{(i)}~Labelling each region by its first-return period yields the
    semantic segmentation ground truth.
    All label maps are properties of the map in (a) alone, and are therefore
    shared by every image painted from the bank in (b).
  }
  \label{fig:data_pipeline}
\end{figure*}

\section{Data Generation}
\label{sec:data}

Training a segmentation model requires a large and diverse corpus of
Poincar\'e maps together with pixel-level ground-truth labels.
We generate both synthetically, using randomized discrete
area-preserving maps $(q_n,p_n)\to (q_{n+1}, p_{n+1})$ followed by an automated labeling pipeline.

Since any Poincar\'e section is itself an area-preserving map, we can
work directly with discrete symplectic maps rather than integrating
continuous equations of motion, avoiding the need to choose a
Hamiltonian, a surface of section, and long integration times, and
replacing many ODE steps with a single map iteration.

Each map is assembled as a composition of sub-steps that are individually
exactly integrable and area-preserving, with the nonlinear sub-steps
parametrized by a varying number of shape parameters and a common
perturbation strength.  The composition is therefore symplectic by
construction: it preserves phase-space area to machine precision, irrespective
of that strength, which instead offers a single knob carrying each map from a
nearly integrable KAM scenario to fully developed chaos.  We use several
constructions of this kind (Appendix~\ref{app:map_families}) to generate a
broad variety of maps, including the classical double-kick form, which
contains the Chirikov standard map~\cite{Chirikov1979} and the H\'enon
map~\cite{Henon1969} as special cases, but also extending to more general forms
that can mimic the flow generated by a non-separable Hamiltonian, as well as
the stroboscopic maps of driven nonlinear resonators.

We also vary the topology of the phase space itself, sampling maps on a bounded
region of the plane, on the cylinder (one periodic coordinate), and on the
torus $\mathbb{T}^2$ (both coordinates periodic).  The plane and the cylinder
both occur among the real systems of Sec.~\ref{sec:deployment}, and the network
must handle each without being told which it is given.

The automated labeling pipeline is illustrated in
Fig.~\ref{fig:data_pipeline}.
Each sampled map (panel~a) is integrated twice, for two different purposes.
The first integration produces the trajectory bank from which the network
inputs are later painted: a fixed set of $N = 400$ orbits from random initial
conditions, followed for $T = 2500$ iterations and stored as pre-binned
\texttt{uint8} pixel indices rather than floating-point coordinates (panel~b; Appendix~\ref{app:bank}).  The labels, by contrast, require
an orbit through every pixel, which a fixed set of initial conditions cannot
guarantee; the second integration therefore grows its initial conditions
adaptively, seeding new orbits wherever the image is still empty, until the
whole $256\times256$ grid has been visited (panel~c).
What remains is to decide, for each pixel, whether it belongs to a regular
region or to the chaotic set.  The natural tool is a chaos indicator applied to
each covering orbit.  The largest Lyapunov exponent is the standard choice but
converges slowly; we use instead the time-averaged MEGNO
indicator~\cite{Cincotta2000}, which distinguishes the linear divergence of
neighbouring regular orbits from the exponential divergence of chaotic ones
after far fewer iterations (panel~d; Appendix~\ref{app:megno}).  Marking every
pixel visited by an orbit so classified gives a first chaotic set.

This is not sufficient on its own.  The chaotic layer bordering an island chain
can be exponentially thin in the perturbation strength, and a layer narrower
than a pixel is almost never entered by the covering orbits: the pixels along
it are claimed instead by the regular orbits on either side.  These are
precisely the layers we cannot afford to miss.  A separatrix layer is the
boundary of a nonlinear resonance: it divides the island chain the resonance
creates, in which orbits librate, from the non-resonant region enclosing it, in
which they circulate.  Without it the connected-component step below would
absorb the chain into its surroundings, and the resonance would leave no trace
in the ground truth.  We therefore do not mark the separatrix by where the
covering orbits happen to fall, but construct it: we locate the hyperbolic
periodic orbit that generates it and trace the invariant manifolds emanating
from that orbit (panel~e; see Appendix~\ref{app:separatrix_heuristic} for
details).

The union of the two---the pixels occupied by chaotic orbits and those traced
out by the manifolds---is the binary ground truth (panel~f), and its
complement---the regular foreground---is decomposed into connected components,
with components that meet across a periodic edge merged so that an island
straddling the boundary is not counted twice (panel~g;
Appendix~\ref{app:chains}).
Components that a single orbit visits in turn belong to one island chain and
receive a common label, giving the instance ground truth (panel~h); labelling
each chain by the number of islands an orbit visits before returning gives the
semantic ground truth (panel~i).

In total we generate ${\sim}50{,}000$ labelled maps for training, spread over
all the map constructions and the three phase-space topologies
(${\sim}23{,}000$ toroidal, ${\sim}22{,}000$ planar, ${\sim}5{,}000$
cylindrical), together with a held-out validation set of ${\sim}1000$ maps
generated from disjoint random seeds.  All maps are rendered at
$256\times256$ pixels.  The network is never shown a fixed image of any of
them: every input is painted afresh from the trajectory bank at training time,
an augmentation we return to in Sec.~\ref{sec:architecture}.

\section{Model Architecture and Pretraining}
\label{sec:architecture}

\begin{figure*}
  \centering
  \includegraphics[width=\textwidth]{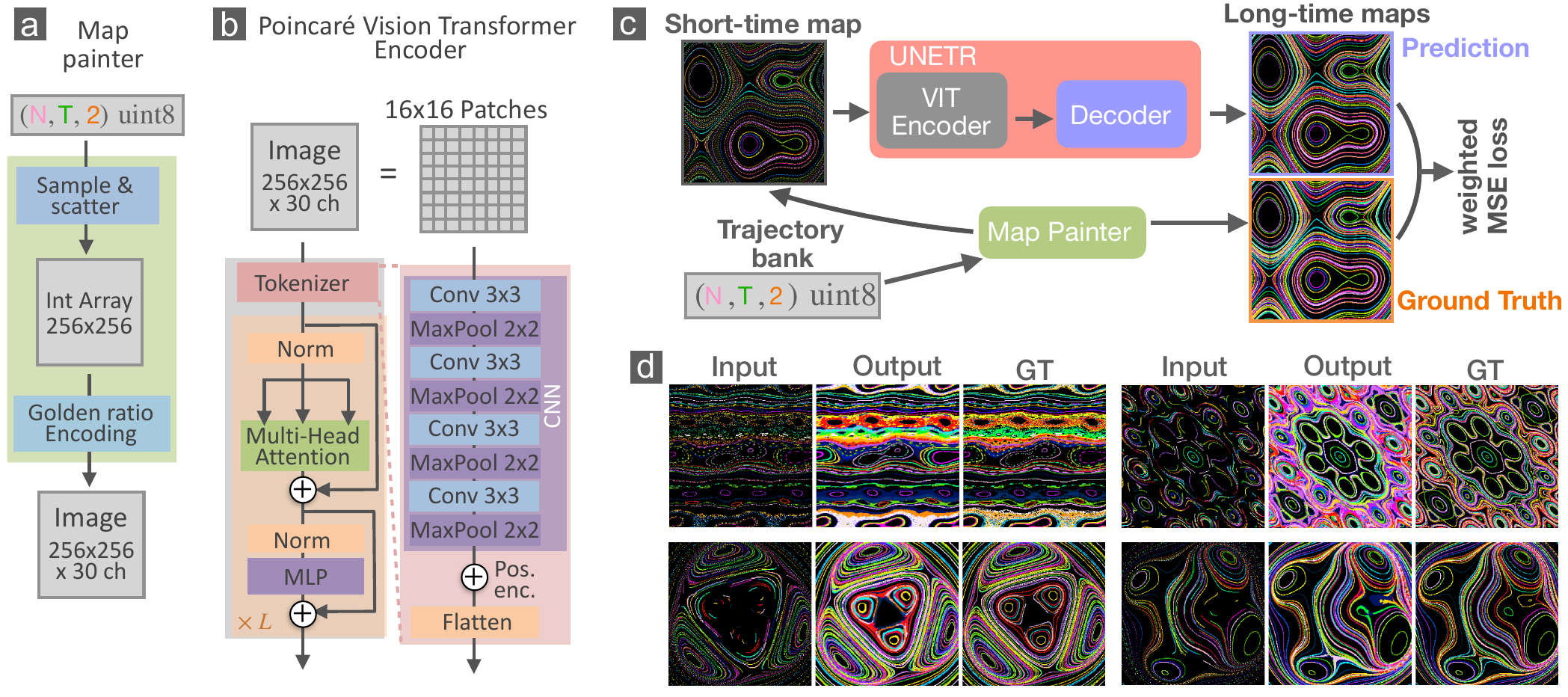}
  \caption{
    Input representation, architecture, and physics-inspired pretraining.
    \textbf{(a)}~The map painter.  A fresh network input is synthesized at
    every training step from the stored trajectory bank of
    Fig.~\ref{fig:data_pipeline}b: a random subset of $k$ orbits and a random
    window of $T_{\mathrm{w}}$ consecutive iterations are scattered onto the
    $256\times256$ grid, giving an integer image whose value at each pixel is
    the index of the orbit that painted it, which the golden-ratio encoding
    turns into a $256\times256\times30$ real-valued image.
    \textbf{(b)}~The Poincar\'e vision-transformer encoder.  The image is
    tokenized by a small convolutional block alternating convolutional and max-pooling layers; learned positional encodings are
    added and the resulting $256$ tokens pass through $L$ pre-norm
    transformer blocks.
    \textbf{(c)}~Long-time map completion.  The painter renders two images
    from the \emph{same} orbits with the \emph{same} colours: a short-window
    input and the full-length target.  The encoder of~(b), together with the
    UNETR convolutional decoder, must predict the latter from the former under
    a region-weighted mean-squared-error loss.
    \textbf{(d)}~Completion on four held-out validation maps: short-window
    input, prediction, and long-time ground truth.  The predicted panels are
    shown decoded to the nearest encoding vector, as orbit identities.  
    The completion of every validation map is shown in the completion gallery of the data deposit~\cite{PVTData2026} and on the project page~\cite{PVTSite}.
  }
  \label{fig:architecture}
\end{figure*}

All four tasks of Sec.~\ref{sec:tasks} are solved by a single architecture: one
encoder, pretrained once, combined with a task-specific decoder and fine-tuned
separately for each task (Fig.~\ref{fig:overview}c).  This section describes how
a Poincar\'e map is turned into a network input, the network itself, and the two
pretraining stages (Fig.~\ref{fig:architecture}); the task-specific decoders,
losses, and fine-tuning protocol are presented together with the results in
Sec.~\ref{sec:results}.

\subsubsection{Input representation.}
Rasterizing the recorded intersections onto the $256\times256$ grid is not by
itself enough, because it discards the information a physicist relies on most:
\emph{which} orbit produced each point.  The islands of one chain are
recognizable precisely because the same orbit reappears in all of them, so
orbit identity must be carried into the image as a per-orbit colour.
Since an image contains up to several hundred orbits, we use more than
the three channels of an RGB image, namely $M=30$, and assign the $j$-th
orbit drawn the $j$-th term of a
deterministic quasi-random sequence in $[0,1]^M$, whose codes remain well
separated however many orbits are drawn (Appendix~\ref{app:encoding}).

The image itself is never stored.  The segmentation ground truth of
Sec.~\ref{sec:data} is a property of the map alone: it does not depend on which
orbits are drawn, on how many there are, or on how long each is followed.  We
exploit this invariance by synthesizing a fresh input at every training step
directly from the trajectory bank (Fig.~\ref{fig:architecture}a), drawing a
random subset of $k$ orbits, a random window of $T_{\mathrm{w}}$ consecutive
iterations, and a random offset along the bank
(Appendix~\ref{app:bank}).  A single labelled map thus yields a number of
distinct inputs that is combinatorially large in the size of its bank: the
choice of which $k$ of the $N = 400$ banked orbits to draw alone gives
$\binom{N}{k}$ possibilities, more than $10^{120}$ once summed over the range of
$k$ we use, before the window length and offset are varied at all.  They span
everything from a few sparsely sampled orbits to a fully populated phase
portrait, and all of them share one ground truth.
Because colours are assigned in draw order while the subset is drawn at random,
the same mechanism also permutes the colours between realizations, so the
network cannot tie a fixed colour to a fixed orbit.

\subsubsection{Architecture.}
The network is a UNETR~\cite{Hatamizadeh2022}: a vision
transformer~\cite{Dosovitskiy2021} whose intermediate representations are fed,
through skip connections, into a convolutional decoder that restores the full
$256\times256$ resolution (Appendix~\ref{app:unetr_detail}).  We use a ViT-L
encoder throughout---$24$ pre-norm blocks of width $1024$ with $16$ attention
heads, ${\sim}3\times10^{8}$ parameters---and a decoder base width of $64$
channels.  The choice of encoder was motivated by the non-local nature of Poincar\'e maps: our instances typically comprise several regular islands scattered across phase space to form a chain. Labelling such a chain correctly requires comparing the colours of distant pixels, a task in which global attention is expected to excel.

The one departure from the standard recipe is the tokenizer
(Fig.~\ref{fig:architecture}b).  A vision transformer normally forms each token
by a learned linear map of the raw $16\times16$ patch; replacing this by a small
convolutional stem trains better on natural images~\cite{Xiao2021}, and we adopt
that structure with one change: max-pooling is the only nonlinearity.  A
Poincar\'e map is extremely sparse---an invariant curve is one pixel wide, and
its colour is the only marker of orbit identity---so a channel-wise maximum,
which preserves any feature active anywhere in its window, serves better than an
operation that blends the dozen curves crossing a patch. Indeed,  this design choice  pays off: comparing two networks trained without pretraining for instance segmentation under
otherwise identical conditions, we find that the network with our tokenizer improves on the
standard linear patch embedding by $11\%$ in the main figure of merit of that
task (Sec.~\ref{sec:results}; details in Appendix~\ref{app:tokenizer}).

\subsubsection{Masked-autoencoder pretraining.}
Our starting point was a generic state-of-the-art pretext task in computer
vision, masked autoencoding~\cite{HeMae2022}: three quarters of the
$16\times16$ patches are hidden, and a shallow transformer decoder---discarded
afterwards---must reconstruct their full $30$-channel pixel content from the
visible ones (Appendix~\ref{app:mae}).

\subsection{Physics-inspired pretraining: long-time map completion}
\label{sec:completion}

Long-time map completion entered this work as a task in its own right
(Sec.~\ref{sec:tasks}): predicting the converged phase portrait from short
integrations replaces the expensive part of a broad parameter survey with a
single forward pass.  For the H\'enon--Heiles system, extending $150$ orbits from
$500$ to $2500$ crossings takes about two minutes on a laptop GPU, against $0.09$~s for
the forward pass (Appendix~\ref{app:continuous_maps}).  Only once it was working did we realize that it could
also serve as a pretraining task, and one better matched to our setting---%
abstract physical visualizations rather than natural images---than the generic
pretext tasks inherited from computer vision.  We therefore adopted it as a
second pretraining stage, on top of the masked autoencoder that was by then
already trained.

The task is defined entirely by the dynamics, and the painter of
Fig.~\ref{fig:architecture}(a) supplies both halves of a training pair: a random
subset of $k \in [30,200]$ orbits rendered over a short window of
$t_{\mathrm{in}} \in [200,500]$ iterations is the input, and \emph{the same}
orbits, in the same colours, rendered over the full $T = 2500$ iterations of the
bank are the target.  Input and target are therefore spatially aligned and share
their colour assignment, which makes the task a well-posed dense
image-to-image regression: the sparse sample of each orbit in the input must be
completed into the full invariant set---a closed curve, or a chaotic
region---that the orbit covers at long times.
The input window is deliberately short relative to the target: capping it at a fifth of the full
length keeps the completion non-trivial.  Painting the pair online also has
a consequence here that it does not have downstream.  While for the segmentation
tasks every image drawn from a given map shares one ground truth, here the
target follows the subset drawn, so each of those combinatorially many subsets
carries a target of its own.  This is what makes the task usable for
pretraining.  A pretext task whose targets can be memorized eventually
saturates: past some point the network records answers instead of extracting
structure, and the representation stops improving.  No such point is reached
here---training and validation losses remain indistinguishable throughout
pretraining, and the error on the regular pixels keeps falling as a
power law of the epoch
(Appendix~\ref{app:completion}, Fig.~\ref{fig:completion_mae}).

A first advantage of the completion task compared to MAE pretraining lies in the fact that it uses the same UNETR architecture used for all downstream tasks.  A
masked autoencoder attaches its decoder to the final encoder layer alone, so the
pretraining loss reaches the intermediate blocks only through their effect on
that last representation.  The UNETR decoder instead draws on the ViT at one
quarter, one half, and three quarters of its depth as well as at its output, so
each of these intermediate representations receives gradient directly and is
encouraged to become useful for dense prediction in its own right---which is how
the downstream models will use it.  The completion decoder is itself discarded;
its role during pretraining is to induce those representations in the encoder.

The last and, arguably, most fundamental advantage of completion pretraining, compared to more standard computer
vision pretext tasks, is that it requires learning how the image is transformed
under the physical dynamics.  Standard approaches manufacture their supervision from a single image---hiding,
distorting, or rearranging part of it and asking the network to undo the
operation---so the resulting training pair holds no more than that one image
did.  Completion's pair holds more: producing its target required iterating the
map, and no operation on the input image could have supplied it.  The labels of the three
downstream tasks are themselves properties of the long-time dynamics---whether
an orbit is chaotic, which chain it belongs to, with what period---and must
likewise be inferred from a short sample of it.  It is therefore natural that
pretraining on completion should help them, and
Sec.~\ref{sec:results_pretraining} shows that it does.

How well completion itself is solved is shown in
Fig.~\ref{fig:architecture}(c,d).  As mentioned above, the task amounts to
discovering, for each displayed trajectory, the invariant set it explores from a
subset of the points sampled on it.  How hard this is depends on which points the dynamics
happens to supply. Empirically we distinguish three qualitatively different cases as outlined below.

The first case covers generic regular orbits with an irrational rotation
number.  In this case successive iterates hop around the invariant curve, and
after a few hundred iterations the orbit has left a sparse but essentially
complete dotted outline of the whole curve; completion then requires simple
interpolation.  This is the most common scenario for regular orbits, see
Fig.~\ref{fig:architecture}(c).

The second case covers a minority of regular orbits that advance only slowly
around their curve, either because the orbit lies close to a
separatrix, where the rotation number tends to zero, as in
Fig.~\ref{fig:architecture}(c), or because the map itself is weakly twisting, as
in three of the maps of Fig.~\ref{fig:architecture}(d).  These appear in the
input as isolated short arcs, and the invariant set has genuinely to be
extrapolated rather than merely filled in.  The network performs this
extrapolation well, recovering complete invariant curves, sometimes to
pixel-level accuracy, from arcs that cover only a fraction of them. 

The third case covers chaotic orbits.  One such orbit will eventually fill its
ergodic component. In a typical map, however, several different orbits explore the same region,
giving rise to a speckled pattern that is intrinsically unpredictable (see the
upper-left map of Fig.~\ref{fig:architecture}(d)).  The network can still reduce
the loss by learning that pattern in a statistical sense, and it does,
recovering the position and colour composition of the chaotic regions.  To
minimize the mean-square loss it can do no better than predict the conditional
mean at each such pixel, so these regions come out smoother than the
ground-truth speckle. Left alone, this irreducible component would dominate the
objective, chaotic pixels being numerous and their residual never falling.  We
therefore weight regular pixels above chaotic ones
(Appendix~\ref{app:completion}), which keeps the gradient on the part of the
image that can be predicted while still requiring the network to locate the
chaotic regions.  The target itself needs no labels, being obtained simply by
iterating the map further; the weighting does, so the stage is not strictly
self-supervised, though the supervision it uses is far weaker than a
segmentation label.

Quantitatively, after $1000$ epochs of pretraining the network reaches a mean
foreground-occupancy intersection over union of $0.74$ over the $997$ held-out
maps from a $500$-iteration input, a fivefold extrapolation, and $0.70$ from a
$200$-iteration input, a $12.5$-fold extrapolation.  The five maps of
Fig.~\ref{fig:architecture}(c,d) are drawn at the latter window and span $0.61$
to $0.79$, averaging $0.68$: selecting them for the harder regime has, if
anything, made the figure slightly less favourable than a random draw would
have been.  Splitting the error by region makes the account above explicit: the
regular contribution falls by more than a factor of two over training and is
still decreasing at the end, while the chaotic contribution is flat throughout
(Appendix~\ref{app:completion}).

Since completion pretraining is initialized from the masked autoencoder, one
may ask whether the latter is still needed.  A shorter completion run started with
a randomly initialized encoder shows that without the masked-autoencoder
stage the network needs $1.6$ to $2.7$ times as many epochs to reach the same
validation error on the regular pixels and the same foreground-occupancy IoU,
and the gap does not close over the $195$ epochs of that run
(Appendix~\ref{app:completion}, Fig.~\ref{fig:completion_mae}).  The
masked autoencoder thus speeds up completion pretraining, but this does not
establish that it is needed.  The run without it was still improving when we
stopped it, at a learning rate still close to its peak, and the downstream
gains of completion pretraining saturate beyond about $600$ epochs
(Sec.~\ref{sec:results_pretraining},
Table~\ref{tab:pretraining_avg_convention}), so that a completion run a few
times longer might reach the same downstream performance without the
masked-autoencoder stage.  Only such a run, followed by fine-tuning, can decide
whether the two pretext tasks are complementary or whether the masked
autoencoder merely accelerates the second.

\section{Training on downstream Tasks}
\label{sec:results}

\begin{figure*}
  \centering
  \includegraphics[width=\textwidth]{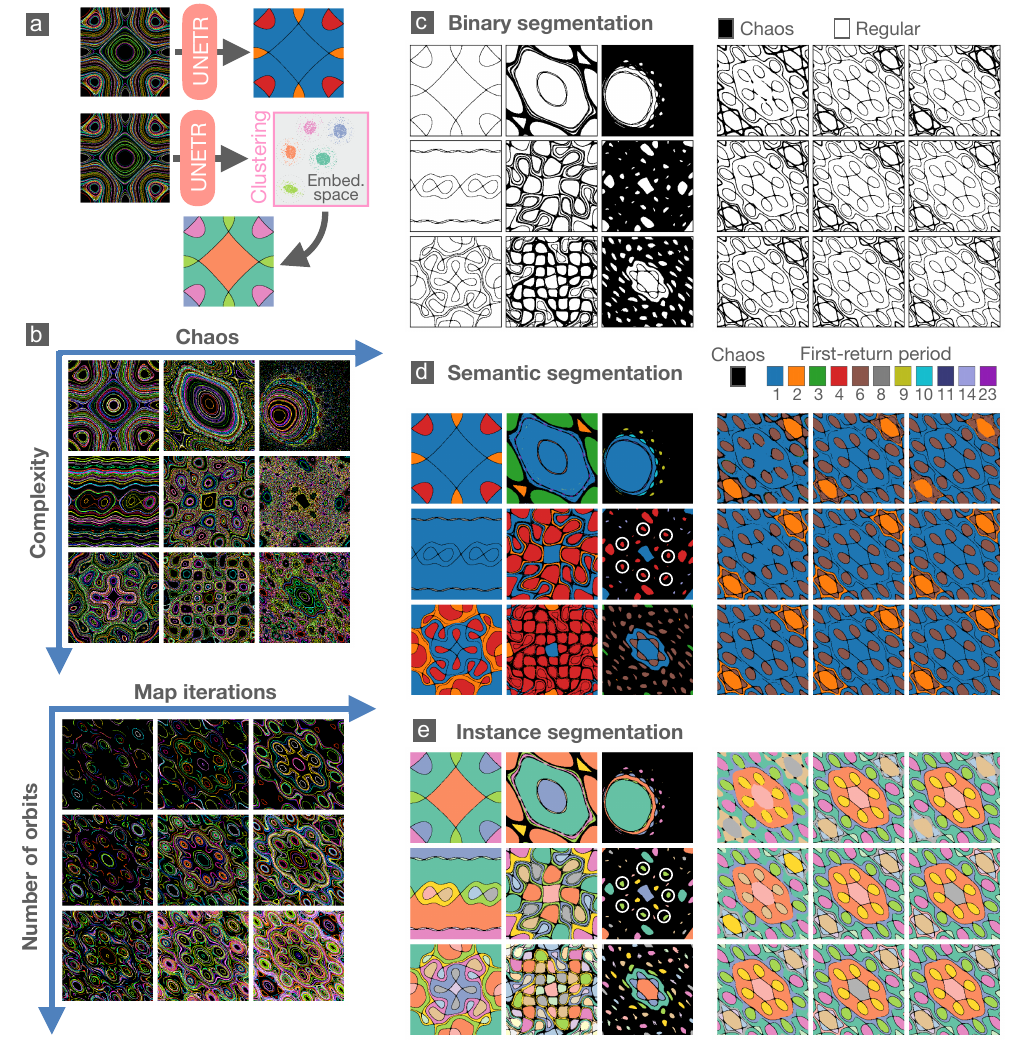}
  \caption{%
    Segmentation predictions on validation Poincar\'e maps.
    \textbf{(a)} Inference pipeline: semantic and binary labels are read
    directly off the decoder output (top), whereas instance labels are obtained
    by clustering the per-pixel embeddings in a post-processing step (bottom).
    Panels~(b)--(e) each show two $3\times3$ blocks.  \emph{Upper block:} nine different
    maps, ordered by increasing chaoticity (left to right) and increasing
    complexity (top to bottom), each painted with the same number of
    orbits and map iterations.  \emph{Lower block:} a single map, painted
    with an increasing number of orbits (top to bottom) and an increasing
    number of map iterations (left to right).
    \textbf{(b)} Input maps.
    \textbf{(c)} Binary segmentation: regular (white) vs.\ chaotic and
    separatrix (black).
    \textbf{(d)} Semantic segmentation: predicted first-return period, colour
    key inset. The two maps whose phase space is not doubly periodic carry the period map
    \emph{derived} from the predicted instance map
    (Appendix~\ref{app:derived}); every other cell
    is the semantic network's direct prediction.
    \textbf{(e)} Instance segmentation: predicted instances, coloured by
    mean-shift clustering of the discriminative embeddings
    (Appendix~\ref{app:instance_inference}).
    All panels show network predictions; no ground truth is displayed.
    The predictions on every validation map, with the ground truth and the per-map scores, are in the galleries of the data deposit~\cite{PVTData2026} and on the project page~\cite{PVTSite}.
  }
  \label{fig:segmentation_results}
\end{figure*}

The three segmentation tasks of Sec.~\ref{sec:tasks} are solved by fine-tuning
the pretrained network of Sec.~\ref{sec:architecture}.   As explained in that section, all three tasks keep the same
UNETR architecture, but only the ViT encoder is carried over from pretraining:
the decoder is task-specific (Fig.~\ref{fig:overview}c), reinitialized at random
and trained from scratch for each task.  For semantic and binary segmentation, the labels are predicted directly by the network, see Fig.~\ref{fig:segmentation_results}(a, top). For semantic segmentation,
the first-return period for periods up to $28$ is one-hot encoded in the $30$-channel decoder output, with one channel reserved for periods $\ge29$ and one channel for the chaotic pixels. Since for binary segmentation the two classes are complementary, we use a single channel in this case. For instance segmentation, different instances of the same class carry no order and thus cannot be tied to a channel. This motivated De~Brabandere \textit{et al.}~\cite{DeBrabandere2017} to introduce the approach we follow here: During training, the network learns to predict per-pixel embeddings that are shaped by a discriminative loss. During inference, the pixel embeddings produced by the network are assigned to the different
instances using a simple clustering algorithm, see Fig.~\ref{fig:segmentation_results}(a, bottom). We depart from that recipe in two places, each quantified in
Sec.~\ref{sec:debrabandere}: the forces
of the discriminative loss are re-weighted by a power of the instance area,
$w_c \propto N_c^{\,p}$, for which we use $p = 1/2$
(Appendix~\ref{app:finetuning}), and a second decoder head (Appendix~\ref{app:unetr_detail}) predicts
the foreground mask, so that a single forward pass returns both the mask and the embedding. 

\subsection{Qualitative results}
\label{sec:qualitative}

Figure~\ref{fig:segmentation_results}(b--e) shows visually how the three networks
perform on a selection of held-out maps and painting parameters.    The input images, subdivided into two blocks, are displayed in panel (b). In the
upper block, the maps themselves are varied while the number of trajectories
($k = 100$) and of map iterations ($T_{\mathrm{w}} = 500$) remain
constant.  Here, the maps are ordered by 
chaoticity  and complexity, as quantified by the
number of instances.  In the lower block, a single
map is held fixed and  the number of trajectories and iterations are varied assuming the values $k \in \{30, 80, 150\}$, and
$T_{\mathrm{w}} \in \{200, 500, 1000\}$.  The orbit counts and the two longer
windows lie inside the range used in training; $T_{\mathrm{w}} = 200$ falls just
below it, and is included deliberately to show how the prediction degrades when
the input is sparser than anything the network was trained on. The corresponding predictions are shown in panels (c--e). For most of the dense maps ($k = 100$ and $T_{\mathrm{w}} = 500$ or higher), the accuracy of the predictions  is so good that any disagreement from the GT (not shown) is hard to spot with the bare eye. While decreasing the plotting density, the quality degrades only slowly.

Compared to the natural scenes segmentation networks are usually
built for, our input images represent in several ways a fundamentally distinct challenge.  The objects are at once
numerous and small: the ten maps shown carry $906$  regular islands,
from $9$ on the simplest to $309$ on the most intricate, half of them under
$20$ pixels. For successful  instance segmentation, the network has not only to delineate each island, grouping neighbouring pixels into the same instance, but also to group islands scattered across the phase space into chains, comparing the orbit colours of distant pixels. A general-purpose segmentation model trained on natural images, Segment Anything~\cite{Kirillov2023}, fails at both steps on the maps of Fig.~\ref{fig:segmentation_results}(e): it misses most of the islands, delineates the ones it finds less precisely than our network, and cannot group islands that belong to the same chain (Appendix~\ref{app:sam}, Fig.~\ref{fig:sam}). Its predictions are also unstable: for a single map they change markedly from one painting to the next, whereas those of our network stay nearly the same across the nine paintings of the lower block [Fig.~\ref{fig:sam}(b)].

Grouping islands into chains is not only needed for instance segmentation: the semantic network, too, has to group them implicitly, since the first-return period of an island is the number of islands in its chain. Even in principle, such inference is only possible as long as all islands belonging to the same chain are visible in the displayed phase space region, which is only guaranteed for maps with doubly-periodic boundary conditions. For this reason, we train and deploy the semantic network only on such maps. Otherwise, we deduce the semantic segmentation from the instance segmentation and short iterations of the system dynamics. The derived route is more reliable in determining the return-period while the direct route excels in delineating small islands but can sometimes miscount them: see, e.g., the seven islands circled, which belong to the same chain but are mislabelled as a four-fold resonance. Counterintuitively, the same semantic network successfully ``counts'' 23 tiny islands ($\sim 15$ pixels on average): in panel~(d), first row, second column, they are the violet chain threading the rim of the large central island. This is remarkable also considering the rarity of this class, with  $0.15\%$ of the instances having period 23 or one in $56$ maps (extrapolated from the validation dataset).

\subsection{Quantitative results}
\label{sec:quantitative}

\begin{figure*}
  \centering
  \includegraphics[width=\textwidth]{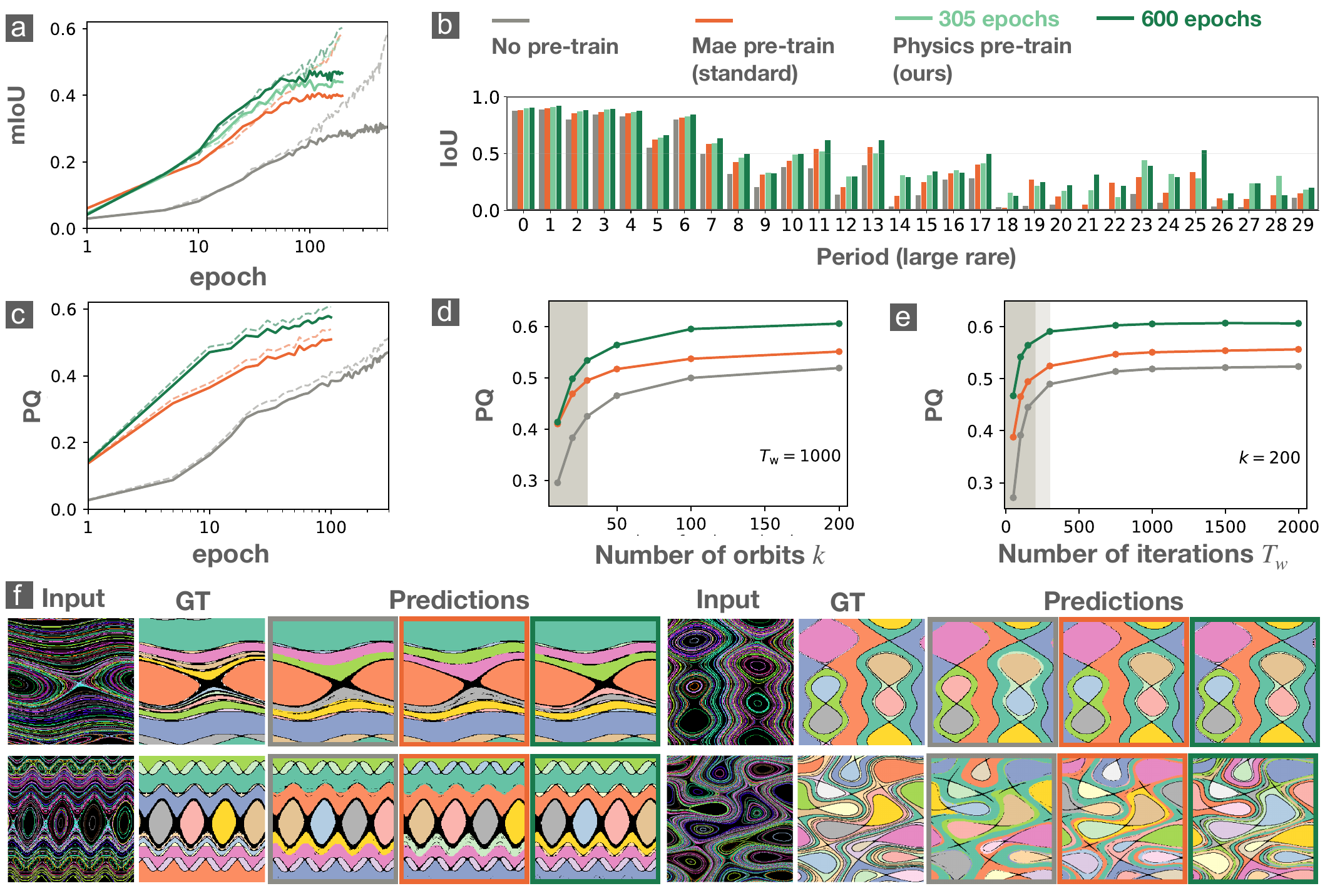}
  \caption{
    Effect of the pretraining task on two downstream tasks.
    Colours identify the pretraining arm throughout: no pretraining (grey),
    standard MAE pretraining (orange), and MAE followed by long-time map
    completion (dark green; in a and b light green uses an earlier checkpoint of the same completion run). All arms share the same
    ViT-L encoder and UNETR decoder.
    \textbf{(a,b)} Semantic segmentation. \textbf{(a)} mIoU  versus fine-tuning epoch
    (solid: validation; dashed: training).
    \textbf{(b)} Per-class validation IoU at the final epoch, with the classes
    ordered by increasing return-period; the larger periods to the right are also the
    rarer ones.
    \textbf{(c--f)} Instance segmentation. \textbf{(c)} Panoptic quality PQ versus fine-tuning epoch (solid: validation; dashed: training).
    The clustering of the embeddings (Appendix~\ref{app:instance_inference}) uses the fast mean shift of Ref.~\cite{DeBrabandere2017}, so that the values lie $0.03$
    to $0.04$ below a mean-shift decoding of the same painting. Panels (a)--(c) are evaluated using  the single painting parameters
    $k = 150$, $T_{\mathrm{w}} = 1000$.
    \textbf{(d,e)} Robustness of the final instance-segmentation models to
    inputs painted outside the training range, quantified by the validation panoptic quality
    $\mathrm{PQ}$ of the full mean-shift clustering: \textbf{(d)} versus the number of
    orbits~$k$ at fixed window length $T_{\mathrm{w}}=1000$, and
    \textbf{(e)} versus the window length~$T_{\mathrm{w}}$ at fixed $k=200$.
    Shading marks paintings outside the fine-tuning range.  In (e) the lighter band,
    $200 \le T_{\mathrm{w}} < 300$, was shown  during both pretrainings but not during fine-tuning, whose floor is
    $T_{\mathrm{w}} = 300$. 
    \textbf{(f)} Instance-segmentation predictions on four validation maps:
    input, ground truth, and the prediction of each arm, framed in the
    corresponding colour.  Three are chosen near the median gain in PQ
    from no pretraining to map completion among the validation maps with $10$ to $30$
    instances; the fourth (bottom right) is a strong case, at the $95$th
    percentile. The predictions  on every validation map are in the pretraining-comparison gallery of the data deposit~\cite{PVTData2026} and on the project page~\cite{PVTSite}. }
  \label{fig:pretraining_ablation}
\end{figure*}

We now turn from the qualitative survey of
Fig.~\ref{fig:segmentation_results} to figures of merit, and use them for ablation studies with the main focus on the effects of the different pretraining stages. 
Two caveats should be stated
before any number is quoted.  First, the ground truth is itself imperfect. By inspecting the validation dataset, we have observed that sometimes  two large regular regions showing qualitatively distinct orbits have been merged into a single instance during the data generation pipeline.  In this case, the network  correctly distinguishes the different instances, but will be penalized in the scoring.
In addition, some
disagreements are not errors of either but an intrinsic ambiguity of the image:
a map too sparsely drawn may not show any trace of a narrow resonance that is present in the GT, which is calculated using a dense map. Appendix~\ref{app:failure_modes} shows representative  cases of these two types of disagreement. 
Second, the maps in our dataset are very diverse in terms of geometry and composition, as testified by the number of instances  varying from $0$ (the $17$ fully chaotic maps among the $995$ of the clean validation set) to $72$, with a median of $7$ and a mean of $10$. Hence, any headline figure will depend strongly on the dataset composition. Even after fixing the dataset, every absolute score depends  on how densely the map was painted.  The
same instance-segmentation model scores a panoptic quality (PQ, defined below) of $0.605$ when evaluated
at the densest painting we use ($k = 150$, $T_{\mathrm{w}} = 1000$) and $0.570$
when averaged over the six paintings of the protocol described below. 
A headline number is therefore a statement about the dataset composition and the validation protocol as much
as about a network.
What survives this is the comparison.  Panels (d) and (e) of
Fig.~\ref{fig:pretraining_ablation} sweep the two painting parameters across
the training range and show the arms degrading in parallel: their ordering
holds throughout and their separation changes little, so neither depends on
where in that range a number is quoted.  We therefore fix one protocol for
all tables and read the numbers as relative rather than absolute. Every
network is scored on six paintings of each validation map, $k \in \{30, 80, 150\}$
orbits by $T_{\mathrm{w}} \in \{500, 1000\}$ iterations, corresponding to the cells of the lower
block of Fig.~\ref{fig:segmentation_results}(b) that lie within the fine-tuning
range, $T_{\mathrm{w}} \geq 300$. A full overview of the network performance on the training distribution, beyond the selected examples shown in the paper, can be gained by inspecting the network predictions on the full validation dataset, which are available, along with the corresponding GT segmentations, as html galleries in the data deposit~\cite{PVTData2026} and on the project page~\cite{PVTSite}.

\subsubsection{Effect of the pretraining task}
\label{sec:results_pretraining}

To analyze the effect of pretraining, we compare  three fine-tuning arms that share the same architecture, data and loss, and differ in where the encoder
weights come from: random initialisation,
masked-autoencoder pretraining (Sec.~\ref{sec:architecture}), and MAE followed
by the long-time map completion task of Sec.~\ref{sec:completion}.  Each is
fine-tuned separately on each of the three downstream tasks.  The schedules are
the one deliberate exception: the  arm without pretraining converges more slowly and is
given two to three times the fine-tuning epochs of the other two on every task,
so the comparison is if anything conservative towards it.
The outcome is consistent across all three, and is summarised in
Fig.~\ref{fig:pretraining_ablation} and Tables~\ref{tab:pretraining_binary},
\ref{tab:pretraining_avg_convention} and \ref{tab:pretraining_instance}: pretrained
encoders learn faster [Fig.~\ref{fig:pretraining_ablation}(a,c)] and converge
to better representations, with the completion stage ahead of MAE alone on
every task and every figure of merit we report.  The completion-pretrained
model also produces visibly cleaner segmentations
[Fig.~\ref{fig:pretraining_ablation}(f)].  Where the three tasks differ is in
how the gain is divided between the two pretraining stages, and that division
is itself informative.

\emph{Binary segmentation.}  Here the network separates regular from chaotic
and separatrix pixels.  It is trained on a cross-entropy together with a
per-image Jaccard term on the regular class (Appendix~\ref{app:finetuning}),
and Table~\ref{tab:pretraining_binary} reports two complementary figures of
merit.

The first is the intersection over union of the \emph{regular} class, the
standard overlap measure of image segmentation,
\begin{equation}
  \mathrm{IoU} = \frac{\mathrm{TP}}{\mathrm{TP} + \mathrm{FP} + \mathrm{FN}}\, .
  \label{eq:iou}
\end{equation}
In this case the true positives $\mathrm{TP}$ are the pixels that the network
and the ground truth both classify as regular, the false positives $\mathrm{FP}$
those the network classifies as regular and the ground truth as chaotic or
separatrix, and the false negatives $\mathrm{FN}$ the reverse.  The score is
formed separately for each map and then averaged over maps, so that every
portrait counts equally (\emph{per-map} IoU); the Jaccard term of the loss is one minus this same
quantity, formed per image.  A perfect segmentation scores one, and a given number
of misclassified pixels costs the more, the smaller the denominator is.  The measure is therefore at its most
discriminating on strongly chaotic maps, where a few small islands are the
whole of the regular class and the denominator is correspondingly small.  It is
at its least in the opposite case, broad regular regions divided by thin
separatrices: there the denominator is large, and the network can still score
well while misclassifying the separatrix pixels as regular.

To measure the network's ability to detect the separatrices we use a second
figure of merit, the fraction of the separatrix pixels it classifies
correctly.  During fine-tuning we incentivize the network to detect such pixels by applying a weight map to the cross-entropy, giving separatrix
pixels four times the weight of other pixels.  
The two pretraining stages contribute very unequally.  MAE pretraining gains
$0.007$ in IoU and $0.011$ in separatrix recall over the arm without
pretraining, while adding the completion stage gains a further $0.019$ and
$0.045$---roughly three and four times as much.  The asymmetry is sharper on
the separatrices than on the areas, which is consistent with the thin
structures being the harder half of the task.  Reconstructing masked patches of
a Poincar\'e map evidently does not, by itself, teach the regular--chaotic
distinction; completing a short arc of an orbit into a long one does, because a
network that can do the latter must already have made that distinction.  The
training losses say the same: at the single painting its own validation uses,
the MAE arm's cross-entropy is in fact slightly \emph{worse} than that of the
arm without pretraining, $0.257$ against $0.232$.

\begin{table}[t]
  \caption{
    Effect of the pretraining task on downstream \emph{binary} segmentation.
    The per-map IoU is that of the
    regular class, with chaotic and separatrix pixels treated as background;
    separatrix recall is the fraction of correctly assigned ground-truth separatrix pixels. The arm without pretraining was trained for $300$ epochs against $150$ for the pretrained ones. Details in Appendix~\ref{app:finetuning}.
    }
  \label{tab:pretraining_binary}
  \begin{ruledtabular}
  \begin{tabular}{lcc}
    Pretraining & per-map IoU & separatrix recall \\
    \hline
    None      & 0.855 & 0.841 \\
    MAE                      & 0.862 & 0.852 \\
    Map completion (1000 ep) & \textbf{0.882} & \textbf{0.897} \\
  \end{tabular}
  \end{ruledtabular}
\end{table}

\emph{Semantic segmentation.}  The first-return period is predicted directly,
with a cross-entropy and a Jaccard term over the $30$ period classes.  Two
averaging conventions are in use here, each defining a different figure of
merit.  The \emph{mIoU}, the standard figure of merit of multi-class semantic-segmentation
benchmarks, computes the IoU of each class from the pixel counts pooled over the
whole validation set and averages it over the $30$ classes.  Every return period thus has an equal
weight, including those appearing in only a handful of maps, so the score captures
how well the network classifies the rare ones.  The \emph{per-map mIoU} is instead
the mIoU of each single map, taken over the classes present in it, averaged over
the validation maps: every map counts
equally, and the score reports the quality of the segmentation on a typical map.
Table~\ref{tab:pretraining_avg_convention} reports both.
The two conventions also separate the two halves of the loss: the cross-entropy
is a mean over pixels and is therefore dominated by the abundant classes, while
the Jaccard term is one minus a soft, batched estimate of the mIoU
(Appendix~\ref{app:finetuning}), which is what supplies a usable gradient for
the rare high periods.
Completion pretraining ($600$ epochs, the encoder of the semantic network) raises the per-map mIoU by $20\%$ over the arm without pretraining and has an even stronger impact on the mIoU, raising it  by $62\%$.  That the second gain is three times the first reflects the extreme class imbalance of our dataset ---the chaotic class together with period one accounts for
$87.5\%$ of all labelled pixels, $23$ of the $30$ classes share $0.5\%$ between
them, and a typical map contains only $3.7$ distinct periods--- as well as the  ability, acquired by the network during the completion pretraining stage,  to generalize from rare examples.  This claim is further substantiated by the
per-class breakdown of Fig.~\ref{fig:pretraining_ablation}(b) showing  much larger gains
for rare, high-period classes in the right half of the
plot, than for  the common low periods in the left half.  All three arms
reach a similar score on the training maps, $0.58$ to $0.60$ under this
convention, so what separates them is not how well they fit the rare classes
but how much of that transfers to maps they have not seen.

\begin{table}[t]
  \caption{Effect of the pretraining on semantic segmentation. We report both mIoU and per-map mIoU.  The three completion rows are checkpoints of one
    completion pretraining run, taken after $305$, $600$ and $1000$ epochs.  The
     arm without pretraining was trained for $500$ epochs against $195$ for the
    pretrained ones.  The relative gains are about three times larger in mIoU than in
    per-map mIoU, because it gives the rare high-period classes the same weight as
    the abundant ones. Details in Appendix~\ref{app:finetuning}.}
  \label{tab:pretraining_avg_convention}
  \begin{ruledtabular}
  {\begin{tabular}{lcc}
    Pretraining & per-map mIoU & mIoU \\
    \hline
    None                   & 0.526 & 0.274 \\
    MAE                                  & 0.588 & 0.365 \\
    Map completion ($305$ ep) & 0.608 & 0.407 \\
    Map completion ($600$ ep) & 0.631 & 0.443 \\
    Map completion ($1000$ ep) & \textbf{0.640} & \textbf{0.447} \\
  \end{tabular}}
  \end{ruledtabular}
\end{table}

The length of the completion pretraining matters as well
[Fig.~\ref{fig:pretraining_ablation}(a,b)]: comparing separate fine-tuning runs, starting from  two checkpoints of the same
completion run taken after $305$ and $600$ epochs, the per-map mIoU and the mIoU
improve by $4\%$ and $9\%$, most of the gain falling on the rare high-period
classes, while starting from the final checkpoint, after $1000$ epochs, adds only about $1\%$
(Table~\ref{tab:pretraining_avg_convention}).

\emph{Instance segmentation.}  Our headline figure of merit for instance
segmentation is the panoptic quality (PQ)~\cite{KirillovPanoptic2019}, the
standard measure when, as here, every pixel is labelled; two further figures of
merit, the mean best IoU and the adjusted Rand index, are reported in
Appendix~\ref{app:other_fom}.
The panoptic quality can be expressed as the product $\mathrm{PQ} = \mathrm{SQ} \times \mathrm{RQ}$ of two figures of merit to isolate different qualities: the recognition quality
(RQ), which measures how well the ground-truth instances are
detected---penalizing both missed and spurious ones---and the segmentation
quality (SQ), which measures how accurately the detected instances are
delineated (Appendix~\ref{app:pq}).
Figure~\ref{fig:pretraining_ablation}(c--f) and
Table~\ref{tab:pretraining_instance} tell a simple story.  First, map
completion is ahead of MAE alone, and MAE ahead of the arm without pretraining,
across the board: throughout fine-tuning
[Fig.~\ref{fig:pretraining_ablation}(c)], at every painting of the input
[Fig.~\ref{fig:pretraining_ablation}(d,e)], and under every figure of merit of
the table.  Second, the gaps barely depend on how densely the input is
painted, as long as the painting stays within the range seen during
fine-tuning, $k \in [30,200]$ orbits and $T_{\mathrm{w}} \in [300,2000]$
iterations (the unshaded regions of
Fig.~\ref{fig:pretraining_ablation}(d,e)).
Third, pretraining improves both factors of PQ: going from no pretraining to map completion,
the recognition quality RQ rises by $17.3\%$ and the segmentation quality SQ by $3.9\%$.  Both gains can be seen with the bare eye in
Fig.~\ref{fig:pretraining_ablation}(f): the map-completion predictions are
visibly sharper, resolving the thin bands between islands, and the separatrix
layers that bound them, where the network without pretraining merges them into broad
regions.  The smaller gain in SQ, if anything, understates the improvement in
delineation, since the instances map completion detects in addition are
predominantly small and, thus,  harder to delineate
(Appendix~\ref{app:pq}).

\begin{table}[t]
  \caption{Effect of the pretraining task on downstream \emph{instance} segmentation. We report the panoptic quality $\mathrm{PQ} = \mathrm{SQ} \times \mathrm{RQ}$ as well as separate figures for the segmentation quality SQ and 
     the recognition quality RQ.  The PQ gain from the MAE pretraining is $11\%$, and completion pretraining adds a further $10\%$.  mBIoU and ARI for the same
    networks are given in Table~\ref{tab:other_fom}.  Details in Appendix~\ref{app:finetuning}.}
  \label{tab:pretraining_instance}
  \begin{ruledtabular}
  {
  \begin{tabular}{lccc}
    Pretraining & PQ & RQ & SQ \\
    \hline
    None       & 0.468 & 0.566 & 0.827 \\
    MAE                      & 0.519 & 0.614 & 0.845 \\
    Map completion (600 ep)   & \textbf{0.570} & \textbf{0.663} & \textbf{0.860} \\
  \end{tabular}}
  \end{ruledtabular}
\end{table}

\subsubsection{Secondary ablation for the instance segmentation task}
\label{sec:debrabandere}

In Table~\ref{tab:debrabandere}, we analyze the effects of two choices of our instance segmentation fine-tuning recipe that depart from the approach of
Ref.~\cite{DeBrabandere2017}.
The first departure is the size weighting of the discriminative loss.
Ref.~\cite{DeBrabandere2017} weights every instance equally, $p = 0$, a natural
choice for natural images, whose instances are of broadly comparable extent and
of equal interest.  The instances of a Poincar\'e map instead span orders of
magnitude in area, from a single region filling much of the window to chains of
islands a few dozen pixels across, and a small chain may not even be visited by
the orbits of a given painting, so that its full share of the gradient could
only be supplied by memorizing the map (Appendix~\ref{app:finetuning}).  The
two extremes have a direct analogue in the semantic task.  With $p = 1$ the
variance term of Eq.~\eqref{eq:disc} is a mean over all foreground pixels, so
that every instance counts in proportion to its area, as every class does in a
cross-entropy; with $p = 0$ every instance counts equally, as every class does in
the Jaccard term.  There we use the two terms together---the cross-entropy so that the
large structures are accounted for, the Jaccard term so that the small ones are
not disregarded---and here the intermediate $p = 1/2$ plays the same role of a
compromise.
Empirically [Table~\ref{tab:debrabandere}(a); Appendix~\ref{app:sizeweight}], area
weighting is the worst choice: it overfits the most and ends lowest in PQ.
The $p = 1/2$ and $p = 0$ networks perform comparably well. We use $p = 1/2$ because it is marginally better in PQ, with a small lead  of
$0.004$ that exceeds the observed run-to-run variation, below $0.001$.

The second departure is the prediction of the  foreground mask using a second head of the same network that predicts the embeddings. Instead,  Ref.~\cite{DeBrabandere2017} trains the network to predict only the embeddings, leaving the prediction of unknown foreground masks to an external network. Our choice is clearly more efficient as it requires a single network. In PQ it trails the two-network approach by $0.005$ [Table~\ref{tab:debrabandere}(b)], whereas it is marginally ahead in mBIoU and ARI (Table~\ref{tab:other_fom}), so the two perform comparably. Comparing with a pipeline using the GT foreground shows that the foreground prediction, rather than the grouping of regular pixels into instances, is the dominant source of error, see Appendix~\ref{app:fghead} for more details.

\begin{table}[t]
  \caption{Effects of the 
    departures from the instance-segmentation recipe of De Brabandere
    \textit{et al.}~\cite{DeBrabandere2017}, whose choices are marked DB.  Each
    block varies one component with everything else fixed.  (a)~Size weighting
    of the discriminative loss, Eq.~\eqref{eq:disc}, $w_c \propto N_c^{\,p}$ with $N_c$  the instance area.
    (b)~Origin of the foreground mask: a separately trained
    segmentation network (our binary model) or the second head of the instance
    network; the last two rows give both networks the ground-truth foreground.  All networks start from the $600$-epoch
    completion-pretrained encoder.   Best value within each block in bold,
    ground-truth rows excluded; mBIoU and ARI for the same networks are given
    in Table~\ref{tab:other_fom}.}
  \label{tab:debrabandere}
  \label{tab:sizeweight}
  \begin{ruledtabular}
  {\begin{tabular}{llccc}
    & Variant & PQ & RQ & SQ \\
    \hline
    \multicolumn{5}{l}{(a) size weighting, $w_c \propto N_c^{\,p}$} \\
    & $p = 0$ (DB)   & 0.566 & 0.662 & 0.855 \\
    & $p = 1/2$      & \textbf{0.570} & \textbf{0.663} & 0.860 \\
    & $p = 1$        & 0.557 & 0.646 & \textbf{0.861} \\
    \multicolumn{5}{l}{(b) foreground mask ($p = 1$)} \\
    & separate network (DB) & \textbf{0.562} & \textbf{0.648} & \textbf{0.867} \\
    & second head    & 0.557 & 0.646 & 0.861 \\
    & ground truth, no head      & 0.706 & 0.762 & 0.926 \\
    & ground truth, second head  & 0.716 & 0.770 & 0.930 \\
  \end{tabular}}
  \end{ruledtabular}
\end{table}
\subsubsection{Inference time}
\label{sec:inference_time}

All our figures of merit are computed with the
embeddings clustered by mean shift (Appendix~\ref{app:instance_inference}).  On
a laptop (Apple M4 Pro; the network on its GPU with MLX, mean shift on a single
CPU core with scikit-learn), painting an input takes $2$~ms, a forward pass of
the ViT-L UNETR $87$~ms ($71$~ms per map in batches of eight, where the throughput saturates), and mean-shift
decoding $0.20$~s on average, so that a map is segmented into instances in
about $0.3$~s; binary and semantic segmentation need no decoding and cost a
single forward pass.  We have also implemented the fast mean shift
of Ref.~\cite{DeBrabandere2017}, which extracts one cluster at a time: it
reduces the decoding time by $83\%$, to $35$~ms, but lowers PQ by $5\%$, from
$0.570$ to $0.542$, because it produces almost twice as many false
positives.  All networks were trained on single AMD Instinct MI300A
accelerators of the Viper supercomputer at the Max Planck Computing and Data
Facility, at about nine minutes per fine-tuning epoch.

\section{Deployment on Real Maps}
\label{sec:deployment}

\begin{figure*}
  \centering
  \includegraphics[width=\textwidth]{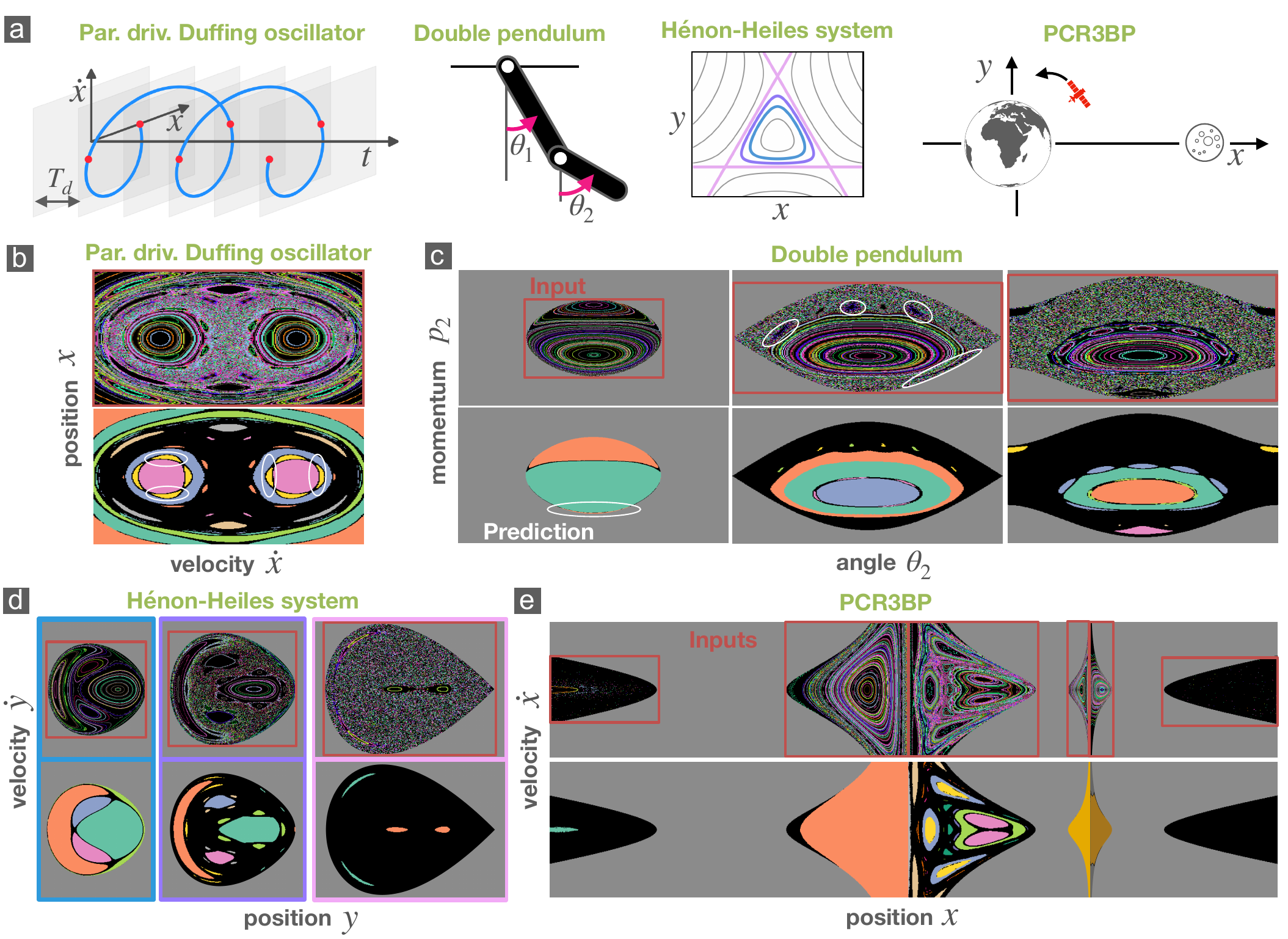}
  \caption{
    Zero-shot deployment on four out-of-distribution real-case scenarios.  Every
    prediction comes from the instance network of
    Fig.~\ref{fig:segmentation_results} with no change of weights, and the embeddings are clustered
    with the two-stage variant of Appendix~\ref{app:instance_inference}.  Equations,
    section conventions and parameters are given in
    Appendix~\ref{app:continuous_maps}.
    \textbf{(a)}~Geometry of the four systems, in panel order: the stroboscopic
    section of the driven Duffing oscillator in the extended phase space, with an iteration recorded every driving period $T_{\mathrm{d}}$; the double
    pendulum; the
    H\'enon--Heiles equipotentials, with the three energies of (d) in colour;
    and the planar circular restricted three-body problem illustrated by a satellite moving in the
    frame co-rotating with the circular orbits of the Earth and Moon about their centre of mass.
    \textbf{(b)}~Driven Duffing oscillator, the one system with no conserved
    quantity and hence no inaccessible region. The white ellipses  mark four mislabelled islands forming an instance on
    their own but merged with four islands belonging to a different one (see
    text).
    \textbf{(c)}~Double pendulum at three energies spanning the
    libration-to-rotation threshold, where the accessible domain turns from a
    bounded lens into the full cylinder.
    \textbf{(d)}~H\'enon--Heiles at three energies across the
    regular-to-chaotic transition, the accessible region growing and becoming
    teardrop-shaped; frame colours as in (a).
    \textbf{(e)}~Planar circular restricted three-body problem, whose
    accessible region is disconnected and is covered by six windows, segmented
    independently and glued for display.  
    In all panels, the red rectangles outline the
    crops rescaled onto the network's $256 \times 256$ input and grey  marks the excluded region.
  }
  \label{fig:ood_deployment}
\end{figure*}

Until now we have trained and evaluated our network on discrete symplectic maps drawn from a somewhat arbitrary distribution, that we have designed combining a small number of parametrized closed-form transformations. This leaves open the question whether our network is a real foundational model able to generalize to any Poincaré map, irrespective of the physical system that generated it.  In this section, we test this hypothesis by deploying  our network on four
continuous-time systems, drawn from four different corners of physics, see Fig.~\ref{fig:ood_deployment}(a).
The first system, a parametrically driven Duffing oscillator~\cite{Kovacic2011}, demonstrates the application of our approach to weakly nonlinear driven resonators---realized in devices from nanomechanical beams~\cite{Mahboob2008} to superconducting circuits~\cite{Grimm2020}---and, more generally, stroboscopic maps.  For this type of maps, the phase of the driving $\phi$
is viewed as an additional coordinate, and the Poincaré map for the section  $\phi = 0$  is obtained by taking a snapshot after every driving period $T_{\mathrm{d}}$.
The three remaining case studies are autonomous systems with two degrees of freedom and different phase space topologies: the double pendulum~\cite{Shinbrot1992} is the textbook example of chaos in an everyday mechanical system and its map is defined on a cylinder; the
H\'enon--Heiles system~\cite{HenonHeiles1964} is one of the founding examples of Hamiltonian chaos; and the planar
circular restricted three-body problem~\cite{Szebehely1967} is the system in which Poincar\'e found
the homoclinic tangle~\cite{Poincare1890}, and remains the basis of spacecraft trajectory design~\cite{Koon2008}.  Unlike the training maps, the maps of these four systems are obtained by integrating the continuous dynamics numerically and recording the intersections of the trajectories with the Poincaré section (see Appendix~\ref{app:continuous_maps} for details on our implementation).

The maps shown to the network differ from the training distribution not only in how they are generated but, more strikingly, in their qualitative appearance.  Except for the driven Duffing oscillator, a constant of motion confines the orbits to an accessible region of the section, the Hill region, whose shape is fixed by the value of the conserved quantity and has nothing in common with the square, cylinder or torus covered by the training maps, see Fig.~\ref{fig:ood_deployment}(c)--(e); in the three-body problem it is not even connected.  We nevertheless deploy the network with no change of weights and with the inference pipeline of Sec.~\ref{sec:results}, with two adaptations.  The first is a pre-processing step that accommodates any scaling of the section coordinates and any shape of the accessible region: the section is cropped to the bounding box of the accessible region---the red rectangles of Fig.~\ref{fig:ood_deployment}---and rescaled onto the network's square input.  The inaccessible part of the bounding box is left empty, so that in the image the network receives nothing distinguishes a pixel outside the accessible region from one inside it that no orbit happened to visit.  The accessible region enters only after the forward pass, to restrict the predicted foreground before the embeddings are clustered (Appendix~\ref{app:pipeline}).  The second adaptation is a refinement of this clustering: an additional mean-shift pass, seeded by the pixels left unassigned by the first, prevents the accidental merging of well-resolved clusters (Appendix~\ref{app:instance_inference}).

A pixel-level ground truth for these maps would require running the labelling pipeline of Sec.~\ref{sec:data} on a flow rather than on a closed-form map, at an estimated cost of about eight hours per H\'enon--Heiles section on a laptop GPU (Appendix~\ref{app:continuous_maps}) against well under a second for one forward pass.  Instead, we check whether the predicted instances are consistent with the dynamics of test trajectories launched from each instance. Possible failure modes like splitting or merging of instances  leave their own signature in this check, and the same orbits can be used to repair the prediction (Appendices~\ref{app:ood_verification} and~\ref{app:ood_repair}).

The network generalizes to all four systems, including maps that are qualitatively different from any it has seen in training: $42$ of the $60$ testable instances pass the check, and no chain is split between instances.  Nearly all the remaining errors concern small or thin islands, which the network either misses, as in the double pendulum at $E=-1$ [Fig.~\ref{fig:ood_deployment}(c) middle], where each of two mirror-image chains of period four (one of them is marked in the figure) loses one island, or labels only in pieces, as for the light-green regular band of the H\'enon--Heiles system at $E=1/12$ [Fig.~\ref{fig:ood_deployment}(d) left], which runs along the whole edge of the accessible region and becomes as thin as a pixel. The missed islands are predicted as background rather than assigned to another instance, and the orbit-based repair recovers most of them.  The one error of a different kind is a merger of two chains of period four, one of which is marked by white ellipses in Fig.~\ref{fig:ood_deployment}(b): the two chains are images of one another under the symmetry $(x,\dot{x})\to(-x,-\dot{x})$ and interleaved around a single ring, so that only the colouring of the orbits distinguishes them from one chain of period eight; the repair splits them.  An example of each kind of error is shown in Fig.~\ref{fig:ood_failures}, and all the counts are given in Appendix~\ref{app:ood_verification}, where Table~\ref{tab:ood_failures} lists every instance that fails the check. We emphasize that this performance extends to the maps with extended inaccessible regions.  In every training map the phase space is the whole square, cylinder or torus, whereas here the orbits are confined to a region of arbitrary shape, whose edge is not marked in the input.  Correct predictions on maps with a feature so different from the training distribution go beyond interpolation between training examples, and show that the network is capable of non-trivial generalization.

Finally, the network handles correctly a feature that we did not anticipate.  Poincar\'e sections are normally chosen transverse to the flow, but in the double pendulum at the lowest energy [Fig.~\ref{fig:ood_deployment}(c) left] the flow becomes tangent to the section $\theta_1=0$ along the boundary of the accessible region.  Since only the crossings with a fixed sign of the velocity, here $\dot{\theta}_1>0$, are recorded, the section of a torus that reaches this boundary is cut into disjoint arcs, and a single regular region appears as several unconnected parts.  The network assigns the thin band along the lower edge of the accessible region to the same instance as the large region above it.  We first took the band for a spurious prediction, but orbits launched inside it spend nearly all of their iterates in the large region above and are regular (their MEGNO indicator tends to $2$): the band belongs to the same family of invariant curves, and the prediction is correct.

To verify the contribution of our completion pretraining to the generalization capabilities of our network, we repeated the deployment with the networks fine-tuned
without pretraining and after MAE pretraining only, on the same examples
(Fig.~\ref{fig:ood_appendix} of Appendix~\ref{app:ood_arms}).  Without map-completion
pretraining, the predictions are qualitatively worse: the networks merge instances that do
not belong together, fail to detect some of the mid-sized islands, and produce less sharp
predictions. In the check described above, 42 of the 60 testable instances of the network with map-completion pretraining pass, against 33 of 53 after MAE pretraining only and 29 of 53 without pretraining (Table~\ref{tab:ood_verification} of Appendix~\ref{app:ood_arms}).

\section{Conclusion and outlook}
\label{sec:conclusion}

We have introduced the Poincar\'e Vision Transformer, a model that reads
Poincar\'e maps the way a physicist does.  Recasting the expert analysis of a
map as binary, instance and semantic segmentation, we trained a single
vision-transformer backbone on synthetic symplectic maps, and introduced a
physics-inspired pretraining task, long-time map completion, that outperforms
both training without pretraining and masked-autoencoder pretraining on all
three tasks.  Its advantage lies in what the training pair contains: masked
autoencoding manufactures its supervision from a single image, whereas
completion asks for what only the dynamics can supply---where each orbit goes
next---and, since every subset of orbits carries a target of its own, it can be
trained for as long as compute allows.  The resulting model segments a map in a fraction of a second and transfers,
without retraining, to Poincar\'e sections of physical systems whose maps
differ qualitatively from everything seen in training: they are obtained by
integrating a continuous flow rather than iterating a closed-form map, and may be
confined to accessible regions of arbitrary shape.
We therefore expect it to apply to two-dimensional Poincar\'e maps in
general, irrespective of the underlying physical system and, owing to
the self-similarity of phase space (Sec.~\ref{sec:tasks}), at any scale.  Such generality is rarely achieved in
supervised learning, where a network's reach usually ends at the edge of its
training distribution.

The most direct application is the one that motivated this work.  AI-driven
research workflows already plan experiments and simulations and steer the
exploration of parameter
space~\cite{Boiko2023,Naegele2026,Vishwasrao2026,Somasekharan2026}, but they
rely predominantly on numerical analysis of the outcome, and extract far less
from a plot than a physicist does.  Our model can serve such
an agent as an eye: called on a freshly computed section, it returns the regular
and chaotic regions, the island chains and their return-periods, in a form that a
language model can reason about.  Its predictions can moreover be checked with
the dynamics that generated the map, at negligible cost---one short trajectory
per island, as in Sec.~\ref{sec:deployment}---so that the model can run
unattended without being trusted blindly. 

At a fraction of a second per map, this makes
systematic surveys of phase-space structure across parameter space practical,
from the dynamic aperture of accelerator lattices~\cite{Papaphilippou2014} to
the stability of planetary systems~\cite{Laskar1989}.  Map completion is useful
in the same setting wherever trajectories are expensive, as in field-line
tracing through a tokamak equilibrium~\cite{Boozer2004}, or impossible to
extend, as for sections reconstructed from experimental data.

A single map can also seed a survey that reaches beyond it.  Every island chain
found by the network surrounds a stable periodic orbit of the flow, which
crosses the section once in each island before closing on itself, and its
islands are separated by the crossings of the unstable periodic orbit that
generates the separatrix.  For both, the segmentation therefore supplies an
initial condition on the section and the number of crossings after which the
orbit closes---the first-return period of the chain---and following a
trajectory for that many crossings estimates its period in time.  These are the
guesses that gradient-based searches for periodic orbits
require~\cite{Suvakov2013,LiLiao2017,Page2024}: one minimizes, over the initial condition
and the period, the distance between the start and the end of the trajectory,
with gradients that a differentiable integrator supplies directly.  Run blindly,
from initial conditions sampled at random, such searches mostly find unstable
orbits, whose stability has to be established after the fact, and do not reveal
which of them organize the phase space.  Seeded by the
network, every orbit comes with its place in the phase portrait---the centre of
an island chain, or the saddle that bounds it---and its stability follows from
the same derivatives that drive the search.  In a two-degree-of-freedom flow
these periodic orbits are not isolated: each belongs to a one-parameter family
parametrized by the energy~\cite{LichtenbergLieberman1992}, which can be
followed by numerical continuation~\cite{Yao2026} without computing any further map.  From a
single section, an agent could thus chart the main periodic orbits of the
system across a finite range of energies, with their periods and stability, and
locate the bifurcations along each family, where its stability changes and new
island chains are born---precisely the energies at which a new map is worth
computing.

Several directions would extend this work towards a broader foundation model.
The most natural is to widen the training distribution within classical
dynamics: dissipative maps, whose attractors replace the invariant tori, and
higher-dimensional maps, explored through two-dimensional slices.  Equally
natural is to widen the set of tasks---for instance to locating the periodic
orbits discussed above---and to test how far the pretrained representation
carries over to new tasks learned from few labelled examples.  A less obvious test is whether it carries over beyond physics, to natural images.  This is plausible: the first layers of a network can be learned from a single image with strong augmentation~\cite{Asano2020}, and vision transformers pretrained on a very limited set of images synthesized from mathematical formulae, fractals, approach the performance of pretraining on millions of natural images~\cite{Nakamura2023}.  Poincar\'e maps are rich in what such images supply---closed contours, nested curves and self-similar structure on many scales---and completion further requires following curves and linking distant pixels of the same orbit.  
A more speculative extension, which would substantially widen the scope of
the model, is to quantum dynamics.  The Husimi and Wigner distributions of
quantum states are the other prominent visualization of phase space, and they
are intimately connected to Poincar\'e maps: they reveal the invariant
structures of the underlying classical dynamics, with regular states
concentrated on its tori, irregular ones spread over its chaotic
sea~\cite{Berry1977}, and scarred ones along its unstable periodic
orbits~\cite{Heller1984}.  Classical and quantum phase-space portraits are thus
two views of the same structures.  This
makes it attractive to train a single, broader foundation model on phase-space
visualizations of every kind, classical and quantum, in which what is learned
from one kind of image informs the reading of the other, e.g.,
interpreting a quantum distribution through the classical structure it reflects.

More broadly, none of our ingredients is specific to Poincar\'e maps: a
generator of synthetic examples with known physics, a painter that renders
combinatorially many images from each, a pretext task taken from the physics
that produced them, and the expert analysis recast as image-to-image tasks.
Whenever an image is produced by a known process, that process can supply
pretext tasks whose targets carry information the image itself does not---a
longer integration, a finer resolution, a denser sampling.  Applied to the other diagnostic plots
that physicists read by eye---bifurcation diagrams, band structures,
spectrograms, phase diagrams---it could bring the visual side of scientific
reasoning within reach of AI systems.

\section*{Use of AI in the research}

The software of this work was developed, from February 2026 onwards, with
the agentic coding assistant Claude Code (Anthropic), using successive Claude Sonnet (versions 5 and 5.5) and Claude Opus models, up to Opus~5.5.
Starting from an existing code base, the agent extended it and wrote new
components---the generation and labelling of the maps, the training and
evaluation of the networks, and the figure scripts---from the author's
specifications, and ran the computations.  The author then had it carry out
a series of physics-informed sanity checks, many of them through plots that
the author inspected; these exposed loopholes and bugs, which were removed
until every check was passed.  A second round of checks accompanied the
writing of the manuscript: the author had the agent read the code and
describe it to them, so that the description given in the paper is faithful
to the implementation and every design choice the agent had made on its own
was reviewed and accepted by the author.

The manuscript was written in close interaction with Claude Code
(Anthropic; Claude Sonnet and Claude Opus models).  For each section the
author agreed a plan with the agent, which drafted the text; the author had
the draft restructured as needed and then revised every paragraph, either
directly or by instructing the agent.  Text drafted by the agent remained
marked in the source until the author had revised it, and the agent's changes
are recorded under a separate identity in the project's version history.  The
author takes full responsibility for the content of this paper.

\section*{Data Availability}
The trained models, the validation data and the prediction galleries
underlying this work are openly available in Edmond, the open research data
repository of the Max Planck Society~\cite{PVTData2026}.  The deposit contains
the weights of every network reported here---the pretrained encoders (masked
autoencoder and long-time map completion) and the fine-tuned binary, semantic
and instance-segmentation networks of every ablation arm; the trajectory banks
of the validation set, from which every evaluation input is painted, together
with their ground-truth labels; and, for every validation map, the input, the
ground truth and the prediction, with the per-map figures of merit from which the
tables are aggregated.  There is one gallery for each task---binary, semantic,
instance and long-time map completion---and a further instance gallery that
compares the three pretraining arms; each is a static
web page in which the maps can be filtered by family and topology and sorted by
any score, or by their distance from the median, so that typical as well as best
and worst cases can be inspected (Figs.~\ref{fig:segmentation_results}
and~\ref{fig:pretraining_ablation}, Tables~\ref{tab:pretraining_binary},
\ref{tab:pretraining_avg_convention}, \ref{tab:sizeweight} and~\ref{tab:pretraining_instance}).
The orbit data, the panels and the code of the out-of-distribution examples of Sec.~\ref{sec:deployment} and of the appendix figure that accompanies them are included, as are the evaluation results behind every table.
The galleries can also be browsed on the project page~\cite{PVTSite}, together with an interactive explorer of the network predictions.

\begin{acknowledgments}

I thank Florian Marquardt for encouragement and fruitful discussions, including suggesting map completion as a task, and Yuki Asano for suggesting the comparison with Segment Anything and the potential use of the pretrained model for natural images.
The networks were trained on the HPC system Viper at the Max Planck Computing and Data Facility.

\end{acknowledgments}

\appendix

\section{Random map families}
\label{app:map_families}

All training maps are built by composing sub-steps that are individually
exactly integrable---linear area-preserving transformations, and nonlinear
perturbations generated by a Hamiltonian whose flow is known in closed form.
The resulting map is symplectic by construction: it preserves phase-space area
to machine precision at any perturbation strength, so no symplectic-integrator
tolerance enters the ground truth.  We use three constructions: the first two
are generic ansatz designed to cover a broad range of phase portraits, while
the third attempts to mimic  the stroboscopic maps of weakly nonlinear resonators.

\subsection{Separable double-kick maps}

The first construction is based on the classical double-kick ansatz for two-dimensional
area-preserving maps,
\begin{align}
  p_{n+1} &= p_n - \epsilon\, f(q_n), \label{eq:map_p}\\
  q_{n+1} &= q_n + \epsilon\, g(p_{n+1}), \label{eq:map_q}
\end{align}
where $f$ and $g$ are smooth nonlinear kick functions and $\epsilon$
is a perturbation parameter that controls the transition from
integrable ($\epsilon\to 0$) to fully chaotic dynamics. We alternate the kick with a linear transformation. Overall, the map can be compactly written  as
$F(\mathbf{x}) = \mathbf{T} + U\, K(\mathbf{x})$, where
$K(\mathbf{x})$ is the nonlinear kick,
$\mathbf{T}$ is a translation, and $U$ is a unit-determinant
linear transformation parametrized as
\begin{equation}
  U =
  \begin{pmatrix}
    e^{a}\cos\alpha  & e^{b}\sin\alpha \\
    -e^{-b}\sin\alpha & e^{-a}\cos\alpha
  \end{pmatrix},
  \label{eq:U_matrix}
\end{equation}
which satisfies $\det U = 1$ by construction.  The eigenvalues obey
$|\lambda_\pm|=1$ (elliptic regime) as long as
$\cosh|a|\,|\cos\alpha| < 1$.  The eccentricity of the resulting
orbits is controlled by the parameter
\begin{equation}
  s = \sqrt{\frac{\cosh^2 b\;\sin^2\alpha}
                  {1 - \cos^2\alpha\;\cosh^2 a}}\,,
\end{equation}
which we bound by a maximum value $s_{\max}$ to avoid extremely
elongated orbits.
For random sampling, the rotation angle $\alpha$ is drawn uniformly on
$(0, 2\pi]$; the parameters $a$ and $b$ are then sampled within the
stability region consistent with $s \leq s_{\max}$.
The identity ($a=b=0$, $\alpha=0$) and pure-rotation ($a=b=0$) limits
are given finite sampling weight as important special cases.

Most well-known area-preserving maps are special cases of this ansatz:
\begin{itemize}
  \item The \emph{Chirikov standard map}~\cite{Chirikov1979}
    corresponds to $U = \mathds{1}$,
    $f(q)=\sin q$, and $g(p)=p/\epsilon$.
  \item The \emph{H\'enon map}~\cite{Henon1969} is recovered with
    $U = \bigl(\begin{smallmatrix} 0 & 1 \\ -1 & 0 \end{smallmatrix}\bigr)$
    (a $90^\circ$ rotation),
    $f(q)=q^2$, $g=0$, and $\mathbf{T}=(1,0)$.
    The \emph{Lozi map}~\cite{Lozi1978} is the same but with
    $f(q)=|q|$.
  \item The \emph{kicked oscillator} (or \emph{web map}) uses an
    arbitrary rotation angle $\alpha$ with $a=b=0$, i.e.\
    $U = \bigl(\begin{smallmatrix} \cos\alpha & \sin\alpha \\
    -\sin\alpha & \cos\alpha \end{smallmatrix}\bigr)$.
  \item \emph{Action-angle maps}---modeling a Poincar\'e section of a
    perturbed integrable system---take the form
    $g(p) = \nu(p)/\epsilon$, where $\nu(p)$ is the unperturbed
    frequency, and $f(q)$ is a periodic zero-average kick.
    This generalizes the kicked rotor to non-uniform rotation numbers.
\end{itemize}
By randomizing the functions $f$ and $g$, the parameters of $U$,
the translation $\mathbf{T}$, and the perturbation strength
$\epsilon$, we can sample a broad distribution of phase
portraits---from nearly integrable KAM scenarios to fully developed
chaos---within a single unified framework.
On the torus, $f$ and $g$ are random superpositions of sinusoidal Fourier
components; on the plane they are splines through randomly placed
Gaussian-process control points, with $\mathbf{T}$ chosen so that a randomly
selected elliptic fixed point lies inside the computational domain $[-1,1]^2$.
In the special case of a trivial linear step, $U = \mathds{1}$ and
$\mathbf{T} = 0$, Eqs.~\eqref{eq:map_p}--\eqref{eq:map_q} are the symplectic
Euler discretization, with time step $\epsilon$, of the flow of the separable
Hamiltonian $H(q,p) = V(q) + W(p)$, with $V' = f$ and $W' = g$.

\subsection{Non-separable oblique maps}

Because the Hamiltonian of this limit is separable, a sum of a function of $q$
and a function of $p$, the resonances it drives are aligned with the coordinate
axes.  To lift this restriction we replace it by a general perturbing
Hamiltonian built from $N_{\mathrm{m}}$ plane-wave modes,
\begin{equation}
  H(q,p) \;=\; \sum_{n=1}^{N_{\mathrm{m}}} c_n \cos\!\left(k_n q + l_n p + \phi_n\right),
  \label{eq:twist_H}
\end{equation}
and the double kick by a symplectic splitting of its flow over the time
$\epsilon$, followed, as before, by the linear step
$\mathbf{x}\to \mathbf{T} + U\mathbf{x}$.  Each mode alone generates an exact
shear,
\begin{equation}
  S_n^{\delta t}:\;
  \begin{pmatrix} q \\ p \end{pmatrix}
  \longmapsto
  \begin{pmatrix} q - c_n l_n \sin(k_n q + l_n p + \phi_n)\,\delta t \\
                  p + c_n k_n \sin(k_n q + l_n p + \phi_n)\,\delta t \end{pmatrix},
\end{equation}
because its argument is conserved along its own flow.  One iteration composes
these shears symmetrically (Strang splitting),
$S_1^{\epsilon/2}\cdots S_{N_{\mathrm{m}}-1}^{\epsilon/2}\,
S_{N_{\mathrm{m}}}^{\epsilon}\,S_{N_{\mathrm{m}}-1}^{\epsilon/2}\cdots S_1^{\epsilon/2}$:
the first $N_{\mathrm{m}}-1$ modes act for $\delta t = \epsilon/2$ on the way in
and again on the way out, the last one for $\delta t = \epsilon$, so that
$2N_{\mathrm{m}}-1$ shears make up the step.  The composition is symplectic for
any $\epsilon$ and approximates the time-$\epsilon$ flow of
Eq.~\eqref{eq:twist_H} to $O(\epsilon^2)$; for $U = \mathds{1}$ and
$\mathbf{T} = 0$ the map is thus a discretization of that flow with time step
$\epsilon$, as the double kick is of the flow of a separable Hamiltonian.  The
map we label is the composition itself, not the flow it approximates, so the
splitting error does not affect the ground truth.

The number of active modes is the larger of two integers drawn uniformly from
$1$ to $8$, which favours maps with many modes; the amplitudes $c_n$ are drawn
uniformly from $[-1,1]$ and the phases $\phi_n$ uniformly, and $\epsilon$ from a
half-normal distribution of scale $0.08$ on the torus and the cylinder and $0.1$
on the plane, floored at a tenth of that scale.  The wavevectors are integers in
$[-4,4]$ on the torus, as periodicity requires; on the cylinder only $k_n$, the
component along the periodic coordinate, is an integer, while $l_n$ is real in
$[-4,4]$; on the plane $(k_n,l_n)$ has a length drawn uniformly from $[1,4]$ and
a random direction.  Because the wavevectors point in arbitrary directions, this
construction produces resonance chains at arbitrary angles---hence the name--- and a visibly richer
set of island configurations than the separable family.

\subsection{ Kerr maps}

The third construction aims 
to mimic the stroboscopic map of weakly nonlinear resonators. More precisely, the constructed maps, which we dub Kerr maps, approximate such maps in the regular limit, $\epsilon\to 0$, which in these systems corresponds to the rotating-wave approximation.  We build two versions of our Kerr maps. The first version  mimics a scenario with an $n$-photon parametric drive, which leads to an $n$-fold rotational symmetry in phase space.  In the second version,  the $n$-photon drive of frequency $n\omega_d$ is accompanied by  a linear drive of frequency  $\omega_d$. The implementation described below is one of many possible choices, and
none of its details is essential: the Kerr maps are meant to produce phase
portraits with the structures typical of driven nonlinear resonators close to
the regular limit, not to reproduce the dynamics of any particular system.
Away from that limit, where the splitting error makes the maps chaotic, their
dynamics should not be expected to resemble that of the stroboscopic maps of
real resonators.

We combine the exact flows generated by the following terms, with
$r^2 = q^2+p^2$:
\begin{itemize}
  \item a radial term $W = \delta\,r^2 + c_K\,r^4 + c_6\,r^6 + c_8\,r^8$,
    comprising the detuning $\delta$, the Kerr nonlinearity $c_K$ and two
    higher-order nonlinearities; it generates a rotation by the
    amplitude-dependent angle $2W'(r^2)\,\delta t$;
  \item different  $m$-photon drives
    $D_m\,\mathrm{Re}\bigl[e^{-i m \phi_m}(q+ip)^m\bigr]$, of amplitude $D_m$ and
    phase $\phi_m$, with $m = 1$ corresponding to a linear drive; the drives used are $m = 1$, $m = n$ and, for $n \geq 5$ only, an additional two-photon drive ($m = 2$, $|D_2| \leq 0.1$); each drive is decomposed exactly into a sum of
    $m$ polynomial shear Hamiltonians of degree $m$ at equally spaced angles, whose exact flows are composed; for
    $m = 1$ the flow is a translation.
\end{itemize}
For even $m$ the $m$ shears generate, besides the $m$-photon drive, the radial
term $\binom{m}{m/2} D_m\,r^m/2$, which is compensated by subtracting it from the
coefficient of $r^m$ in $W$.  With $\Phi_X^{\delta t}$ the flow of the term $X$
over the time $\delta t$, the symmetric splitting step is
\begin{equation}
  S^{\delta t} = \Phi_W^{\delta t/2}\,\Phi_{D_1}^{\delta t/2}\,\Phi_{D_2}^{\delta t/2}\,
                 \Phi_{D_n}^{\delta t}\,
                 \Phi_{D_2}^{\delta t/2}\,\Phi_{D_1}^{\delta t/2}\,\Phi_W^{\delta t/2},
\end{equation}
and $\tilde S^{\delta t}$ is the same step without the linear drive $D_1$.  The two
Kerr maps are
\begin{align}
  M_1 &= R\!\left(\tfrac{2\pi}{n}\right) \bigl(S^{\epsilon/N}\bigr)^{N}, \\
  M_2 &= C^{\,n-k}\,\Phi_{D_1}^{\epsilon}\,C^{\,k}, \qquad
  C = R\!\left(\tfrac{2\pi}{n}\right) \bigl(\tilde S^{\epsilon/(nN)}\bigr)^{N},
\end{align}
where $R(\theta)$ is the rotation by $\theta$, the advance of the rotating
frame over one period of the $n$-photon drive, $k = \lfloor n/2 \rfloor$, and
$N = 1$ for $n = 1$ and $N = 4$ otherwise.  $M_1$ samples the motion once per
period of the $n$-photon drive, $M_2$ once per period of the linear drive.

The order $n$ ranges from $1$ to $6$ and is fixed for each generation run; it
is roughly uniformly distributed in the first version, while the second
contains few maps with $n = 1$ and is weighted towards small $n$.  The leading
radial coefficient, $c_K$ for $n \leq 3$, $c_6$ for $n = 4,5$ and $c_8$ for
$n = 6$, has magnitude
$1/2$ and a random sign; the other two are uniform in $[-1,1]$.  $|\delta|$ is
uniform in $[0,1]$, with the sign opposite to that of $c_K$ in $90\%$ of the
maps.  $D_1$ and $D_n$ are uniform in ranges that shrink with $n$, and
$\epsilon$ is half-normal with an $n$-dependent scale
of order $0.15$.  The maps live on the plane, and some orbits escape to
infinity (Appendix~\ref{app:megno}).

\subsection{Phase-space topology}

Within the double kick and oblique families, we vary the topology of the phase space:
\begin{itemize}
  \item \emph{Plane.}  Neither coordinate is periodic; the image shows
    the window $[-1,1]^2$, which orbits may leave and re-enter, and some orbits
    escape to infinity (Appendix~\ref{app:megno}).
  \item \emph{Cylinder.}  The position $q$ is periodic and the momentum $p$ is
    not, as for a pendulum or a driven rotor.  The linear part $U$ is then
    restricted to those matrices that preserve the cylinder.
  \item \emph{Torus.}  Both coordinates are identified modulo their period.  The
    linear part $U$ is restricted to a discrete set of $SL(2,\mathbb{Z})$
    matrices compatible with the periodicity---the identity, the $90^\circ$
    rotation, the shear
    $\bigl(\begin{smallmatrix} 1 & 1 \\ 0 & 1 \end{smallmatrix}\bigr)$, and the order-six element
    $\bigl(\begin{smallmatrix} 1 & 1 \\ -1 & 0 \end{smallmatrix}\bigr)$.
\end{itemize}
Of these, the plane and the cylinder occur among the real systems of
Sec.~\ref{sec:deployment}, and the network is given no indication of which
topology it is looking at.

\section{Ground-truth labelling}
\label{app:labelling}

The labels of Sec.~\ref{sec:data} are produced in four steps, which the
subsections below follow in order: a coverage search that places an orbit
through every pixel, the classification of each orbit as regular or chaotic,
the reconstruction of the separatrices, and the decomposition of the regular
foreground into instances which may form island chains. In the process, we determine the return period of each chain.

\subsection{Coverage search}
\label{app:coverage}

The labelling requires an orbit through every pixel, which no fixed set of
initial conditions can guarantee: a randomly chosen orbit may be confined to a
single island and leave most of the image untouched.  We therefore grow the set
of initial conditions adaptively.  Starting from a coarse $10 \times 10$ grid of
initial conditions, every orbit is followed for $10^4$ iterations and the pixels
it visits are marked on an occupation grid.  New initial conditions are then
seeded in pixels that are still empty, and the process repeats until the whole
$256 \times 256$ grid is occupied, up to a limit of $20$ rounds.  Maps that are
not fully covered within that limit are discarded rather than labelled from a
partial cover; so are maps whose separatrix reconstruction leaves gaps
(Appendix~\ref{app:separatrix_heuristic}).  The number of orbits the search
returns therefore varies from map to map.

\subsection{MEGNO chaos classification}
\label{app:megno}

Each orbit returned by the coverage search is labelled regular or chaotic
using the Mean Exponential Growth factor of Nearby Orbits
(MEGNO)~\citep{Cincotta2000}, a time-averaged chaos indicator that converges
much faster than a direct finite-time Lyapunov exponent and cleanly separates
the linear (regular) from the exponential (chaotic) divergence of neighbouring
trajectories.  Alongside each orbit $x_n$ we evolve a tangent vector $\delta_n$
under the linearised (tangent) map.  Writing
$a_n = \ln\!\big(\lVert\delta_n\rVert/\lVert\delta_{n-1}\rVert\big)$ for the
per-step logarithmic stretching, the MEGNO indicator and its running time
average are
\begin{equation}
  Y(t) \;=\; \frac{2}{t}\sum_{n=1}^{t} n\,a_n,
  \qquad
  \bar Y(t) \;=\; \frac{1}{t}\sum_{k=1}^{t} Y(k).
  \label{eq:megno}
\end{equation}
For a regular (quasi-periodic) orbit $\bar Y(t)\!\to\!2$ as $t\!\to\!\infty$ and
the associated growth rate vanishes, whereas for a chaotic orbit $\bar Y$ grows
linearly in $t$ at a rate proportional to the largest Lyapunov exponent.  We
integrate every coverage orbit for a long horizon
($T_{\mathrm{M}} = 2.8\times10^{5}$ iterations) and classify it as chaotic when
the time-averaged growth rate derived from $\bar Y$ exceeds a small threshold
$\sim\!1/T_{\mathrm{M}}$.  To
suppress the finite-time transient, the indicator is read off at the
\emph{near-return} iterate---the step within the final few percent of the
trajectory that comes closest to the initial condition.

The chaotic region is then obtained by \emph{occupation}: for every orbit
classified as chaotic, all grid pixels its trajectory visits are marked, and the
chaotic binary image is the union of these footprints over all chaotic orbits
(Fig.~\ref{fig:data_pipeline}c,d).  Because the coverage-search step guarantees
that every pixel is visited by at least one orbit, the complementary (unmarked)
pixels constitute the regular foreground that is subsequently decomposed into
island chains.

On the plane, not all orbits are bounded: some leave the window and escape to
infinity.  Classifying such orbits as regular or chaotic is subtle.  The
distinction between regular and chaotic motion refers to the long-time
behaviour on bounded invariant sets, which an escaping orbit does not have.
Before escaping, an orbit may linger for a long time in a chaotic layer, a
manifestation of transient chaos~\cite{LaiTel2011,Aguirre2009}, or it may leave
directly; telling these cases apart requires an analysis of the transient
dynamics that is beyond the scope of this work.  We therefore adopt a labelling
convention.  When an orbit escapes fast enough, its tangent vector, whose growth
is driven by that of the orbit, overflows in floating-point arithmetic within
the horizon $T_{\mathrm{M}}$, and the MEGNO indicator becomes non-finite.  Such
an orbit is assigned to the chaotic class, although strictly speaking it is
neither regular nor chaotic.  The convention is consistent with the meaning of
the chaotic class in this work, which collects everything that is not a regular
island in the window; only the pixels the orbit visits inside the window before
leaving are marked.  Orbits that drift outward too slowly to overflow keep a
finite indicator and are classified by its value like any other orbit.

\subsection{Separatrix candidate identification}
\label{app:separatrix_heuristic}

\paragraph{Why the chaos indicator is not enough.}
Marking the pixels visited by orbits that MEGNO classifies as chaotic
(Appendix~\ref{app:megno}) recovers the chaotic sea and any layer wide enough to
be sampled, but misses the thin chaotic layers along the separatrices of the
nonlinear resonances.  Their width is exponentially small in the perturbation
strength and can fall well below a pixel, and with about one covering orbit per
pixel such a layer is rarely sampled.  Yet these are the layers that matter
most: each separates a librating island chain from the non-resonant region
around it, and without it the decomposition of Appendix~\ref{app:chains}
absorbs the chain into its surroundings.  We therefore reconstruct the
separatrices rather than observe them, in three steps: covering orbits that pass
close to a saddle are flagged, each flagged orbit is refined by Newton's method
to a periodic point that is kept only if it is hyperbolic, and the separatrices through the accepted saddles are traced along their unstable manifolds.

\paragraph{Candidate orbits.}
An orbit passing close to a saddle creeps past it, spending many iterations in
a small neighbourhood, before moving quickly along its manifolds, so that its
visits concentrate on a few pixels.  We measure this concentration by
\begin{equation}
  P_i \;=\; \frac{M_i}{T_i / B_i},
  \label{eq:sep_heuristic}
\end{equation}
the number of visits $M_i$ to the most-visited pixel of orbit $i$ divided by the
mean number of visits per visited pixel, with $T_i$ the number of iterations the
orbit spends in the domain and $B_i$ the number of distinct pixels it visits.  A
regular orbit covering its invariant curve uniformly has $P_i \approx 1$, an
orbit near a saddle $P_i \gg 1$. We take as candidate saddle points the most-visited pixels of orbits with $P_i$ above a finite threshold   $P_{\mathrm{th}}$.

\paragraph{Newton refinement.}
We take each candidate as a starting point and
estimate the period $n$ as the smallest number of iterations after which its
orbit returns to within four pixels of it (searching up to $200$ iterations), so
that the saddle sought is a fixed point of $F^n$.  Newton's method is applied to
$F^n$ from that point, and the refined point $x^\star$ is accepted if Newton
converges, $\lVert F^n(x^\star) - x^\star\rVert < 10^{-3}$ pixels.  In rare
cases Newton converges to a stable (elliptic) fixed point instead, which has no
manifolds to trace; it is excluded by requiring the linearization to be
hyperbolic, which corresponds to the condition $\lvert \operatorname{tr} DF^n(x^\star)\rvert > 2$.
More often, the candidate does not lie near a fixed point at all: the
return distance $\lVert F^n(x) - x\rVert$ has a strictly positive minimum,
Newton does not converge, and the candidate is discarded.

Since a discarded candidate costs only one Newton solve, whereas a missed
saddle removes a boundary and the defect propagates into the instance and
semantic labels, the threshold $P_{\mathrm{th}}$ is deliberately set low at $P_{\mathrm{th}}=7$. Visual inspection of a small set
of test maps showed this value to be low enough to detect every separatrix
bounding islands larger than a few pixels.

\paragraph{Manifold tracing and checks.}
Accepted saddles that coincide within a pixel are merged, since the $n$ points
of a period-$n$ orbit are otherwise found $n$ times.  For each remaining saddle
the separatrix is traced by iterating its eight neighbouring
pixel centres forward for $3\times10^{4}$ steps. The pixels these orbits visit form the separatrix
image of that saddle, which must satisfy two conditions.  First, it must be
closed: a manifold traced to completion closes into a continuous curve, so every
separatrix pixel has at least two of its eight neighbours on the separatrix (on
the plane, the edge of the domain counts as a neighbour).  Second, it must
enclose at least as many regions of more than three pixels as the period-$n$
orbit has points inside the domain, one for each island of the chain; this
excludes the separatrices of resonances whose islands are too small to be
resolved at the resolution of the image, which would merely divide the two sides
of the resonance.  If either condition fails, the trace is extended in steps of
$3\times10^{4}$ iterations, up to four times its original length, and tested
again.  A separatrix that is still open is truncated, and the map is discarded
rather than labelled with a broken boundary; a closed separatrix that encloses
too few regions is left out of the separatrix image.

\subsection{From the binary image to island chains and periods}
\label{app:chains}

The binary ground truth of Appendices~\ref{app:megno}
and~\ref{app:separatrix_heuristic} marks every pixel as chaotic or regular.  Subsequently the
instance and semantic label maps are both derived from it using also the system dynamics.

\paragraph{Theoretical picture.} 

A regular orbit lies on an invariant torus, which intersects the Poincar\'e
section in $n$ disjoint closed curves.  The map permutes these curves
cyclically, so that an orbit returns to the curve it started on after $n$
iterations: $n$ is the return period, our semantic label.  Invariant tori come
in continuous families of tori with the same~$n$, delimited by separatrices.
An instance in our classification is the union of the
intersection curves of one such family.  In this ideal picture an instance
consists of $n$~connected regions, which the map permutes cyclically.  For
maps on the torus the image shows the whole phase space, so every region lies
inside it (interior components).  For maps on the plane or the cylinder the
image shows only a window of phase space: some regions are cut by its edge
(border components), and some lie entirely outside it.  The binary image
departs from this picture in two ways: pixelization can cut a region into
several connected components, and a separatrix that is not resolved joins
regions of different instances.  The connected components of the image are
therefore only a first guess; the dynamics decides which of them form an
instance, and what its period is.

\paragraph{Initial decomposition.}
The regular foreground is labelled with a standard $4$-connected
connected-component algorithm, with one modification: components touching across
a periodic edge are merged so that an island
straddling the seam of the torus or the cylinder is one component rather than
two.  Components are sorted by decreasing area and those of at most $4$~pixels
are discarded, since a handful of pixels carries too little information for the
orbit-based tests below.

\paragraph{Return period estimation.}
Each surviving component is processed in turn, in order of decreasing
area; on the plane, all interior components are processed before the border
components.
From a point in its interior we
integrate an orbit of $3\times10^{4}$ iterations.
We use different approaches to choose a starting point. The default   is the
component's centroid.  This is replaced by a median pixel of the component when the
centroid falls outside it. For border components  the orbit starts from the pixel deepest inside the component,
  measured from both the separatrix and the image edge (distance transform).  The period is then estimated as the first recurrence: the smallest $n$
for which the $n$-th iterate returns to the component it started in,
searched within the first $300$~iterations.  Because an
orbit can land on a separatrix pixel at the edge of its island, a return is
counted when any pixel within one pixel of the orbit point belongs to the
starting component.  As a consistency check, the estimate is accepted only if the orbit also
returns at $2n$ and $3n$ and makes no return in between; for $n>100$, where
the first $300$~iterations do not reach~$3n$, only the return at $2n$ is
checked. The consistency check can also reveal some errors of the initial
decomposition, e.g., due to discretization or a missing separatrix.  For a border component the check is relaxed.  Its intersection curve is cut by
the edge of the window, so an orbit of period~$n$ returns to the same curve
every $n$~iterations but not necessarily to the part of it inside the
component, and it may leave the window altogether: the first return can then
differ from~$n$, and the check fail, although the decomposition is correct.
For a border component we therefore take the shortest first return over up to
three starting points, check it only at~$2n$, accept $n=1$ unconditionally,
and read a failed check as the orbit having left the window rather than as an
error of the decomposition.  Orbit points outside the window are ignored
throughout.

\paragraph{Grouping connected components.}

A component that starts a new instance is grouped in four steps: its orbit
collects components, these are repaired, the period is estimated again, and
only then are labels assigned.

In most cases the orbit visits the $n$ islands of a chain in a fixed cyclic
order and returns to its starting component after $n$ iterations, and the
components it visits are the candidates for the instance.  They are first
repaired, since pixelization can cut an island into several components: the
orbit is used as evidence that they belong together.  A mask is built from the
components the orbit visits and the pixels it lands on, and re-decomposed into connected components this time using an 
$8$-connected algorithm.  Each connected component of the
mask replaces the components of the original decomposition it contains, and
the orbit pixels in it become part of the new component.  For a small island,
a mask component may even be made only of orbit pixels; it is joined to the
component the orbit started from.  The $8$-connected algorithm can join fragments that had been separated in the initial $4$-connected algorithm. Nevertheless, it may still miss components that the orbit does not reach.

As a consistency check, the period is then estimated again on the repaired
labels, and this second estimate is the one that is assigned.  The components
of the repaired mask receive a common instance label, the period becomes their
semantic label, and they are not further processed.  If the period cannot be
established, the component remains unassigned; it receives no label unless the
orbit of a component processed later attaches it to an instance.  The estimate
made before the repair is not used for the labels: it decides whether a
component is processed at all (it is skipped if no return is found) and, when
its orbit enters components that are already labelled, how it is treated (see
below).

In many cases, all components belonging to one instance are obtained from a
single orbit.  Because the border components are processed last (see above), a
border component reached by the orbit of an interior component of the same
instance is labelled with it, and its less reliable period estimate is never
needed.  Because the orbit pixels become part of the components, the instance
and semantic maps also label some pixels that the binary image marks as
chaotic or separatrix, a median of $1.6\%$ of them over the validation maps;
the same holds for the orbit pixels of a component that joins an existing
instance (see below).

\paragraph{Orbits entering labelled components.}
When a component is processed, its orbit may enter components that are already
labelled.  This happens in two situations: the component belongs to an
instance labelled earlier, or its orbit  reaches a pixel belonging to a neighbouring instance. 

In the first situation, the component belongs to a multi-component instance
but was not reached by the first orbit of that instance, and remains
unassigned until it is itself processed.  This may occur if the component is a
border component, or if it results from an error of the initial decomposition
due to discretization.  The first orbit lies on a single torus, so in each
island it lands only on the pixels along one closed curve; a fragment that
contains only other tori of the family is never reached, and for a border
component the curve may leave the window where the component lies.  The orbit
of the missed component then enters the labelled instance with the same
period, and the component joins that instance.  A border component can also
belong to an instance whose other components all lie outside the window. In this case, it may happen that the period cannot be reliably established.  Such components are still grouped:
components visited by a common orbit form one instance, whose period is marked
as unknown.

In the second situation the component does not belong to one of the instances its
orbit enters. This typically happens because of discretization. For narrow islands a pixel nominally assigned to an island may already contain initial conditions within the chaotic layer.  An orbit starting from such an initial condition moves within the layer,
which may be thinner than a pixel, and lands on pixels assigned to the
neighbouring instance, since these pixels also contain points of the layer. To avoid wrong mergers of instances we introduce the following
safeguards. If the
period of the component differs from that of the instance its orbit
enters, the island is processed as a new instance, but components
already labelled with a different period are kept out of the repair, which
would otherwise join the two across the separatrix.  If the orbit enters two
different instances, it cannot belong to a single one, and the component
receives neither an instance nor a period label.

In both situations, a
component whose period fails the consistency check receives no label.
The exception is a border component whose check fails:
since the failure may only mean that the orbit left the window, it is repaired
and its period estimated again, as for a new instance.

\paragraph{Limitations.}
Maps on the cylinder did not receive the same treatment of border components as
maps on the plane, which could decrease the quality of their ground truth. More generally, the safeguards do not catch every
case.  For example, the period check does not protect neighbouring components of the same
period, so that an orbit in the chaotic layer can join narrow regular strips on
both sides of a separatrix into a single instance.  Such errors in the ground truth
contribute to the disagreements between ground truth and prediction as discussed
in Appendix~\ref{app:failure_modes}.

\section{Trajectory banks and data augmentation}
\label{app:bank}

The ground-truth segmentation of a Poincar\'e map is a property of the
underlying area-preserving map alone: it is fixed by the phase-space
structure---the location of the island chains, separatrices, and chaotic
sea---and is independent of \emph{which} orbits are drawn, of how many
there are, and of how many points each contributes. We exploit this
invariance as a cheap and powerful data augmentation that complements the
two geometric augmentations described at the end of this appendix.

For each labeled map we integrate, \emph{once}, a bank of $N=400$
trajectories from random initial conditions for $T=2500$ iterations,
$\{(q^{(i)}_n,p^{(i)}_n)\}$ with $i=1,\dots,N$ and $n=0,\dots,T$, and
compute the label maps as described in Sec.~\ref{sec:data}. At training
time we then synthesize a \emph{fresh} input image from this bank at every
step, all realizations sharing the same ground truth. Each realization is
specified by three integers drawn uniformly per image: the number of
orbits $k\in[k_{\min},k_{\max}]=[30,200]$, the window length (number of map
iterations) $T_{\mathrm{w}}\in[T_{\min},T_{\max}]=[300,2000]$, and the window
offset $t_0\in[0,\,T-T_{\mathrm{w}}]$. We select a random size-$k$ subset of the
bank and render, for each selected orbit, the points
$\{(q_n,p_n)\}_{n=t_0}^{t_0+T_{\mathrm{w}}}$. Varying $k$ and
$T_{\mathrm{w}}$ changes how densely phase space is sampled---from a few
partially traced tori to a fully populated chaotic sea---while the
segmentation labels remain unchanged.

Colour is assigned in draw order: the $j$-th selected orbit receives the
$j$-th vector of the 30-channel golden-ratio encoding. Since the subset is
drawn at random, this is equivalent to a random permutation of the colour
vectors across orbits, so the network cannot associate a fixed colour with
a fixed trajectory and must instead rely on the spatial structure of the
orbits and on colour continuity along each orbit. This augmentation thus
subsumes a pure colour-permutation augmentation as a special case, and
composes with the translations along the periodic coordinates and the $90^\circ$ rotations
applied to the rendered image. Because the network input is a fixed
$256\times256$ raster, each bank is stored as pre-binned 8-bit pixel
indices rather than floating-point coordinates, so a bank occupies little
more than a handful of pre-rendered images  while supporting a
practically unlimited number of distinct realizations.

Two further augmentations act on the geometry of the image.  The first is a
translation along the periodic coordinates: on the torus both $q$ and $p$ are
shifted, on the cylinder only the periodic coordinate $q$, and the plane is
left unchanged.  Because the shifted coordinate is periodic, the translation
is a cyclic shift of the rendered image and of its labels by a random whole
number of pixels; it yields the Poincar\'e map of an equivalent map with
identical dynamics and correspondingly shifted ground truth, at no
computational cost.  The second is a rotation of the rendered image by a
random multiple of $90^\circ$, which is exact for all three topologies up to
the relabelling of the axes.  The translation is applied first, while the
periodic axis is still aligned with the pixel grid.

\section{Architecture details}
\label{app:architecture}

\subsection{Orbit colour encoding}
\label{app:encoding}

\paragraph{Golden-ratio encoding.}
Each orbit drawn into an image is tagged with a vector in $[0,1]^M$, $M=30$,
and every pixel it visits is set to that vector; unvisited pixels are zero.
The $j$-th orbit in draw order receives the $j$-th term of a Kronecker
sequence with generalized-golden-ratio frequencies~\cite{Roberts2018},
\begin{equation}
  (\mathbf{c}_j)_d = \Big( \tfrac{1}{2} + j\,\alpha_d \Big) \bmod 1,
  \qquad
  \alpha_d = \varphi_M^{-d},
  \quad d = 1,\dots,M ,
  \label{eq:golden}
\end{equation}
where $\varphi_M$ is the unique positive root of
$x^{M+1} = x + 1$.  The choice of this particular sequence is somewhat
arbitrary.  What matters is that the tags of the $k$ orbits in an image are
well separated whatever $k$ is, since $k$ is redrawn at every training step
(Appendix~\ref{app:bank}), and that the assignment is deterministic, so that
the input and the target of the completion task of Sec.~\ref{sec:completion}
give the same tag to the same orbit. The number of channels $M$ involves
a trade-off: a larger $M$ separates the tags of distinct orbits better, so that
orbit identity is easy to read off locally, but it widens the network input and
the completion target.  Within this trade-off the value $M=30$ is again
somewhat arbitrary.  With it, the two closest of the $400$ tags of a full bank
are separated by about a quarter of the diagonal $\sqrt{M}$ of the hypercube.

When several orbits visit the same pixel---which happens throughout a chaotic
sea---the pixel keeps the tag of the orbit that painted it last, so the
speckled interleaving of colours described in Sec.~\ref{sec:tasks} is
reproduced in the encoding.  In regular regions, where curves rarely overlap
at this resolution, each pixel carries the tag of the single orbit lying on it.

\subsection{Tokenizer}
\label{app:tokenizer}

The tokenizer maps the $256\times256\times30$ image to $16\times16 = 256$
tokens of dimension $D = 1024$ through
\begin{center}
\begin{tabular}{llcc}
\hline
stage & operation & spatial & channels \\
\hline
1 & Conv $1{\times}1$, MaxPool $2{\times}2$ & $256 \to 128$ & $30 \to 64$ \\
2 & Conv $3{\times}3$, MaxPool $2{\times}2$ & $128 \to 64$  & $64 \to 128$ \\
3 & Conv $3{\times}3$, MaxPool $2{\times}2$ & $64 \to 32$   & $128 \to 256$ \\
4 & Conv $3{\times}3$, MaxPool $2{\times}2$ & $32 \to 16$   & $256 \to D$ \\
\hline
\end{tabular}
\end{center}
followed by flattening and the addition of a learned positional embedding, one
per token.  All convolutions are purely linear---no normalization and no
activation---so that max-pooling is the only nonlinearity in the block.  The
convolutional stack has $2.7\times10^{6}$ parameters at $D = 1024$ (plus
$2.6\times10^{5}$ for the positional embeddings), against $7.9\times10^{6}$
for the single stride-$16$ convolution it replaces.

The reason for this design is the sparsity of the input.  A linear map of the
patch---whether the single stride-$16$ convolution of the standard tokenizer or
the linear part of a convolutional stem---compresses $16\times16\times30$ values
into one token of dimension $1024$, combining the tags of every orbit crossing
the patch into a weighted sum from which no individual curve can be read back.
A channel-wise maximum instead preserves the largest component present
anywhere in the pooling window, so a curve that occupies a single pixel still
leaves a signature at the token scale.  Max-pooling also makes the
representation tolerant to the exact pixel an orbit lands on, which is
appropriate here: the physical content of the image is the geometry of the
invariant sets, not the sub-pixel placement of individual intersections.

\paragraph{Tokenizer ablation.}
The comparison quoted in Sec.~\ref{sec:architecture} trains two instance-segmentation
networks from random initialization that differ only in the tokenizer: our
convolutional stem, with max-pooling as its only nonlinearity, and the standard
learned linear map of the $16\times16$ patch.  Both use a ViT-L encoder, drop-path
rate $0.1$, $300$ fine-tuning epochs and area weighting ($p = 1$) in
Eq.~\eqref{eq:disc}, and are scored as Table~\ref{tab:pretraining_instance} is, on
the six in-range paintings of the $995$ clean validation maps.  Our stem reaches
$\mathrm{PQ} = 0.444$ ($\mathrm{RQ} = 0.536$, $\mathrm{SQ} = 0.828$), the linear
embedding $\mathrm{PQ} = 0.401$ ($\mathrm{RQ} = 0.490$, $\mathrm{SQ} = 0.818$); the
gain is almost entirely in recognition, $25{,}932$ against $22{,}882$ detected
instances at a nearly equal number of false positives ($9{,}823$ against $9{,}578$).

\subsection{UNETR decoder}
\label{app:unetr_detail}

\begin{figure}
  \centering
  \includegraphics[width=\columnwidth]{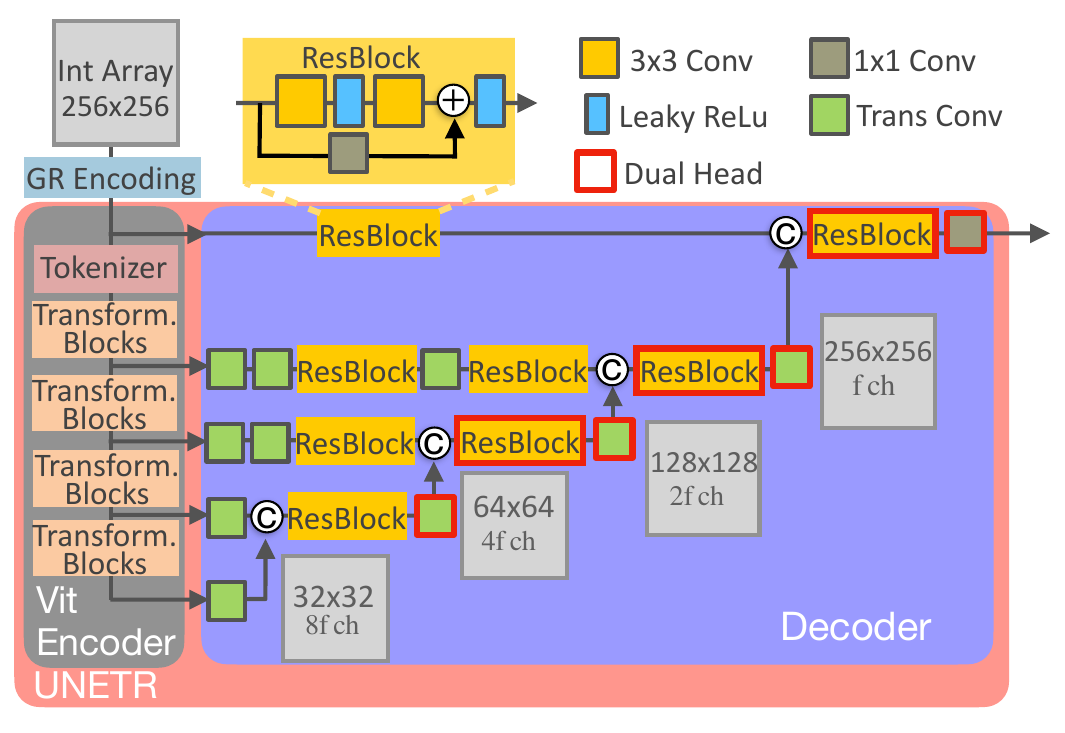}
  \caption{%
    Detailed layout of the UNETR architecture.
    Intermediate representations from the $L$-layer ViT encoder at layers
    $L/4$, $L/2$, and $3L/4$ are projected to the target spatial
    resolution via transposed convolutions (green) and fed as skip
    connections into the CNN decoder through channel-wise
    concatenation.  Each decoder stage upsamples by~$2{\times}$ and
    fuses the skip with a residual block
    (Conv~$3{\times}3$ + InstanceNorm + LeakyReLU, yellow).
    A final $1{\times}1$ convolution maps the feature volume to the
    output class logits.
    The layers outlined in red are duplicated in
    the dual-head variant used for instance segmentation
    (embedding branch + background branch).
  }
  \label{fig:unetr_detail}
\end{figure}

Our implementation started from the UNETR example of the JAX AI Stack
documentation~\cite{JaxAIStackUNETR}, which departs in minor details from
Ref.~\cite{Hatamizadeh2022}; the architecture used here is the one described
in this appendix.
The decoder (Fig.~\ref{fig:unetr_detail}) restores the $16\times16$ token grid
to the $256\times256$ input resolution in four stages, $16 \to 32 \to 64 \to 128
\to 256$, each doubling the resolution and halving the number of channels, from
$8F$ at the coarsest stage to $F$ at full resolution, where $F$ is the base
width.  Each stage is fed a skip connection.  The three coarse ones come from
the outputs of encoder blocks $6$, $12$ and $18$ of the $24$-block ViT-L
(i.e.\ $L/4$, $L/2$, $3L/4$), reshaped to the token grid and brought to the
resolution of their stage by transposed convolutions; the last one comes from
the input image itself, through a convolutional block at full resolution, and
supplies the pixel-level detail that the $16\times16$ tokens cannot carry.  The
final $1\times1$ convolution sets the output: $30$ channels for map completion
(the encoding of the long-time image), one logit for binary segmentation, and
$30$ class logits for semantic segmentation.  For instance segmentation the
decoder splits after its first stage: the $32\times32$ features are shared, and
two branches with their own remaining stages and output convolutions produce the
$D = 10$ dimensional embedding and the foreground logit
(Appendix~\ref{app:finetuning}).

\section{Masked-autoencoder pretraining}
\label{app:mae}

\begin{figure}
  \centering
  \includegraphics[width=\columnwidth]{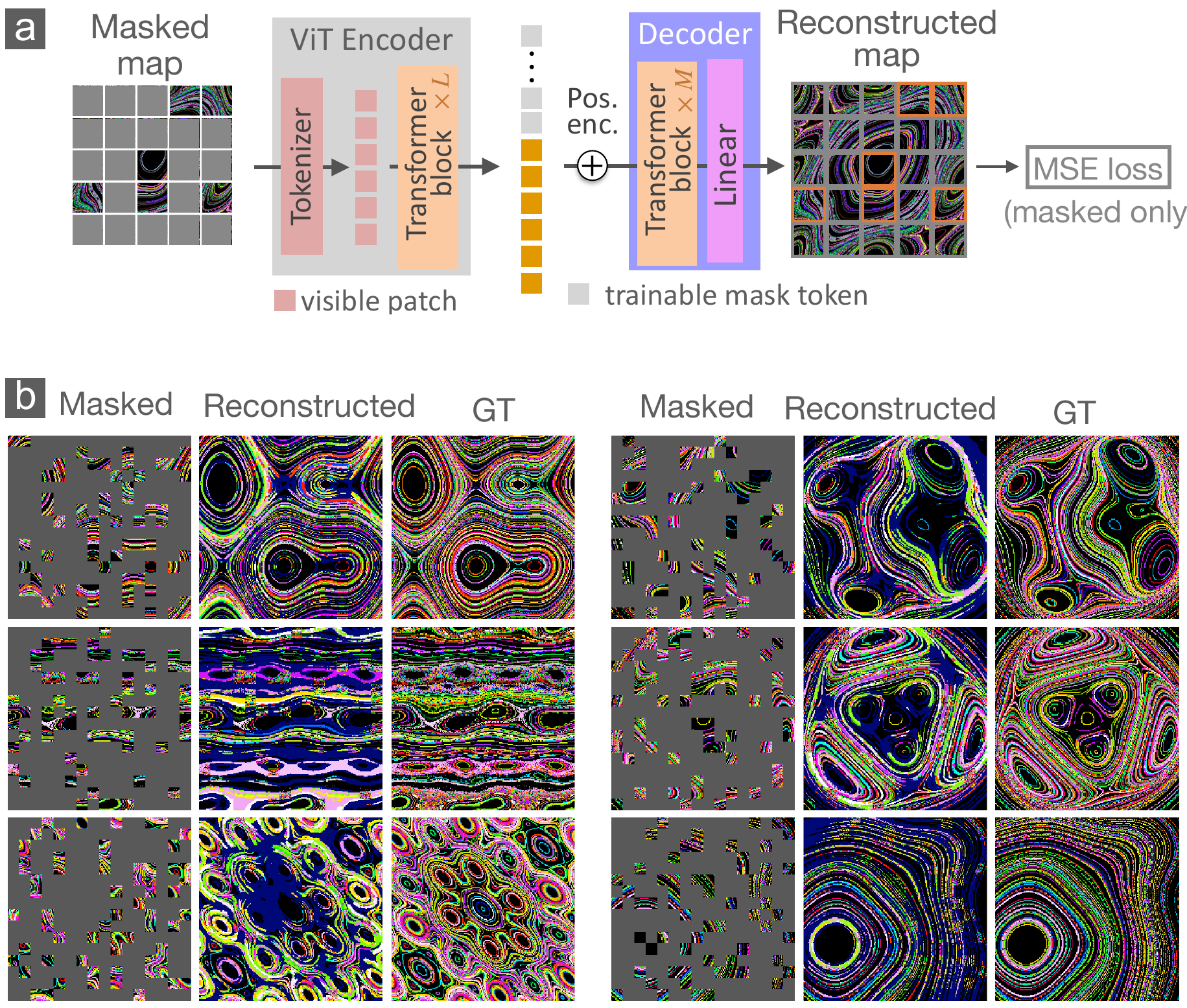}
  \caption{
    Masked-autoencoder pretraining.
    \textbf{(a)}~Sketch of the pretraining pipeline.  $75\%$ of the patches are masked
    (grey) and zeroed before the tokenizer; the ViT encoder processes the
    visible tokens only.  A trainable mask token is inserted at every masked
    position, the positional embeddings are added, and a shallow transformer
    decoder followed by a linear layer predicts the content of every patch.
    The mean squared error is taken over the masked patches only (orange).
    \textbf{(b)}~Masked input, reconstruction and ground truth for six
    validation maps, doubly periodic (left) and on the plane (right), from the
    masked autoencoder at epoch $1750$, whose encoder initializes the MAE arm of
    Fig.~\ref{fig:pretraining_ablation}.  Inputs are painted with $k = 150$
    orbits over $T_{\mathrm{w}} = 1000$ iterations and coloured by orbit as in
    Fig.~\ref{fig:segmentation_results}.  In the reconstruction the visible
    patches are copied from the input, and every pixel of a masked patch is
    given the colour of the orbit whose tag is nearest to the predicted vector,
    or left empty when the zero vector is nearer or the prediction has norm
    below $0.5$, the rule used for map completion in Fig.~\ref{fig:architecture}.
    Many orbits in the reconstructions are dark blue, the colour of the
    first orbit.  Its tag has by far the smallest norm [$1.4$, against $2.6$
    for the next smallest and a median of $3.2$; Eq.~\eqref{eq:golden}], so it
    is the tag nearest to any prediction of reduced norm.  We attribute this to
    the squared-error loss: where the decoder cannot tell which of several
    orbits passes through a pixel, or whether any does, the loss is minimized by
    a weighted average of the candidate tags and the zero vector, which has a
    reduced norm.
  }
  \label{fig:mae_scheme}
\end{figure}

\paragraph{Objective.}
The first pretraining stage follows the masked-autoencoder recipe
of Ref.~\cite{HeMae2022}.  Of the $256$ patches of an input image, a fraction
$\rho = 0.75$ is selected uniformly at random and hidden; the encoder sees only
the remaining $64$ tokens, to which a shared learned mask token and the
positional embeddings of the hidden patches are then appended, and a shallow
transformer decoder predicts the content of the hidden patches.  The
reconstruction target is the full pixel content of each masked patch in the
$30$-channel encoding, i.e.\ a $16\times16\times30$ block per patch, and the
loss is the mean squared error over masked patches only.  The decoder is
discarded after pretraining; only the encoder, including its tokenizer and
positional embeddings, is carried forward.
  Figure~\ref{fig:mae_scheme} sketches the scheme and shows reconstructions of
validation maps.

\paragraph{Masking before tokenization.}
In the original formulation the encoder is applied to the visible tokens alone,
which is exact because each token is a function of one patch only.  Our
tokenizer (Appendix~\ref{app:tokenizer}) is not patch-local: its three
$3\times3$ convolutions give each token a receptive field that extends up to $14$
pixels beyond its patch, so a token adjacent to a masked patch would carry
information about it and the pretext task would be partly trivial.  We
therefore zero the masked patches in the $256\times256$ raster \emph{before}
applying the tokenizer, and only then discard the masked tokens.  The
bookkeeping cost is that the tokenizer runs on the full image rather than on a
quarter of it; the encoder blocks, which dominate the cost, still see only the
visible tokens.

\paragraph{Hyperparameters.}
Encoder: ViT-L, depth $24$, width $1024$, $16$ heads, MLP dimension $4096$.
Decoder: depth $4$, width $1024$, $8$ heads.  Optimizer AdamW with base
learning rate $7\times10^{-5}$, weight decay $0.05$,
$\beta = (0.9, 0.95)$, batch size $128$.  Inputs are painted from the same
seven trajectory banks used for completion pretraining
(Appendix~\ref{app:bank}), with the window length drawn from
$T_{\mathrm{w}} \in [200, 2000]$ and $k \in [30, 200]$ orbits.  The run is
scheduled for $2000$ epochs; the checkpoint used to initialize the completion
stage of Sec.~\ref{sec:completion} is the one at epoch $1350$, while the
MAE-only arm of the ablation of Sec.~\ref{sec:results_pretraining} is
fine-tuned from the later checkpoint at epoch $1750$.  The MAE-only arm has
therefore received $400$ more epochs of masked-autoencoder pretraining than the
encoder the completion stage started from, so the comparison is if anything
conservative towards it.

\section{Long-time map completion: loss, schedule, and diagnostics}
\label{app:completion}

\paragraph{Objective.}
The completion network maps the short-window image to the long-time image in the
same $M=30$-channel golden-ratio encoding space, through the UNETR architecture of
Sec.~\ref{sec:architecture}, whose final $1\times1$ convolution outputs the $M$
encoding channels directly, with no output nonlinearity.
The loss is a per-pixel weighted mean squared error between predicted and target
encodings.  Pixels that no orbit visits in the target carry unit weight, while
visited pixels are upweighted by an overall factor $W_{\mathrm{fg}} = 10$; among
the latter we further distinguish regular from chaotic pixels with relative
weights $(w_{\mathrm{reg}}, w_{\mathrm{chaos}})$, rescaled such that
$w_{\mathrm{reg}}^2 + w_{\mathrm{chaos}}^2 = 2\,W_{\mathrm{fg}}^2$ so that equal
ratios recover the unsplit foreground weighting.  We use a raw ratio of $4{:}1$,
i.e.\ effective weights $13.7$ and $3.4$.
Downweighting the chaotic region is essential: the long-time colour of an
individual chaotic pixel is irreducibly unpredictable, and with equal weighting
this noise dominates the gradient and masks the learnable signal carried by the
regular curves.  The chaotic weight is nevertheless kept well above zero, so the
extent and density of the chaotic sea are still learned.
The regular/chaotic mask needed for this weighting is taken from the labels
already computed in Sec.~\ref{sec:data}; it enters \emph{only} through the loss
weights and is never presented to the network, which must therefore still infer
the location of the chaotic sea from the input alone.

\begin{figure}
  \centering
  \includegraphics[width=\columnwidth]{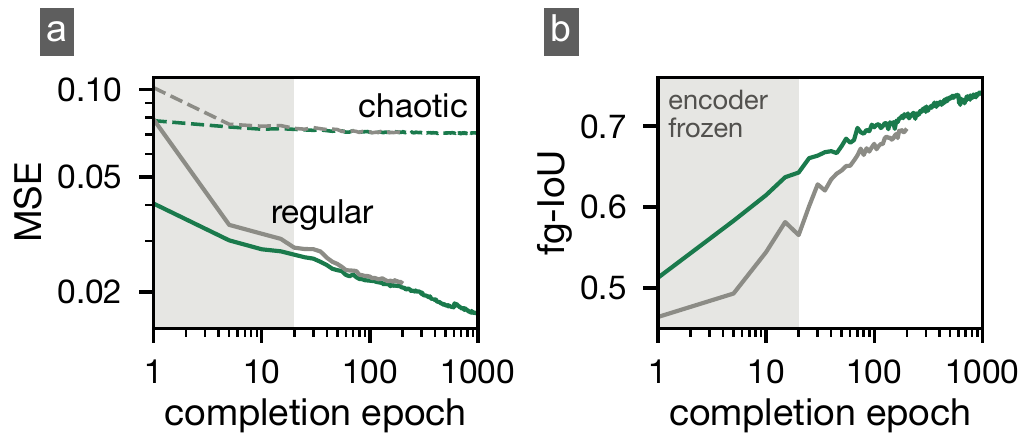}
  \caption{Long-time map completion with (green) and without (grey) the
    masked-autoencoder initialization of the encoder; apart from the
    initialization and the frozen phase (shaded), the two runs are identical.
    \textbf{(a)}~Validation mean squared error on the regular (solid) and
    chaotic (dashed) pixels of the target.
    \textbf{(b)}~Validation foreground-occupancy IoU.
    The run without MAE initialization was stopped at epoch $195$ of the same
    $600$-epoch schedule.  The step at epoch $600$ marks the restart of the
    MAE-initialized run when it was extended beyond its original schedule
    (see text).
    Training curves are not shown: they lie on top of the validation curves,
    within about $1\%$ in regular-region error and $0.005$ in fg-IoU.}
  \label{fig:completion_mae}
\end{figure}

\paragraph{Schedule.}
The completion stage is initialized from the masked-autoencoder encoder, with the
decoder  randomly initialized.  Fine-tuning a pretrained encoder together
with a random decoder distorts the pretrained features early on, so we adopt a
linear-probe-then-finetune schedule: the encoder is held frozen for the first
$K = 20$ epochs, during which the decoder trains at the full learning rate on top
of the fixed representation; the encoder is then unfrozen and the learning rate is
re-warmed over $20$ epochs before a cosine decay to $10^{-6}$ over the remaining
epochs.  Unlike the downstream fine-tuning runs we use a \emph{uniform} learning
rate across depth rather than layer-wise decay: layer-wise decay permanently
throttles the early encoder blocks, whereas the purpose of this stage is
precisely to let the encoder adapt to the dynamics.  We train for $600$ epochs
with AdamW (base learning rate $2.5\times10^{-4}$, weight decay $0.05$,
$\beta = (0.9, 0.95)$) and batch size $64$, over all seven trajectory banks, and
then extend the run to $1000$ epochs in two further segments, to $900$ and to
$1000$.  Each segment restarts from the saved weights with a fresh optimizer
state and the cosine schedule recomputed for the new end point.  At epoch $600$
this raises the learning rate from $3\times10^{-6}$ to $6\times10^{-5}$, which
costs the loss a visible step, recovered within about $100$ epochs
(Fig.~\ref{fig:completion_mae}); at epoch $900$ the recomputed learning rate is
only $6\times10^{-6}$ and the restart leaves no visible trace.  The checkpoint
used for Fig.~\ref{fig:architecture} and quoted in Sec.~\ref{sec:completion} is
the one at epoch $1000$.

\paragraph{Diagnostics.}
Progress is monitored with a deterministic evaluation---a fixed $500$-iteration
prefix as input against the full $2500$-iteration target---using two metrics
alongside the loss.  The first is a foreground-occupancy intersection over union
(fg-IoU), which declares a pixel occupied when the predicted encoding vector has
norm above $0.5$ and compares this support with the true long-time footprint; it
is more informative than the MSE where predicted colours become muddy, as they do
inside chaotic seas.  The second splits the foreground MSE into its regular and
chaotic parts.
Over the run, fg-IoU rises from $0.51$ at initialization to $0.643$ at the end of
the frozen phase and, after the encoder is unfrozen, continues smoothly to
$0.739$ by epoch $600$ and $0.741$ by epoch $1000$; the absence of any transient
at unfreezing confirms that the re-warmup
protects the pretrained features, and the size of the subsequent gain shows that
adapting the encoder---not merely training a decoder on frozen MAE
features---is where most of the benefit lies.
The split MSE (Fig.~\ref{fig:completion_mae}a, green) makes the nature
of the residual error explicit.  The regular-region error falls by more than a
factor of two, from $0.040$ to $0.017$ at epoch $1000$, and on the logarithmic
axes of the figure it follows an approximately straight line: it decreases as
a power of the epoch, the typical behaviour of a training curve that has not
saturated.  The exponent is about $0.1$ and steepens to about $0.16$ between
epochs $200$ and $600$, as the learning rate decays.  The chaotic-region error,
by contrast, is nearly flat, falling only from $0.078$ to $0.071$.  The plateau
in the aggregate loss is therefore not a capacity or optimization limit but the
unpredictability of individual chaotic orbits, and indeed the network renders
chaotic regions smoother than they are, approaching a conditional-mean density
rather than reproducing specific orbits.

The network does not overfit at any point of the $1000$ epochs.  The training
error on the regular pixels stays within about $1\%$ of the validation error,
and slightly above it, and the training and validation fg-IoU differ by less
than $0.005$; in Fig.~\ref{fig:completion_mae} the two would be
indistinguishable, so only the validation curves are shown.  This is the
behaviour anticipated in Sec.~\ref{sec:completion}: the target is the
long-time continuation of the particular orbits drawn, which changes with
every redraw, so there is nothing to memorize.  It contrasts with the
downstream segmentation tasks, whose targets are fixed per map and which do
overfit (Appendix~\ref{app:finetuning}).

\paragraph{Role of the masked-autoencoder initialization.}
To test whether the masked-autoencoder stage is still useful once completion
pretraining is in place, we repeated the completion run from a randomly
initialized encoder, with the same data, loss, optimizer and $600$-epoch
schedule; the only other change is that the encoder is not frozen at the start,
since freezing a random encoder serves no purpose.
Figure~\ref{fig:completion_mae} compares the two runs over the first $195$
epochs, where we stopped the run without MAE initialization.
The initialization does not affect the chaotic-region error, which converges to
the same value in both runs and is set by the unpredictability of the chaotic
orbits.  On the regular pixels, by contrast, the MAE-initialized run is ahead
throughout.  It reaches the regular-region error of the other run at epoch $190$
already at epoch $115$, and its fg-IoU at epoch $70$; at epoch $190$ the two stand
at $0.0210$ and $0.0217$ in regular-region error and at $0.715$ and $0.693$ in
fg-IoU.  The lead is not a head start that the run without MAE initialization
gradually makes up: the number of epochs by which it lags grows over training,
from about $10$ at epoch $50$ to about $75$ at epoch $190$, and the fg-IoU gap
stays at about $0.02$.  The masked-autoencoder stage is not free, however: one
of its epochs costs about $40\%$ of a completion epoch, so the $1350$ epochs that
precede the completion stage are worth about $540$ completion epochs.

\section{Downstream fine-tuning: optimization and losses}
\label{app:finetuning}

This appendix collects the machinery behind Sec.~\ref{sec:results}: the
optimization recipe shared by the three downstream tasks, the loss each one is
trained with, and the loss curves of the instance network.  The decoding of
the instance embeddings is described in Appendix~\ref{app:instance_inference},
the evaluation protocol and the figures of merit in
Appendix~\ref{app:evaluation}, and the secondary ablations of the instance
network in Appendix~\ref{app:ablations}.

\paragraph{Optimization.}
All three models start from the completion-pretrained ViT-L encoder---the
checkpoint after $1000$ epochs for binary segmentation, after $600$ epochs for
semantic and instance segmentation---and use a decoder of base width $64$ channels.  They share one recipe: AdamW
($\beta = 0.9, 0.95$) with weight decay $0.05$, batch size $64$, base learning
rate $2.5\times10^{-4}$ with $15$ warmup epochs followed by a flat phase and a
cosine decay to $10^{-6}$, layer-wise learning-rate decay $0.75$ across the
encoder blocks, and bfloat16 mixed precision.  Only the ViT weights are
transferred from the pretrained checkpoint; the decoder and the output layer are
randomly initialized.

\paragraph{Binary segmentation.}
The output is one logit per pixel and the target is the union of the chaotic sea
and the separatrices.  The loss is a pixel-weighted binary cross-entropy plus a
smoothed Jaccard term.  Separatrix pixels enter the cross-entropy with weight
$w_{\mathrm{sep}} = 4$, since they are a thin minority whose omission costs
almost nothing per pixel while destroying the very boundaries the task exists to
find.  The Jaccard term is computed on the \emph{regular} class rather than the
chaotic one, per image rather than pooled over the batch, and smoothed,
\begin{equation}
  L_{\mathrm{jacc}}
  = 1 - \Bigl\langle
      \frac{I_{\mathrm{reg}} + s}{U_{\mathrm{reg}} + s}
    \Bigr\rangle_{\mathrm{images}} ,
  \qquad s = 128~\mathrm{px},
  \label{eq:binary_jaccard}
\end{equation}
where, with the regular class as the positive class,
$I_{\mathrm{reg}} = \mathrm{TP}$ and
$U_{\mathrm{reg}} = \mathrm{TP} + \mathrm{FP} + \mathrm{FN}$, as in
Eq.~\eqref{eq:iou}.  To make the term differentiable, the pixel counts are
replaced by their soft versions: writing $p_i$ for the predicted probability
that pixel $i$ is regular and $y_i \in \{0,1\}$ for the ground truth,
$\mathrm{TP} = \sum_i p_i y_i$, $\mathrm{FP} = \sum_i p_i (1-y_i)$ and
$\mathrm{FN} = \sum_i (1-p_i)\, y_i$, the sums running over the pixels of one
image. The direction matters: scored on the chaotic
class, a regular island of $80$ pixels missed entirely in a $256\times256$ map
moves the overlap by about a part in a thousand, so no weighting of such a term
would make the network attend to it.  The smoothing caps the gradient a region
can demand in proportion to its size---a fully missed structure of $N$ pixels
costs $N/(N+s)$, so a one-pixel speck costs $0.008$ and an $80$-pixel island
$0.385$---which keeps stray pixels from counting as heavily as real structure.
The result is insensitive to $s$ in this range: $s = 64$ and $s = 128$ finish
within noise of each other.

\paragraph{Semantic segmentation.}
The output is $30$ logits per pixel: the chaotic class, the first-return
periods $1$ to $28$, and the  periods $\ge 29$ folded into the last class.
The loss is the cross-entropy plus a batch-level soft Jaccard term of weight
$4$, which counteracts the extreme class imbalance, for example a period-$17$ chain
occupies a few hundred pixels of one map in a hundred.
That term is built exactly as the mIoU of Sec.~\ref{sec:results} is, and
departs from it in two respects only: intersections and unions are accumulated
over the pixels of one batch rather than over the whole validation set, a batch
being the largest unit available while training, and they are accumulated from
the softmax probabilities rather than from the hard class assignment, so that
the term is differentiable.  Writing $p_{ic}$ for the predicted probability of
class $c$ at pixel $i$, $y_{ic}$ for the one-hot ground truth and $C_B$ for the
classes present anywhere in the batch,
\[
  L_{\mathrm{jacc}} = 1 - \frac{1}{|C_B|} \sum_{c \in C_B}
      \frac{\sum_i p_{ic}\, y_{ic}}
           {\sum_i \bigl( p_{ic} + y_{ic} - p_{ic}\, y_{ic} \bigr)} \, .
\]
Minimising it therefore maximises a soft estimate of the very quantity the
mIoU column of Table~\ref{tab:pretraining_avg_convention} reports.  Classes
absent from the batch are excluded rather than scored zero, so a period missing
from the current draw neither rewards nor penalises the network.

\paragraph{Instance segmentation.}
The two decoder branches output a $D = 10$ dimensional embedding
$\mathbf{e}_i \in \mathbb{R}^{D}$ and a foreground logit per pixel.  The logit
separates the regular regions from the chaotic sea under the same weighted
cross-entropy as the binary task.  The embedding is trained with a generalization of the
discriminative loss of Ref.~\cite{DeBrabandere2017}.  Let $C$ be the number of
instances labelled in the map, $N_c$ the pixel count of instance~$c$,
$\boldsymbol{\mu}_c = N_c^{-1}\sum_{i \in c}\mathbf{e}_i$ its mean embedding,
$[x]_{+} = \max(0,x)$, and $w_c$ a per-instance weight normalized to
$\sum_c w_c = 1$.  Then
\begin{equation}
\begin{aligned}
  L_{\mathrm{var}} &= \sum_{c} w_c \,\frac{1}{N_c} \sum_{i \in c}
     \bigl[\,\lVert \boldsymbol{\mu}_c - \mathbf{e}_i \rVert
     - \delta_{\mathrm{v}}\bigr]_{+}^{2} ,\\[2pt]
  L_{\mathrm{dist}} &= \frac{1}{1 - \sum_{c} w_c^{2}}
     \sum_{c_A \neq c_B} w_{c_A} w_{c_B}
     \bigl[\,2\delta_{\mathrm{d}} - \lVert \boldsymbol{\mu}_{c_A}
     - \boldsymbol{\mu}_{c_B} \rVert\bigr]_{+}^{2} ,\\[2pt]
  L_{\mathrm{reg}} &= \sum_{c} w_c \lVert \boldsymbol{\mu}_c \rVert ,
\end{aligned}
\label{eq:disc}
\end{equation}
and the embedding loss is $\alpha L_{\mathrm{var}} + \beta L_{\mathrm{dist}}
+ \gamma L_{\mathrm{reg}}$. For our training we choose  $\alpha = \beta = 1$, $\gamma = 10^{-4}$,
$\delta_{\mathrm{v}} = 0.7$ and $\delta_{\mathrm{d}} = 1.5$. The pull term $L_{\mathrm{var}}$ collects  hinged attractive forces that become exactly zero once the corresponding pixel embedding lies
within $\delta_{\mathrm{v}}$ of its own instance mean; the push term $L_{\mathrm{dist}}$ collects hinged repulsive forces between each pair of instances, which become zero once the
two instance means are $2\delta_{\mathrm{d}}$ apart. The specific choice of the hyperparameters $\delta_v$ and $\delta_d$ is explained in Appendix~\ref{app:instance_inference}. The weights $w_c$ are the one place where we depart from
Ref.~\cite{DeBrabandere2017}.  We weight each instance by a power
of its area, $w_c \propto N_c^{\,p}$. Setting $p=0$  recovers the loss function of  Ref.~\cite{DeBrabandere2017}, so
Eq.~\eqref{eq:disc} is a one-parameter generalization of it. We use $p = 1/2$
throughout, a compromise between equal weight for every instance ($p = 0$) and
equal weight for every pixel ($p = 1$). Sec.~\ref{sec:debrabandere} and
Appendix~\ref{app:sizeweight} compare the three empirically.  Decoding at inference is described in
Appendix~\ref{app:instance_inference}.

\begin{figure}[t]
  \centering
  \includegraphics[width=50mm]{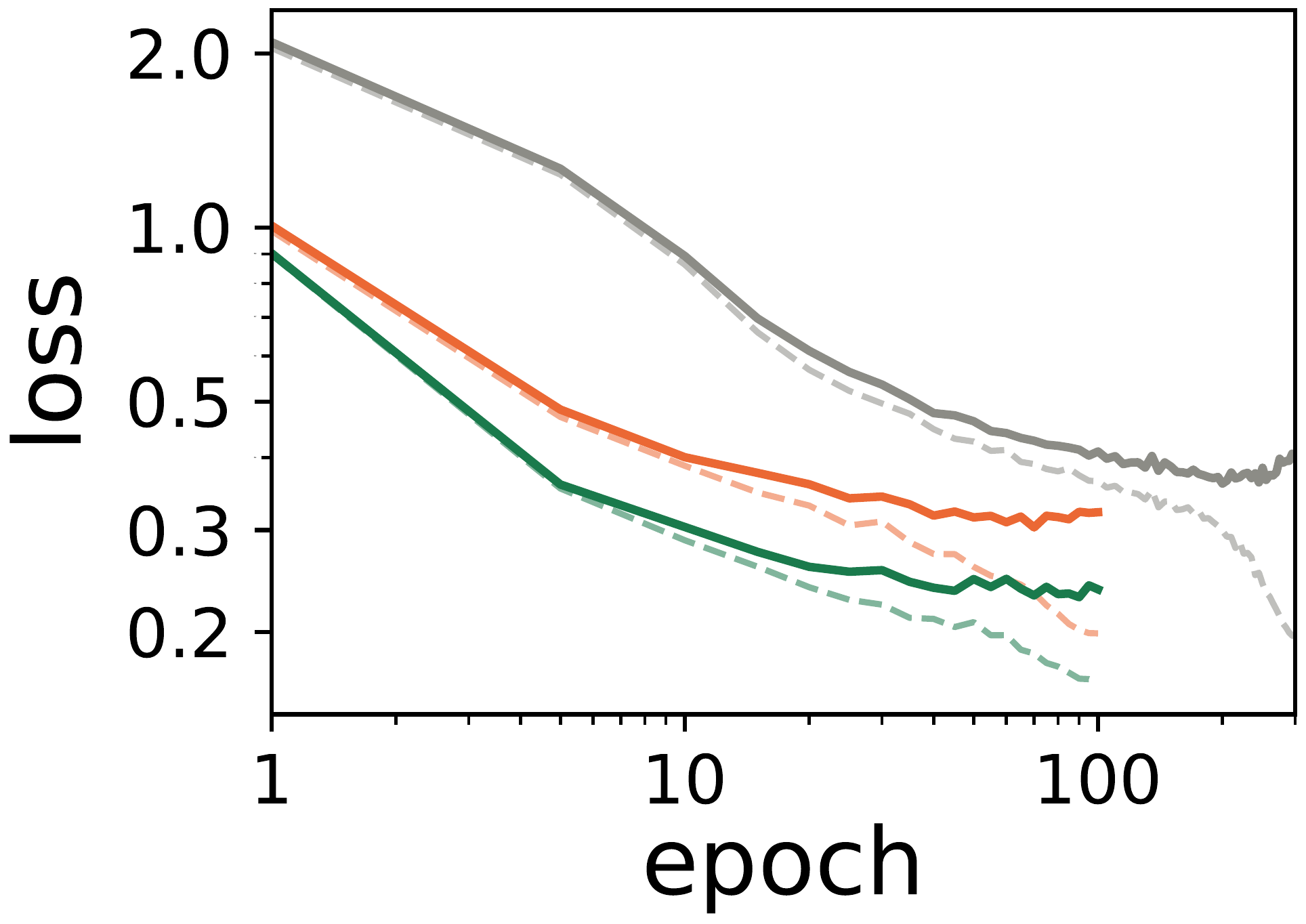}
  \caption{Loss during instance-segmentation fine-tuning of the three arms of
  Fig.~\ref{fig:pretraining_ablation}: no pretraining (grey, $300$ epochs), MAE
  (orange) and map completion (green), both $100$ epochs.  Total loss, i.e.\ the
  discriminative loss plus the foreground cross-entropy, with $p = 1/2$.  It is
  evaluated in evaluation mode (no drop-path, no augmentation) on the single
  painting $k = 150$, $T_{\mathrm{w}} = 1000$, for the validation maps (solid) and
  for training maps with the composition of the validation set (dashed).}
  \label{fig:instance_loss}
\end{figure}

\paragraph{Loss curves.}
Unlike the completion pretraining (Appendix~\ref{app:completion}), the
fine-tuning overfits in the loss, because its targets are fixed per map.
Fig.~\ref{fig:instance_loss} shows the loss of the instance network for the
three pretraining arms of Fig.~\ref{fig:pretraining_ablation}.  In every run
the training loss is still decreasing at the end of the schedule, whereas the
validation loss levels off after $200$, $70$ and $90$ epochs (no pretraining,
MAE, completion) and then rises by $0.04$, $0.02$ and $0.01$.  The PQ of
Fig.~\ref{fig:pretraining_ablation}(c) does not follow the loss and keeps
increasing, and the ordering of the arms by loss is the same as by PQ.  The
curves of the map-completion arm in both figures come from a second run of the
recipe of the network used in the paper: the first run did not store
the PQ, and we repeated it, which also gives a sample of the run-to-run
variation.%

\section{Instance decoding}
\label{app:instance_inference}

\begin{figure}[t]
  \centering
  \includegraphics[width=\columnwidth]{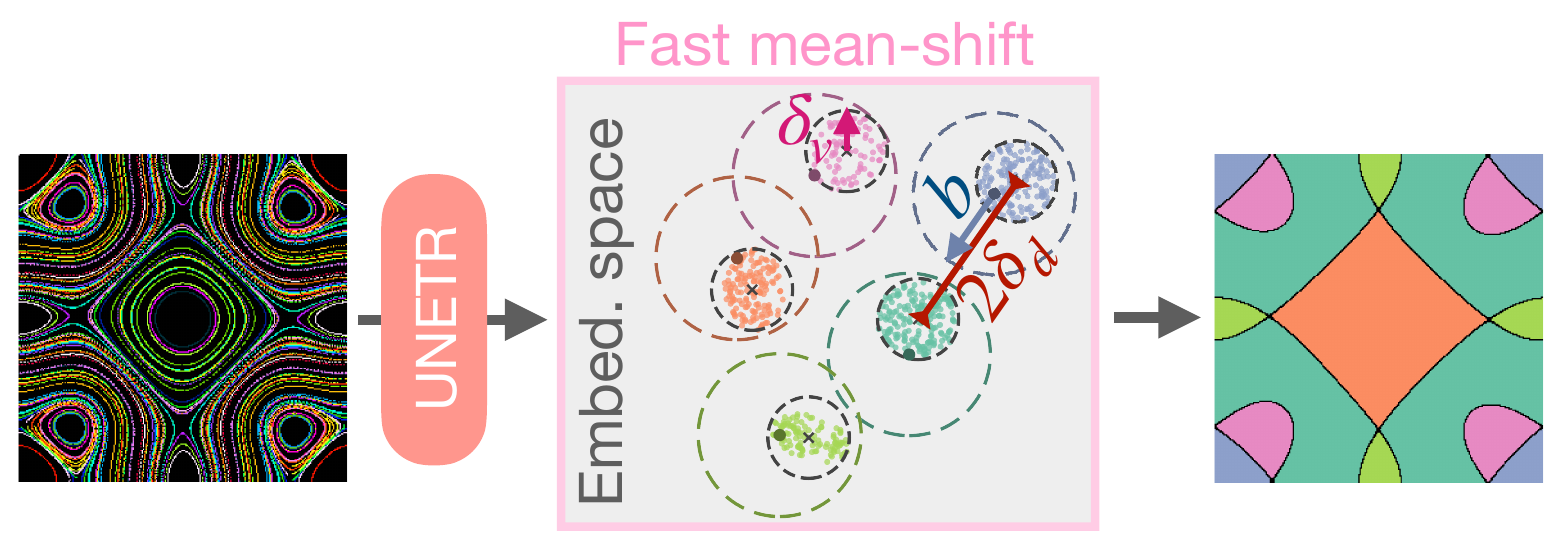}
  \caption{Instance decoding with the fast mean shift of
    Ref.~\cite{DeBrabandere2017}.  The UNETR maps the input painting (left) to
    an embedding vector for every foreground pixel (centre, sketched in two
    dimensions and coloured by instance).  In the ideal case of zero
    discriminative loss, every embedding (dots) lies within a ball of radius $\delta_{\mathrm{v}}$ (small dashed circles) centred on
    its instance mean (crosses), and the means of distinct
    instances are at least $2\delta_{\mathrm{d}}$ apart.  A ball of radius
    $b = 2\delta_{\mathrm{v}}$ (large dashed circles) about a random seed (larger, darker dots) then
    contains its whole instance and no other, so a single shift carries the
    seed to the instance mean, and the pixels within $b$ of it form one
    instance of the predicted map (right).}
  \label{fig:instance_decoding}
\end{figure}

The instance-segmentation head outputs, for each pixel~$i$, an embedding
vector $\mathbf{e}_i\in\mathbb{R}^D$ and a foreground logit.  A pixel is
foreground if $\sigma(\text{logit}) > 0.5$, with $\sigma$ the sigmoid
function, and the embeddings of the foreground pixels are then grouped into
instances.  The number of instances is unknown and varies from map to map,
which rules out clustering algorithms that require it, but the scale of a
cluster is set by the margins of the discriminative loss,
Eq.~\eqref{eq:disc}.  Mean-shift clustering~\cite{Comaniciu2002}, whose only
parameter is a length, the bandwidth~$b$, suits exactly this situation.  We
describe three variants of it, from the fastest to the slowest.  All use a flat
kernel, so that a shift replaces a point $\mathbf{x}$ by the mean of the
embeddings within $b$ of it, and the same bandwidth $b = 2\delta_{\mathrm{v}}
= 1.4$.  The second variant is used for the results of
Sec.~\ref{sec:results}, the third for the deployment of
Sec.~\ref{sec:deployment}.

\paragraph{Fast mean shift.}
This is the approach of Ref.~\cite{DeBrabandere2017}. It recovers one instance at a time
(Fig.~\ref{fig:instance_decoding}).  A foreground pixel that is not yet
assigned is drawn at random as a seed and shifted to convergence over the
unassigned embeddings alone; the pixels within $b$ of the converged point form
one instance and are removed from the pool.  The procedure repeats until at
most $100$ pixels remain. This threshold,  taken over from the released implementation of Ref.~\cite{DeBrabandere2017}, helps prevent building false positive clusters from pixels falling just outside the bandwidth of  real clusters. Finally, every foreground pixel is assigned to its nearest mode.

The choice of the margins and of the bandwidth is clearest in the ideal case of
zero loss.  Every embedding then lies within $\delta_{\mathrm{v}}$ of its
instance mean, and distinct means are at least $2\delta_{\mathrm{d}}$ apart, so
two embeddings of the same instance are at most $2\delta_{\mathrm{v}}$ apart and
two embeddings of different instances at least
$2(\delta_{\mathrm{d}} - \delta_{\mathrm{v}})$.  If
$\delta_{\mathrm{d}} > 2\delta_{\mathrm{v}}$, as for our
$\delta_{\mathrm{v}} = 0.7$ and $\delta_{\mathrm{d}} = 1.5$, every embedding is
therefore closer to all embeddings of its own instance than to any embedding of
another~\cite{DeBrabandere2017}.  A ball of radius $b$ about a seed of instance
$A$ then contains every embedding of $A$ if $b \geq 2\delta_{\mathrm{v}}$, and
none of any other instance if $b < 2(\delta_{\mathrm{d}} - \delta_{\mathrm{v}})$.
In particular, for
\begin{equation}
  2\delta_{\mathrm{v}} \;\leq\; b \;\leq\; \delta_{\mathrm{d}}
  \label{eq:bandwidth-window}
\end{equation}
the first shift lands exactly on the mean $\boldsymbol{\mu}_A$ and the next
leaves it there: the fast mean shift converges in one step, and the pixels
within $b$ of $\boldsymbol{\mu}_A$ are exactly those of $A$.  We take
$b = 2\delta_{\mathrm{v}}$, the smallest bandwidth in this window and thus the
least able to bridge two instances once the loss is no longer zero; the
released implementation of Ref.~\cite{DeBrabandere2017} uses the other end,
$b = \delta_{\mathrm{d}}$.

The trained embedding does not reach the ideal case: on a representative map
the largest recovered cluster has a $95$th-percentile radius of $1.9$, against
the $\delta_{\mathrm{v}} = 0.7$ the loss aims for.  Real clusters are thus two
to three times wider than the margins intend, and the one-step guarantee no
longer holds.  The fast mean shift then has a structural weakness: because it
removes every emitted ball and continues until almost no pixel is left, it
raises the embeddings in sparse regions, which belong to no dense cluster,
into instances of their own.  It also seeds at random, so its partition
changes from run to run.

\paragraph{Mean shift with bin seeding.}
Standard mean shift avoids both problems.  Every seed is shifted to
convergence over \emph{all} foreground embeddings, converged points closer
than $b$ are merged into one mode, and every foreground pixel is assigned to
its nearest mode; no cluster is discarded.  Embeddings in sparse regions thus
join an existing instance rather than creating one.  Seeding from every
embedding finds every mode but is expensive, $26$~s per map on average for the
maps of Sec.~\ref{sec:deployment}.  Bin seeding, the standard acceleration,
instead starts the search from the nodes of a grid of spacing $b$ that are the
nearest node of at least three embeddings, which also makes the decoding
deterministic.  Bin seeding is a standard option of mean shift, available for instance in scikit-learn~\cite{Pedregosa2011}, and in the main text we refer to this decoder simply as mean shift.  Within the window of
Eq.~\eqref{eq:bandwidth-window} the bandwidth matters little: under the protocol of
Table~\ref{tab:pretraining_instance}, the type-aware ARI (Appendix~\ref{app:ari}) is
$0.800$ at $b = 2\delta_{\mathrm{v}}$ and $0.799$ at $b = \delta_{\mathrm{d}}$, and PQ
$0.570$ and $0.566$.

\paragraph{Fast against binned mean shift.}
We compared the two decoders on the same embeddings, those of the $p = 1/2$
map-completion model, under the protocol of
Table~\ref{tab:pretraining_instance}.  The fast mean shift costs $35$
against $203$~ms per map and recovers slightly more of the true instances
($37{,}263$ against $35{,}175$ matched at $\mathrm{IoU} > 0.5$), but it emits
almost twice as many spurious ones ($19{,}261$ against $9{,}963$), so that PQ
falls from $0.570$ to $0.542$ and the type-aware ARI from $0.800$ to $0.730$
(Sec.~\ref{sec:inference_time}, Table~\ref{tab:other_fom}).  The disagreement is
confined to these spurious detections: scored by the mBIoU
(Appendix~\ref{app:other_fom}), which does not penalize extra instances, the
fast mean shift is even slightly ahead ($0.656$ against $0.649$).  We therefore
report mean shift with bin seeding throughout.  Decoding is not the
bottleneck, since the forward pass alone costs $87$~ms per map, and only a
decoder that does not raise the residue into instances makes the count of
recovered instances meaningful.  The fast mean shift is used only for the
per-epoch evaluations during training, such as the PQ curves of
Fig.~\ref{fig:pretraining_ablation}(c), which it places $0.03$ to $0.04$ below
mean shift with bin seeding. Our experiments confirmed that  the stopping rule
 at $100$ remaining pixels is beneficial. Without such rule, under the same protocol, the fast mean shift recovers about $2\%$ more of the true instances ($38{,}000$ against $37{,}217$) but emits $85\%$ more
spurious ones ($35{,}929$ against $19{,}371$), and PQ falls from $0.541$ to $0.479$.

\paragraph{Refining pass.}
Bin seeding can miss an instance.  A grid node can lie up to
$(b/2)\sqrt{D} \approx 2.2$ from the embeddings it stands for, farther than
the bandwidth itself, so a search started there may climb to a neighbouring
mode, and a small cluster may receive no seed at all.  Its pixels are then
assigned to a distant mode, and two instances that are well separated in
embedding space receive one label.  A second pass repairs this.  It is seeded
at the pixels left unassigned by the first, those farther than $b$ from every
mode found in it, keeps the new modes with at least three embeddings within
$b$, merges modes closer than $b$, and only then assigns every pixel to its
nearest mode, which again removes no cluster.  On the thirteen maps of
Fig.~\ref{fig:ood_deployment}, counting each window of the three-body problem as a map, the refining pass raises the decoding time from
$0.08$ to $0.69$~s per map on average, still about forty times less than
seeding from every embedding.  On the validation maps it leaves the PQ
unchanged within $0.001$ while adding about $4\%$ more false positives, so the
results of Sec.~\ref{sec:results} use the first pass alone.  In the deployment
of Sec.~\ref{sec:deployment}, where it recovered small resonances in several
examples, the refining pass is applied throughout.

All timings are averages on one core of an Apple M4 Pro.

\section{Deriving the period from an instance map}
\label{app:derived}

\begin{figure}[t]
  \centering
  \includegraphics[width=\columnwidth]{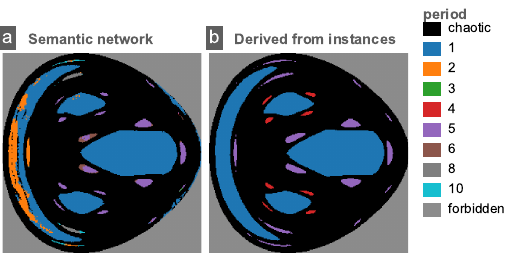}
  \caption{First-return period on the H\'enon--Heiles section at $E = 1/8$
    of Fig.~\ref{fig:overview}.
    \textbf{(a)}~Direct prediction of the semantic network, which is trained
    only on doubly periodic maps.
    \textbf{(b)}~Period derived from the predicted instance map and the
    dynamics, as shown in Fig.~\ref{fig:overview}.  Grey marks the
    energetically forbidden region.}
  \label{fig:hh_semantic}
\end{figure}

The semantic network can only count the islands of a chain that are all
visible, which is guaranteed only on doubly periodic maps, so it is trained and
applied on those alone (Sec.~\ref{sec:results}).  Even where every
island is visible, we found that it generalizes poorly to the plane topology, see Fig.~\ref{fig:hh_semantic}. On every
other map we therefore derive the period map from the predicted instance map and the dynamics.  This
derived route supplies the semantic panel of Fig.~\ref{fig:overview}(c), a
section of the H\'enon--Heiles system at $E = 1/8$, and the two cells of
Fig.~\ref{fig:segmentation_results}(d) whose phase space is not doubly
periodic.  It is also the route by which the systems of
Sec.~\ref{sec:deployment} would be labelled by period.

An instance is invariant under the map, so its return time is trivially one; the
period must come from the instance's sub-structure.  We label the connected
islands of each predicted instance, launch a probe orbit from inside one of
them, and record the sequence of islands the orbit visits.  The period is the
smallest lag $N$ for which that sequence is consistent with itself at
\emph{every} observed pair, i.e.\ the orbit is in the same island at step $k$ and
step $k+N$ wherever both are recorded.  Scanning lags rather than following
consecutive iterates is what makes the estimator work under open boundaries: the
orbit spends most of its time outside the window, so consecutive in-window pairs
are rare while pairs separated by the period survive.  When the whole accessible phase space is shown, a second, purely spatial
estimator---the number of islands in the instance---provides an independent
check.  Our implementation can use it, together with probe orbits
launched from several islands of the instance, as a sanity check: a period is
then assigned only when the two estimators agree or when probe orbits from two
or more islands return the same value, and it is refused when two islands
return different periods, since an instance carrying two periods merges two
chains.  These sanity checks are useful whenever no ground truth is available.
Finally, we note that the dynamics enter
only through a function advancing pixel coordinates, which is supplied by the
discrete map, or by integrating the flow and intersecting it with the section
for the continuous systems of Sec.~\ref{sec:deployment}.

\section{Disagreements due to the ground truth or the input}
\label{app:failure_modes}

\begin{figure}
  \centering
  \includegraphics[width=\columnwidth]{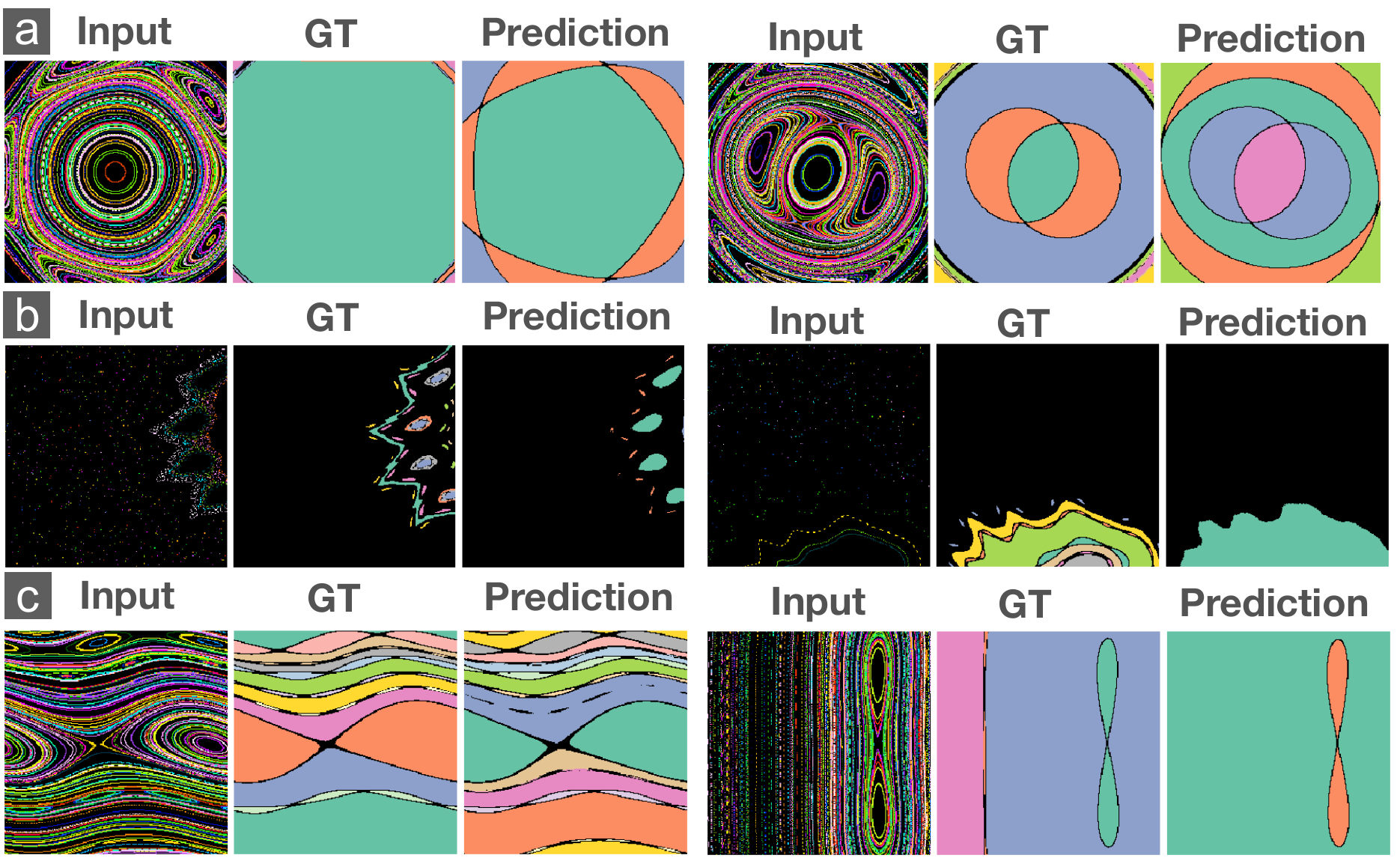}
  \caption{
    Disagreements due to the ground truth (GT) or the input, on validation maps.
    \textbf{(a)}~Two examples in which the GT pipeline failed to identify a separatrix, which causes an
    obvious mistake in the instance-segmentation GT.  The network correctly identifies the missing instances.
    \textbf{(b)}~Two examples in which the painting is too sparse to give any clue about the fine structure
    of the GT.  The network predictions are consistent with the input it sees.
    \textbf{(c)}~Even for dense paintings, very narrow island chains that are not visited by any orbit, and are
    thus not visible to the bare eye in the input, may affect the segmentation substantially (right).  The network
    sometimes also detects very tiny island chains that are not present in the GT because of the filtering of
    small islands (bottom chain in the left panel).}
  \label{fig:gt_examples}
\end{figure}

Not every disagreement between prediction and ground truth is an error of the
network.  Figure~\ref{fig:gt_examples} shows disagreements of two kinds in
which the network does its job and the disagreement comes from elsewhere: the
ground truth is wrong, or the input does not contain the information needed to
reproduce it.  Panels~(a) and~(c) show the dense painting ($k=150$,
$T_{\mathrm{w}}=1000$), panels~(b) the sparsest painting of the validation
protocol ($k=30$, $T_{\mathrm{w}}=500$).

\paragraph{Wrong ground truth.}
In panel~(a) the ground-truth pipeline missed a separatrix, so that regions
whose orbits are qualitatively distinct, and clearly separated in the input,
were merged into a single instance; the network separates them.  Conversely,
the network sometimes finds tiny island chains that are absent from the ground
truth because small islands are filtered out during its generation (bottom
chain in the left panel of~(c)).

\paragraph{Missing information in the input.}
In panel~(b) the painting is too sparse to show the fine structure of the
ground truth, which is computed from a dense covering of phase space; the
prediction is consistent with what the network sees.  Even a dense painting
can miss a very narrow island chain that no painted orbit visits (panel~(c),
right): the chain is invisible in the input, yet it changes the ground-truth
segmentation substantially.

Disagreements of both kinds count against the network in every figure of
merit of Sec.~\ref{sec:results}.

\section{Evaluation protocol and metrics}
\label{app:evaluation}

\subsection{Painting the evaluation inputs}
\label{app:protocol}

All tables use the protocol of Sec.~\ref{sec:quantitative}.  Binary
segmentation is scored on all $997$ validation maps, semantic segmentation on
the $451$ doubly periodic ones, the topology the semantic network is trained
on, and instance segmentation on a clean set of $995$ maps, which excludes two
maps with defective ground truth; the sweeps of
Fig.~\ref{fig:pretraining_ablation}(d,e) use all $997$.  A segmentation score
depends on how densely the input was painted, and over the training ranges that
dependence has not saturated, so a score means little without the painting it
was measured at.  The $T_{\mathrm{w}} = 200$ column of the grid of
Fig.~\ref{fig:segmentation_results}(b) is excluded from the scores because it
falls below the fine-tuning floor of $300$; the figure shows it to illustrate
the degradation outside the training range.  This leaves six paintings.  For
instance segmentation, the counts entering PQ are pooled over the maps of each
painting, and PQ, RQ and SQ are then averaged over the six paintings.

The training curves and per-class scores of
Figs.~\ref{fig:pretraining_ablation}(a--c) are the trainers' own per-epoch
evaluations at the single painting $k = 150$, $T_{\mathrm{w}} = 1000$, so their
final values sit a little above the corresponding table entries.  For the
instance curve of panel (c) this is offset by its decoding with the fast mean
shift (Appendix~\ref{app:instance_inference}), which reads $0.03$ to $0.04$
lower in PQ than mean shift with bin seeding on the same painting.  Every input
is painted from a fixed per-map seed, so each number is deterministic and
train--validation comparisons are like for like.

\subsection{Panoptic quality}
\label{app:pq}

The panoptic quality~\cite{KirillovPanoptic2019} rests on a matching between
predicted and ground-truth instances.  A ground-truth instance $g$ and a
predicted instance $p$ are matched when $\mathrm{IoU}(g,p) > 1/2$; since the
predicted instances are disjoint, as are the ground-truth ones, at most one
predicted instance can pass this threshold with a given ground-truth instance,
and vice versa, so the matching is unique~\cite{KirillovPanoptic2019}.  The
matched pairs form the set $\mathrm{TP}$ of true positives, the unmatched
ground-truth instances the false negatives $\mathrm{FN}$, and the unmatched
predicted instances the false positives $\mathrm{FP}$.  Only regular instances
are scored; the chaotic sea is not an instance.  Then, $\mathrm{PQ}$ is the product of the recognition quality,
\begin{equation}
  \mathrm{RQ} = \frac{|\mathrm{TP}|}
       {|\mathrm{TP}| + \tfrac{1}{2}|\mathrm{FP}| + \tfrac{1}{2}|\mathrm{FN}|},
\end{equation}
and the segmentation quality,
\begin{equation}
   \mathrm{SQ} = \frac{1}{|\mathrm{TP}|}\sum_{(g,p)\in\mathrm{TP}}\mathrm{IoU}(g,p).
  \label{eq:pq}
\end{equation}
RQ is
the harmonic mean of the precision
$|\mathrm{TP}|/(|\mathrm{TP}|+|\mathrm{FP}|)$ and the recall
$|\mathrm{TP}|/(|\mathrm{TP}|+|\mathrm{FN}|)$, and so charges for missed and for
spurious instances alike; the segmentation quality is the mean IoU of the
matched pairs alone.  As is standard, the counts and the sum of IoUs are
accumulated over all maps of the validation set before the ratios are formed,
so that every matched instance carries the same weight regardless of its size
and of the map it belongs to.

Because SQ averages over the matched pairs only, it depends on which instances
are detected.  This matters for the comparison of
Table~\ref{tab:pretraining_instance}: from no pretraining to map completion the
number of true positives grows by $24\%$, and the growth is concentrated on
small and medium instances.  The recall over ground-truth instances of
$100$ to $400$ pixels rises from $0.185$ to $0.306$, and over those of $400$ to
$1600$ pixels from $0.420$ to $0.593$, against $0.928$ to $0.960$ above $6400$
pixels.  Small instances are the hardest to delineate, so these additional
matches lower SQ, and its gain of $3.9\%$ is a conservative measure of the
improvement in delineation.

\subsection{Mean best IoU and adjusted Rand index}
\label{app:other_fom}

\begin{table}[t]
  \caption{
    Mean best IoU and type-aware ARI for every instance network of
    Tables~\ref{tab:debrabandere} and~\ref{tab:pretraining_instance}, and for
    the fast mean shift of Appendix~\ref{app:instance_inference}, under the
    same protocol.  Best value within each block in bold, ground-truth rows
    excluded.}
  \label{tab:other_fom}
  \begin{ruledtabular}
  \begin{tabular}{lcc}
    Network & mBIoU & ARI \\
    \hline
    \multicolumn{3}{l}{pretraining ($p = 1/2$)} \\
    \quad none                 & 0.570 & 0.723 \\
    \quad MAE                     & 0.606 & 0.761 \\
    \quad map completion          & \textbf{0.649} & \textbf{0.800} \\
    \multicolumn{3}{l}{size weighting} \\
    \quad $p = 0$                 & \textbf{0.659} & 0.795 \\
    \quad $p = 1/2$               & 0.649 & \textbf{0.800} \\
    \quad $p = 1$                 & 0.636 & 0.794 \\
    \multicolumn{3}{l}{foreground mask ($p = 1$)} \\
    \quad separate network        & 0.635 & 0.785 \\
    \quad second head             & \textbf{0.636} & \textbf{0.794} \\
    \quad ground truth, no head   & 0.749 & 0.924 \\
    \quad ground truth, second head & 0.757 & 0.927 \\
    \multicolumn{3}{l}{clustering ($p = 1/2$)} \\
    \quad fast mean shift         & \textbf{0.656} & 0.730 \\
    \quad binned mean shift             & 0.649 & \textbf{0.800} \\
  \end{tabular}
  \end{ruledtabular}
\end{table}

Besides PQ we computed two further figures of merit, which the main text does
not use.  The mean best IoU (mBIoU), also known as the average best
overlap~\cite{Uijlings2013}, is the mean over ground-truth instances of the best
IoU any predicted instance achieves with them, formed per map and then averaged
over maps; it belongs to the family of best-overlap scores that includes the
symmetric best Dice reported for leaf segmentation~\cite{DeBrabandere2017}.  It
weights every ground-truth instance equally but does not charge for spurious
predictions, which cost nothing unless they take area from a real instance.  The
type-aware adjusted Rand index (ARI), defined below, scores the agreement of the
two partitions over pixel pairs, map by map; counting pairs, it is dominated by
the largest instance in each map, a $50$-pixel instance counting for roughly
$10^{-4}$ of a $5000$-pixel one.

Table~\ref{tab:other_fom} gives both for every network of the main text.  The
pretraining comparison is unaffected: the ordering no pretraining, MAE, map
completion holds under both.  In the comparisons of
Table~\ref{tab:debrabandere} and of the two clustering algorithms, however, mBIoU favours the variant that produces
more predicted instances---uniform weighting over $p = 1/2$, and the fast
mean shift over binned mean shift---because it rewards the additional true detections and
ignores the additional false ones.  This bias, which best-overlap scores share,
is why we use PQ as our figure of merit.

\paragraph{Type-aware adjusted Rand index.}\label{app:ari}
We evaluate instance segmentation quality using the Adjusted Rand
Index (ARI)~\cite{Hubert1985}.  Given a ground-truth partition and a
predicted partition of all~$n$ pixels, one forms the
$R \times C$ contingency table~$n_{ij}$ counting pixels assigned to
ground-truth cluster~$i$ and predicted cluster~$j$.
The background (chaotic sea) is treated as cluster~0 in both
partitions.
The Rand Index
measures the fraction of pixel pairs that are either co-clustered in
both partitions or separated in both.  The ARI corrects for chance by
subtracting the expected value under a hypergeometric null model:
\begin{equation}
  \mathrm{ARI}
  = \frac{\sum_{ij}\binom{n_{ij}}{2}
          - \bigl[\sum_i\binom{a_i}{2}\bigr]
            \bigl[\sum_j\binom{b_j}{2}\bigr]
            \big/\binom{n}{2}}
         {\tfrac{1}{2}\bigl[\sum_i\binom{a_i}{2}
          + \sum_j\binom{b_j}{2}\bigr]
          - \bigl[\sum_i\binom{a_i}{2}\bigr]
            \bigl[\sum_j\binom{b_j}{2}\bigr]
            \big/\binom{n}{2}}\,,
  \label{eq:ari}
\end{equation}
where $a_i = \sum_j n_{ij}$ and $b_j = \sum_i n_{ij}$ are  the sizes of the ground-truth and predicted clusters.  The
ARI is the linear rescaling of the Rand index that is~0 in expectation for
random partitions with these cluster sizes and~1 for a perfect
match~\cite{Hubert1985}.

Equation~\eqref{eq:ari} is invariant to any relabeling of the
clusters, so it does not distinguish the background (chaotic sea) from
an instance: a co-clustered pair of pixels is credited whether they
are grouped together as background in one partition and as an instance
in the other.  The natural alternative, scoring the foreground
alone, fails on a common class of Poincar\'e maps: those whose regular
foreground forms a single instance, such as one island in a chaotic sea or a
nearly integrable map without separatrices.  Restricted to the foreground,
the ground-truth partition of such a map is a single cluster, whose expected
chance agreement equals its maximum, so the ARI is exactly zero for every
prediction that is not itself a single cluster, and undefined for one that
is.  A segmentation that is perfect apart from a few stray pixels assigned to
a second instance---$8$ of $1024$ in a test case---therefore scores $0$,
whereas the same defect on a map with two instances costs only $0.015$.
Including the background as cluster~0 removes the degeneracy, since the
ground-truth partition then always has at least two clusters, and the
near-perfect prediction above scores $0.998$.  Naive inclusion, however, introduces the
opposite failure: a prediction that floods a largely chaotic map with
a single spurious instance forms two matching large regions (true
background versus predicted instance) and is scored as a near-perfect
match.

We therefore adopt a \emph{type-aware} ARI.  Each cluster is assigned
a type---chaotic (cluster~0) or regular (any instance label)---and a
co-clustered pixel pair counts as agreement only when its
ground-truth and predicted types coincide.  Concretely, we restrict
both the observed and the expected agreement in Eq.~\eqref{eq:ari} to
contingency cells of matching type by inserting the indicator
$s_{ij}=\delta_{\tau(i)\tau(j)}$, where $\delta_{lk}$ is the Kronecker delta, $\tau(0)=0$, and
$\tau(j)=1$ otherwise:
\begin{equation}
  \mathrm{ARI}_{\mathrm{type}}
  = \frac{\sum_{ij}s_{ij}\binom{n_{ij}}{2}
          - \sum_{ij}s_{ij}\binom{a_i}{2}\binom{b_j}{2}\big/\binom{n}{2}}
         {\tfrac{1}{2}\bigl[\sum_i\binom{a_i}{2}+\sum_j\binom{b_j}{2}\bigr]
          - \sum_{ij}s_{ij}\binom{a_i}{2}\binom{b_j}{2}\big/\binom{n}{2}}\,.
  \label{eq:ari-type}
\end{equation}
Dropping the cross-type cells---ground-truth background paired with a
predicted instance, or vice versa---removes their contribution from
both terms of the numerator.  The metric coincides with the standard
ARI to within ${\sim}0.02$ whenever the predicted
foreground/background split is correct, still rewards correctly
identifying the chaotic sea (background matched to background), and
assigns a near-zero score to background floods and other
type-confusions.  It remains invariant to permutations of the instance
labels within each type.  When both partitions consist of a single cluster of the
same type---a fully chaotic map predicted to contain no instance, or a map
that is one regular instance predicted as such---Eq.~\eqref{eq:ari-type} is
undefined, and we score this correct prediction as~1.

\section{Secondary ablations of the instance network}
\label{app:ablations}

This appendix gives further details on the two ablations of the instance
network summarized in Sec.~\ref{sec:debrabandere} and
Table~\ref{tab:debrabandere}.  Both concern choices in which our recipe departs
from Ref.~\cite{DeBrabandere2017}: the weighting of the instances in the
discriminative loss (Appendix~\ref{app:sizeweight}) and the prediction of the
foreground by a second head of the instance network
(Appendix~\ref{app:fghead}).  All networks compared here start from the
$600$-epoch completion-pretrained encoder and are scored with the protocol of
Appendix~\ref{app:protocol}.

\subsection{Size weighting}
\label{app:sizeweight}

The discriminative loss, Eq.~\eqref{eq:disc}, weights each instance by a power
of its area, $w_c \propto N_c^{\,p}$ (Appendix~\ref{app:finetuning}).
Weighting by area ties the influence of an instance to the evidence that the
painting provides for it, but it can be pushed too far.  With $p = 1$ the
variance term is simply a mean over all foreground pixels, and the smallest
instances, which are real structure whenever the painting does sample them,
receive too little of the gradient to be learned.  We compared uniform
weighting ($p = 0$, as in Ref.~\cite{DeBrabandere2017}), area weighting
($p = 1$) and the intermediate $p = 1/2$
[Table~\ref{tab:debrabandere}(a)].

Table~\ref{tab:debrabandere}(a) gives PQ, RQ and SQ for the three
weightings, and Table~\ref{tab:other_fom} the mBIoU and ARI.  Area weighting
is the worst choice under every figure of merit except SQ.  Uniform weighting
and $p = 1/2$ are close: $p = 1/2$ leads by $0.004$ in PQ and by $0.005$ in
ARI, the two are level in RQ, and only the mBIoU, which is blind to false
positives, prefers $p = 0$.  We use $p = 1/2$ throughout the paper.

The lead of $p = 1/2$ over $p = 0$ is small but exceeds the run-to-run
variation, which we estimate from two pairs of runs of the same recipe, the
$p = 1/2$ network with a repeat and two $p = 0$ networks: their PQ, averaged
over the six paintings, differ by $0.0001$ and $0.0006$.  The same lead, $0.003$
to $0.004$, is found at $200$ fine-tuning epochs and with or without discarding
clusters of fewer than $20$ pixels in the decoding.  The PQ of the single
painting evaluated during training fluctuates by $0.01$ and more between runs
and is not a measure of the final scores.

\begin{figure}[t]
  \centering
  \includegraphics[width=\columnwidth]{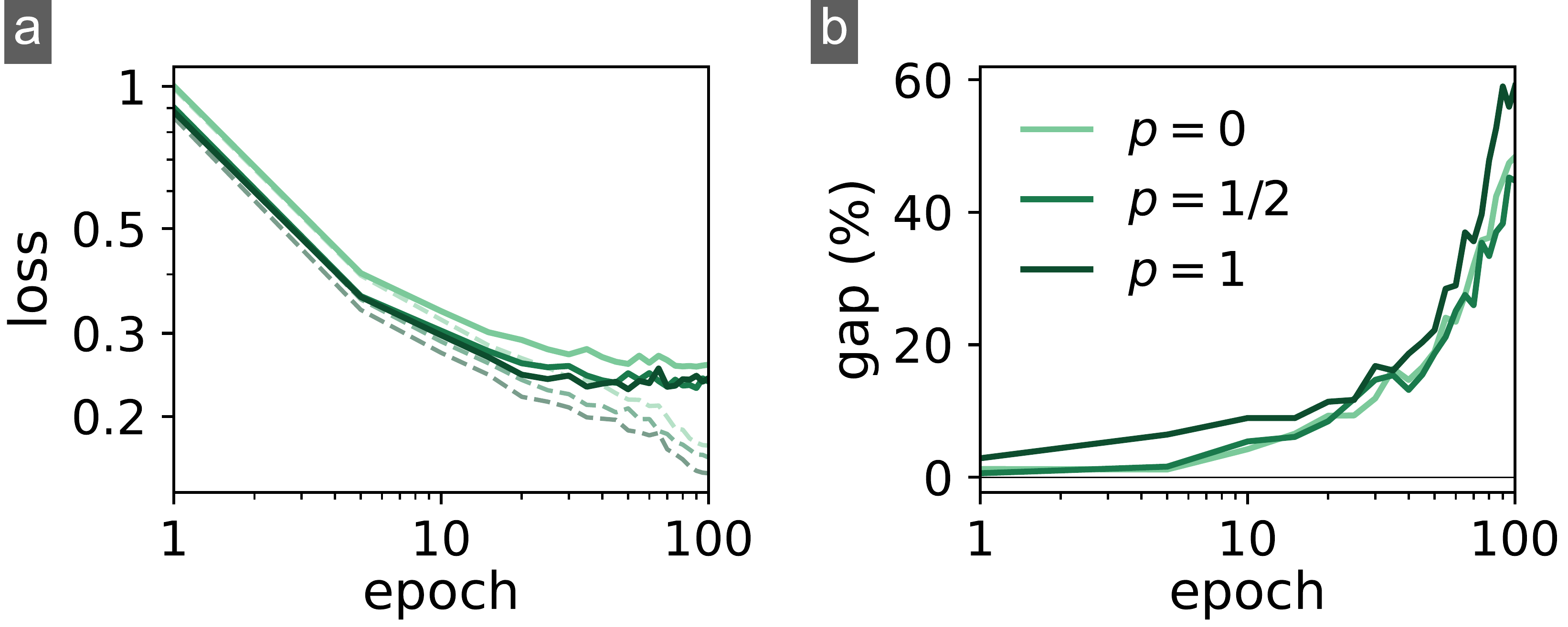}
  \caption{Overfitting of the three size weightings $w_c \propto N_c^{\,p}$ of
    the instance loss, all fine-tuned from the $600$-epoch completion-pretrained
    encoder.  \textbf{(a)}~Total loss, evaluated as in
    Fig.~\ref{fig:instance_loss}, for the validation maps (solid) and for
    training maps with the composition of the validation set (dashed).  The loss
    depends on $p$, so its absolute values are not comparable between the
    weightings.  \textbf{(b)}~Relative gap between validation and training
    loss, $L_{\mathrm{val}}/L_{\mathrm{train}} - 1$.  The $p = 0$ and $p = 1/2$
    curves come from repeats of the networks of Table~\ref{tab:debrabandere}
    with the same recipe.}
  \label{fig:sizeweight_loss}
\end{figure}

Section~\ref{sec:debrabandere} states that area weighting overfits the
most, and Fig.~\ref{fig:sizeweight_loss} substantiates this.  In the loss,
all three weightings overfit within the $100$-epoch schedule.  The relative gap between
validation and training loss grows alike for the three weightings up to about
epoch $60$; over the remaining epochs that of $p = 1$ pulls ahead, and at the
end of training it reaches $59\%$, against $45\%$ for $p = 1/2$ and $48\%$ for
$p = 0$.  The figure of merit does not follow the loss, however: the validation PQ lies
only $0.03$ below the training PQ and is still rising at the end of training
[Fig.~\ref{fig:pretraining_ablation}(c)].

\subsection{Foreground head}
\label{app:fghead}

Section~\ref{sec:debrabandere} compares the two-head network with a network
without the foreground head whose foreground is supplied by our
binary-segmentation network, as in Ref.~\cite{DeBrabandere2017}, in terms of
PQ, RQ and SQ [Table~\ref{tab:debrabandere}(b)].  Table~\ref{tab:other_fom}
completes the comparison with the mBIoU and the ARI.  Under both the second
head is marginally ahead (mBIoU $0.636$ against $0.635$, ARI $0.794$ against
$0.785$), whereas in PQ the separate network leads by $0.005$: the two
pipelines perform comparably, and ours needs one network and one forward pass
instead of two.  Given the ground-truth foreground, the two-head network is
ahead under all three figures of merit, so training the head jointly does not
degrade the embeddings, and the ARI rises by about $0.13$, in line with the
gain of $0.16$ in PQ: most of the remaining error lies in the separation of
regular from chaotic pixels, not in the grouping of regular pixels into
instances.

\section{A general-purpose segmentation model}
\label{app:sam}

\begin{figure*}
  \centering
  \includegraphics[width=\textwidth]{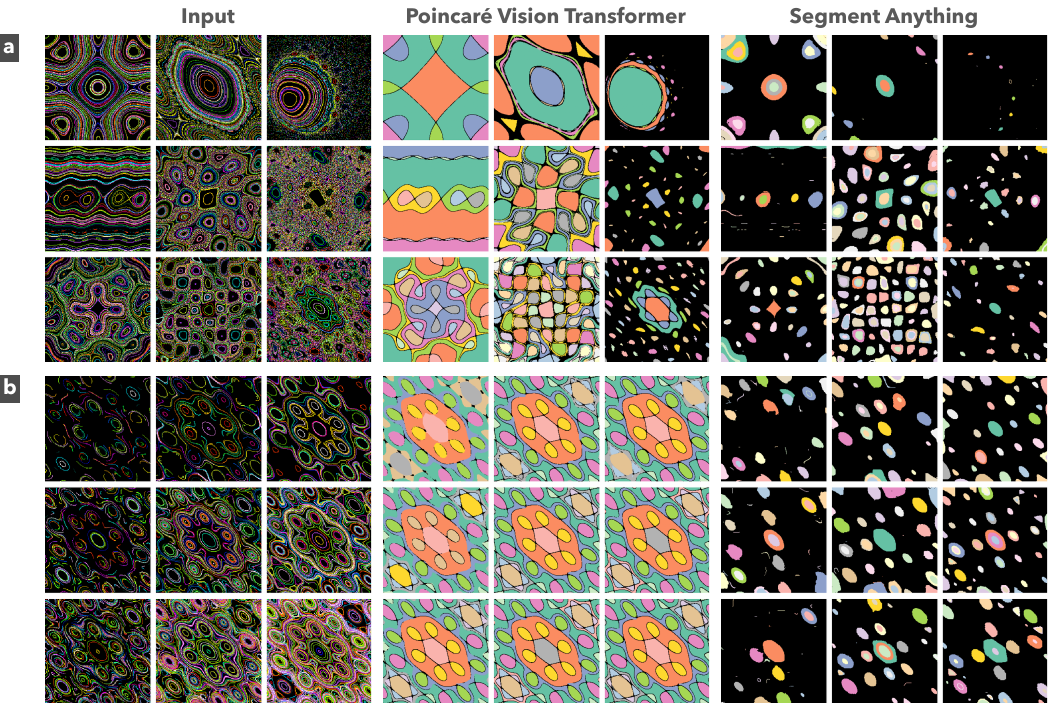}
  \caption{
    The instance segmentation of Fig.~\ref{fig:segmentation_results}(e) repeated with Segment Anything~\cite{Kirillov2023}.
    \textbf{(a)}~The nine maps of Fig.~\ref{fig:segmentation_results}, painted with $100$ orbits of $500$ iterations.
    \textbf{(b)}~A single map painted with $30$, $80$ and $150$ orbits (top to bottom) of $200$, $500$ and $1000$ iterations (left to right).
    Columns: the input, as in Fig.~\ref{fig:segmentation_results}(b); the Poincar\'e Vision Transformer (instance network), as in Fig.~\ref{fig:segmentation_results}(e); and Segment Anything in its automatic mode, applied to the input as displayed, with every mask shown.  Colours identify instances within a panel only.}
  \label{fig:sam}
\end{figure*}

Segment Anything~\cite{Kirillov2023} is a foundation model for promptable image segmentation, trained on more than $10^{9}$ masks in $1.1\times10^{7}$ natural images.  In its automatic mode it is prompted with a regular grid of $32\times32$ points and returns every mask that passes its confidence and stability thresholds.  We apply the largest released model (ViT-H), with the default thresholds, to the input maps as displayed in Fig.~\ref{fig:segmentation_results}(b), upsampled four times to the $1024$-pixel input of the model, and convert its overlapping masks into an instance map by painting them from the largest to the smallest.  Figure~\ref{fig:sam} compares the result with the instance network on the maps of Fig.~\ref{fig:segmentation_results}(e).

Segment Anything misses most of the islands.  On the eighteen panels of Fig.~\ref{fig:sam}, only $14\%$ of the $1799$ islands of the ground truth, and $20\%$ of the $1231$ islands of at least $20$ pixels, are matched by one of its masks with an IoU above $1/2$; large regular regions, in which nested invariant curves fill the image without a visible boundary, are missed entirely.  The islands it does find are delineated fairly well, with a mean IoU of $0.72$, but less precisely than by the network, and some are split into nested rings that follow the colours of their orbits.  More fundamentally, Segment Anything cannot group islands that belong together: every island it finds is an instance of its own, since grouping the islands of a chain requires the dynamics, which the image does not show.  Its prediction for one map also changes from one painting to the next [Fig.~\ref{fig:sam}(b)], whereas that of the network is nearly the same in all nine cells.  On twenty validation maps painted with $k=150$ orbits of $T_{\mathrm{w}}=1000$ iterations, Segment Anything reaches $\mathrm{PQ}=0.04$ ($\mathrm{SQ}=0.73$), against $0.54$ ($\mathrm{SQ}=0.86$) for the instance network on the same maps.  Keeping only the masks that lie at least half in the ground-truth foreground raises it only to $0.05$: what is missing are the islands themselves and their grouping into chains, not the separation of regular from chaotic pixels.%

\section{Reference Poincar\'e maps from continuous Hamiltonian systems}
\label{app:continuous_maps}

The four systems of Sec.~\ref{sec:deployment} are integrated directly, and we
describe the H\'enon--Heiles case~\cite{HenonHeiles1964} in full.   The three others
are then described by their departures from it. The Hamiltonian for the H\'enon--Heiles system
\begin{equation}
  H = \tfrac{1}{2}(p_x^2 + p_y^2)
      + \tfrac{1}{2}(x^2 + y^2)
      + x^2\!y - \tfrac{1}{3} y^3
  \label{eq:hh}
\end{equation}
is separable as $H = T(\mathbf{p}) + V(\mathbf{q})$, and we take the
surface of section to be $\{x=0,\, p_x>0\}$.

\paragraph{Symplectic integrator.}
Each sub-Hamiltonian in the splitting $H = T + V$ can be integrated
exactly.  Over a time $\tau$, $T$ generates the linear drift
$\Phi^{T}_{\tau}\colon (\mathbf{q},\mathbf{p}) \mapsto
(\mathbf{q} + \tau\,\mathbf{p},\, \mathbf{p})$, and $V$ the momentum kick
$\Phi^{V}_{\tau}\colon (\mathbf{q},\mathbf{p}) \mapsto
(\mathbf{q},\, \mathbf{p} - \tau\,\nabla V(\mathbf{q}))$.
We compose these exact sub-steps into a single time-$\Delta t$ step
using the fourth-order symplectic scheme of
Yoshida~\cite{Yoshida1990},
\begin{equation}
  \Phi_{\Delta t}
  = \Phi^{T}_{c_1\Delta t}\,\Phi^{V}_{d_1\Delta t}\,
    \Phi^{T}_{c_2\Delta t}\,\Phi^{V}_{d_2\Delta t}\,
    \Phi^{T}_{c_3\Delta t}\,\Phi^{V}_{d_3\Delta t}\,
    \Phi^{T}_{c_4\Delta t},
  \label{eq:yoshida4}
\end{equation}
with coefficients
\begin{equation}
  w_1 = \frac{1}{2 - 2^{1/3}},\qquad w_0 = 1 - 2\,w_1,
\end{equation}
$\mathbf{c} = (w_1/2,\,(w_0+w_1)/2,\,(w_0+w_1)/2,\,w_1/2)$ and
$\mathbf{d} = (w_1,\, w_0,\, w_1)$.  Because each $\Phi^{T,V}_{\tau}$ is an
exact symplectic sub-map, the composition~\eqref{eq:yoshida4} is
symplectic by construction, with local truncation error
$\mathcal{O}(\Delta t^5)$.  The energy therefore oscillates around a
modified Hamiltonian $\tilde H = H + \mathcal{O}(\Delta t^4)$ without
secular drift---the property that makes long-time Poincar\'e sampling
reliable, independently of the raw numerical precision.

\paragraph{Section crossings.}
At each integration step we detect a sign change of $x$ combined with
$p_x > 0$.  To place the recorded crossing on the section, we take the
pre-step state and advance it by a single Yoshida step of length
$\Delta t^{*} = -x/p_x$.  This is a Newton step for the root of $x(t)$.  A second such step, from the state reached, leaves the
point a distance of order $\Delta t^{4}$ from the section, far below a pixel. The main trajectory is continued from
the \emph{unrefined} post-step state so that all trajectories in a
batch remain on a common time grid; this is essential for the batched
GPU implementation, since it avoids the need for per-trajectory
adaptive time steps.

\paragraph{Initial conditions and implementation.}
For a target energy $E\le1/6$ (up to the escape threshold) we sample
initial conditions on the section by rejection sampling
$(y,p_y)$ within a bounding box and setting
$p_x = \sqrt{2E - p_y^{2} - y^{2} + \tfrac{2}{3}\,y^{3}}$,
retaining only points where the radicand is positive, that is, points in
the accessible region of the section.  The box is $-1 \le y \le 1$,
$|p_y| \le 1.2\sqrt{2E}$.  Its upper edge $y = 1$ lies at the saddle of the
potential, where $y^{2} - \tfrac{2}{3}y^{3}$ has its maximum $1/3$: for
$E \le 1/6$ the accessible region of the section splits there into a bounded
part, $y < 1$, and the escape channel, $y > 1$, (at $E=1/6$ the two touch at the saddle point), and the box encloses the bounded
part entirely, since $|p_y| \le \sqrt{2E}$ on it.  This sampling box is looser
than the box shown to the network, which is the tight bounding box of the
bounded part, $y_{-} \le y \le y_{+}$ and $|p_y| \le \sqrt{2E}$, with
$y_{\pm}$ the two roots of $y^{2} - \tfrac{2}{3}y^{3} = 2E$ that enclose
$y = 0$ (Appendix~\ref{app:pipeline}).  Sampling uniformly in
$(y, p_y)$ spreads the initial conditions evenly in the area preserved by the
map.
All operations are batched over initial conditions and executed on GPU. With $\Delta t = 0.05$ and single-precision arithmetic, the batch advances in
lockstep, so the integration time is set by the number of crossings rather than by
the number of orbits: at $E = 1/8$ a crossing takes about $0.07$~s for $150$ orbits
and $0.1$~s for $10^{3}$ to $2\times10^{3}$ orbits on an Apple M4 Pro GPU
($2\times10^{3}$ orbits followed for $200$ crossings in $20$~s).  Extending $150$
orbits from $500$ to $2500$ crossings, as in the completion target of
Sec.~\ref{sec:completion}, thus takes about $140$~s.  The labelling pipeline of
Sec.~\ref{sec:data} would follow some $10^{3}$ covering orbits for the
$2.8\times10^{5}$ crossings of the chaos indicator (Appendix~\ref{app:megno}), about
eight hours, before the tangent dynamics the indicator also requires are counted.
The per-trajectory energy drift remains bounded at
$|\Delta H| \lesssim 5\times 10^{-6}$, consistent with float32 machine
precision and the symplecticity of the scheme.
Figure~\ref{fig:ood_deployment}(d) uses $E = 1/12$, $1/8$ and $1/6$, each
painted from $50$ orbits followed for $800$ crossings.

\paragraph{Double pendulum.}
With $\Delta = \theta_1 - \theta_2$, the Hamiltonian is $H = T + V$, with
kinetic and potential energy
\begin{equation}
  \begin{gathered}
  T = \tfrac{1}{2}\,\mathbf{p}^{\mathsf{T}} M(\Delta)^{-1}\,\mathbf{p},\\
  V = -(m_1+m_2)\,g\,\ell_1\cos\theta_1 - m_2\,g\,\ell_2\cos\theta_2 ,
  \end{gathered}
  \label{eq:dp}
\end{equation}
where $\mathbf{p} = (p_1, p_2)$ and $M(\Delta)$ is the mass matrix of the
Lagrangian $L = \tfrac{1}{2}\,\dot{\boldsymbol{\theta}}^{\mathsf{T}}
M\,\dot{\boldsymbol{\theta}} - V$, with entries
$M_{11} = (m_1+m_2)\,\ell_1^2$, $M_{12} = M_{21} = m_2\,\ell_1\ell_2\cos\Delta$
and $M_{22} = m_2\,\ell_2^2$.
The kinetic term depends on the angle difference through $M$, so the Hamiltonian does not
split as $T(\mathbf{p}) + V(\mathbf{q})$ and the scheme above does not apply.
We integrate Hamilton's equations---obtained by automatic differentiation of
$H$, the hand-written momentum derivatives being unwieldy---with a fixed-step
fourth-order Runge--Kutta scheme in double precision, and monitor the energy of
every orbit instead of relying on symplecticity: over the integration lengths
used the drift stays below $10^{-8}$ for regular orbits and $10^{-4}$ for
chaotic ones, far below the scale of any structure in the section.  The section
is $\{\theta_1 = 0,\, \dot\theta_1 > 0\}$, recorded in $(\theta_2, p_2)$.
A crossing is detected when $\theta_1$, wrapped to $(-\pi, \pi]$, changes
sign from negative to positive within an integration step; the jump of the
wrapped angle at $\theta_1 = \pm\pi$ is not counted.  The crossing point is
obtained by linear interpolation of the state between the two integration
points that bracket it, at the fraction of the step where the interpolated
$\theta_1$ vanishes.  Unlike the refinement used for the H\'enon--Heiles
system, this places the point on the chord between the two states rather than
on the trajectory, with an error of order $\Delta t^{2}$, far below a pixel.
Since the minimum of $T$ over $p_1$ at fixed $p_2$ is
$p_2^2/(2 m_2 \ell_2^2)$, the accessible region on this section is
\begin{equation}
  p_2^2 \le 2 m_2 \ell_2^2
  \bigl[E + (m_1+m_2) g \ell_1 + m_2 g \ell_2 \cos\theta_2\bigr] ,
\end{equation}
which is bounded in $\theta_2$ below
$E_{\mathrm{wrap}} = g\,[m_2\ell_2 - (m_1+m_2)\ell_1]$ and wraps the full
cylinder above it; for $\ell_1 = \ell_2$ the threshold is independent of the
mass ratio, and equals $-1$ in the units of
Fig.~\ref{fig:ood_deployment}(c).  We use $\Delta t = 0.02$.
Figure~\ref{fig:ood_deployment}(c) uses $m_2/m_1 = 0.1$ and
$\ell_1 = \ell_2 = g = 1$, at the three energies $E = -1.10$, $-1.00$ and
$-0.94$ bracketing $E_{\mathrm{wrap}} = -1$, each painted from $100$ orbits
followed for $700$ crossings.

\paragraph{Planar circular restricted three-body problem.}
In the frame co-rotating with the primaries, with their separation, total
mass and angular velocity set to unity and the primaries of masses $1-\mu$ and
$\mu$ at $(-\mu, 0)$ and $(1-\mu, 0)$, the Hamiltonian is
\begin{equation}
  H = \tfrac{1}{2}\bigl(p_x^2 + p_y^2\bigr) + y\,p_x - x\,p_y
      - \frac{1-\mu}{r_1} - \frac{\mu}{r_2} ,
  \label{eq:pcr3bp_h}
\end{equation}
with $r_1^2 = (x+\mu)^2 + y^2$ and $r_2^2 = (x-1+\mu)^2 + y^2$.  The velocities
are $\dot x = p_x + y$ and $\dot y = p_y - x$, and the conserved Jacobi
constant is $C_J = 2\Omega(x,y) - (\dot x^2 + \dot y^2)$, with the effective
potential
\begin{equation}
  \Omega(x,y) = \tfrac{1}{2}(x^2+y^2) + \frac{1-\mu}{r_1} + \frac{\mu}{r_2}
                + \tfrac{1}{2}\mu(1-\mu) .
  \label{eq:pcr3bp_omega}
\end{equation}
The section is $\{y = 0,\, \dot y > 0\}$, recorded in $(x,\dot x)$; on it
$p_x = \dot x$, so $(x, \dot x)$ is a canonical pair.  The accessible region is
$2\Omega(x,0) - C_J - \dot{x}^2 \ge 0$, and the missing velocity is recovered
as $\dot y = \sqrt{2\Omega(x,0) - C_J - \dot x^2}$.

The Coriolis term $y\,p_x - x\,p_y$ is linear in the momenta, so again
$H \neq T + V$.  An exact splitting is nevertheless available: the quadratic
part of Eq.~\eqref{eq:pcr3bp_h} generates a linear flow, integrated exactly as a
$4\times4$ matrix exponential, and the gravitational part is a kick, so the two
compose under the same Yoshida scheme~\eqref{eq:yoshida4} and the integration
remains symplectic.  The obstacle is instead the close approaches to the
primaries, at which the $1/r$ potential defeats any fixed-step method.  We
remove them by Levi-Civita regularization~\cite{LeviCivita1920}, applied in two
charts: each orbit is regularized about the primary it is nearer to---the other
primary entering the smooth remainder of the potential---with the chart
switched at the midline between them, and the equations integrated in the
fictitious time conjugate to the regularizing transformation, with step
$\mathrm{d}s = 0.003$.  The Jacobi constant then drifts by less than $10^{-9}$
across the orbit bank (median ${\sim}10^{-11}$), including orbits passing
within $0.06$ of a primary, where the unregularized scheme drifts by
${\sim}3\times10^{-6}$ at the same step size.

Figure~\ref{fig:ood_deployment}(e) uses $\mu = 0.03$ and $C_J = 3.45$, each
window painted from $120$ orbits followed for $1500$ crossings.  At this
$C_J$, above the value $3.347$ at $L_1$, the accessible region in position space
consists of three disconnected zones, around the Earth, around the Moon and
outside both.  On the section, where $y = 0$, the Earth and Moon zones are each
further divided into two halves by the collision singularity at the primary,
and the exterior zone into two disconnected pieces on either side of the
system.  The six windows are segmented independently and glued for display. The accessible region is unbounded in $x$ in the exterior zone and unbounded in
$\dot x$ at each primary, so each window is a crop of its zone rather than the
whole of it.

\paragraph{Parametrically driven Duffing oscillator.}
This system is not autonomous and has one degree of freedom, so neither a
conserved quantity nor a surface of section enters at all: the section is
stroboscopic, the state being recorded once per drive period, and no crossing
detection is needed.  The equation of motion is that of a parametrically driven
Duffing oscillator,
\begin{equation}
  \ddot{x} + \bigl[\omega_0^2 + F\cos(\omega_{\mathrm{d}} t)\bigr] x
  + \gamma x^3 = 0 ,
  \label{eq:duffing}
\end{equation}
which we integrate with a fixed-step fourth-order Runge--Kutta scheme at $128$
substeps per drive period, in double precision.  Once $\omega_0$ and $\gamma$ are set to unity by rescaling $x$ and $t$,
the map depends on two dimensionless parameters, the scaled detuning
$\mu = \omega_{\mathrm{d}}(\omega_{\mathrm{d}} - 2)/F$ and the drive strength
$\epsilon = F/\omega_{\mathrm{d}}^{2}$.  In the weak-driving limit
$\epsilon \to 0$, a rotating-wave approximation, which keeps only the resonant
terms, makes the dynamics integrable and dependent on $\mu$
alone~\cite{Peano2012}; $\epsilon$ measures the non-resonant terms that break
integrability and so governs the transition from regular to chaotic motion.
Figure~\ref{fig:ood_deployment}(b) uses $\mu = 0.70$ and $\epsilon = 0.27$
($\omega_{\mathrm{d}} = 2.4571$, $F = 1.6043$), large enough for the nonlinear
resonances around the two period-two states of the parametric oscillator to be
resolved at $256\times256$ resolution.  It is painted from $140$ orbits followed for
$2000$ drive periods, with initial conditions drawn uniformly from the
displayed window $x \in [-1.45,\,1.45]$, $\dot x \in [-3.56,\,3.56]$.

\section{Deployment pipeline}
\label{app:pipeline}

\begin{figure}
  \centering
  \includegraphics[width=\columnwidth]{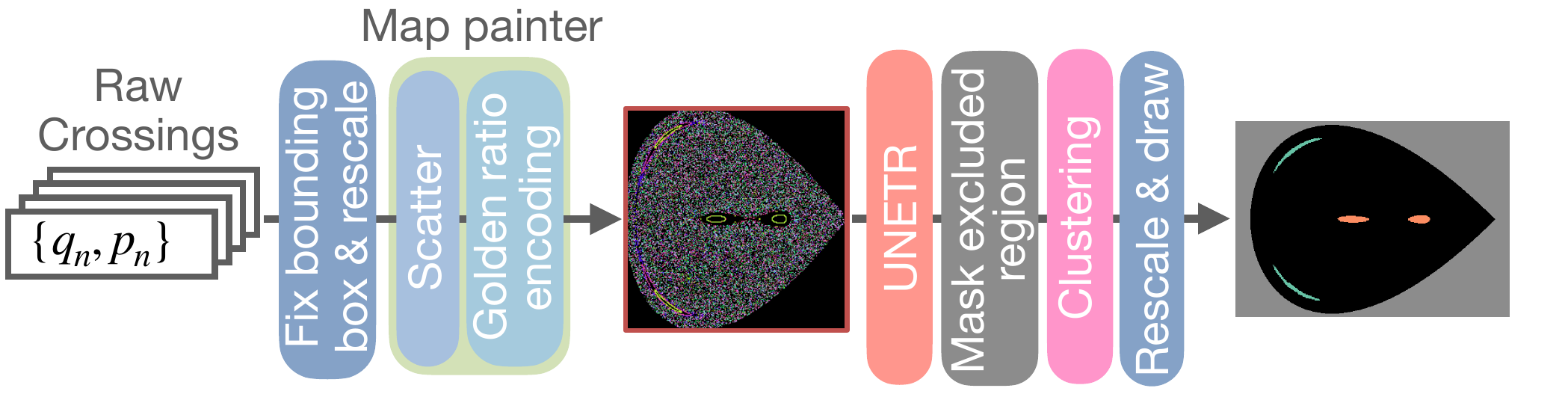}
  \caption{
    Inference on a real system, from the recorded crossings to the segmented
    map: section crossings $\{q_n,p_n\}$ (Appendix~\ref{app:continuous_maps}),
   cropping of the region of interest to its bounding box and anisotropic
    rescaling, painting (scattering onto the $256\times256$ grid and
    golden-ratio encoding), one forward pass of the instance network, masking of
    the excluded region (the part of the bounding box outside the accessible
    region), clustering of the embeddings, and rescaling of the labels to the
    original coordinates, with the excluded region drawn in grey.  The steps are described in the text.}
  \label{fig:inference_pipeline}
\end{figure}

Figure~\ref{fig:inference_pipeline} collects the steps that take a real system
to a segmented map.  Nothing in the chain is specific to the system: the only
place its physics enters is the integration that produces the crossings
(Appendix~\ref{app:continuous_maps}) and the accessible region, used to
mask the network output and to draw the final map.

\paragraph{Pre-processing.}
The integration returns a list of crossings in the system's own coordinates,
and two steps bring it to the form the network expects.  The region of
interest of the section, the whole accessible region when it is bounded and a
window of it otherwise (Appendix~\ref{app:continuous_maps}), is first cropped
to its bounding box and rescaled anisotropically to the square $[-1,1]^2$,
which disposes at once of the arbitrary units of the section coordinates and of
the empty margin that a region of interest of low aspect ratio would otherwise
leave around itself.  The painter of Fig.~\ref{fig:architecture}(a) then
scatters the points onto the $256\times256$ grid and gives each orbit
its golden-ratio colour (Appendix~\ref{app:encoding}), producing the
$30$-channel image.  The excluded region carries no marking of any kind in
that image, being indistinguishable from a pixel no orbit reached.

\paragraph{Network and decoding.}
The image is passed once through the instance-segmentation network of
Sec.~\ref{sec:results}, with the weights it was left with after fine-tuning on
the synthetic maps.  Its per-pixel embeddings are clustered into instances by the mean shift decoding with bin seeding followed by the refining pass
described in Appendix~\ref{app:instance_inference}.  Here the accessible
region enters the decoding: the foreground the network returns is restricted to it
before the clustering.  The instance labels are finally mapped back to the
original coordinates, which is what the panels of
Fig.~\ref{fig:ood_deployment} show, with the excluded region drawn in grey.

\section{Verification and ablation  on real maps}

\label{app:ood_ablation}

\subsection{Verification of the predictions}
\label{app:ood_verification}

The predictions on the real maps have no ground truth, so Sec.~\ref{sec:deployment} checks them against the dynamics alone: orbits are launched inside each predicted instance, and their behaviour is compared with what a correct instance would imply.  This appendix gives the criteria and the results for the network with map-completion pretraining, with the instance maps decoded as described in Appendix~\ref{app:pipeline}; the networks without pretraining and with MAE pretraining only are compared in Appendix~\ref{app:ood_arms}.

\paragraph{Islands and fragments.}
Every instance is split into connected components. For cylindrical geometry,  connected components touching  the seam of the cylinder are joined across it.  A component is classified as an \emph{island} if it has at least $5$ pixels and at least a fifth of the area of the largest component of the instance; the others are classified as \emph{fragments}.  The islands of a chain are mapped onto one another by an area-preserving map and have the same area, but the predicted islands differ in their delineation, and an island narrower than a pixel, such as some of those recovered by the repair of Appendix~\ref{app:ood_repair}, is represented by a long trail of pixels whose area can exceed that of the other islands of the chain several times.  A fifth of the largest component admits both and still excludes the small fragments.

\paragraph{Return period.}
From the deepest interior pixel of the largest island we integrate the dynamics of the system for $300$ iterates of the Poincar\'e map, with the integrators of Appendix~\ref{app:continuous_maps}.  The first return to the island, counted when any pixel within one pixel of the orbit point belongs to it, defines the period $p$.  The period is accepted if the orbit returns at every multiple of $p$ and at no other iterate; if the test fails, up to three starting points are tried, those inside islands that the edge of the image does not cut first.  Where the image cuts the island, as at the edge of the window of the Duffing oscillator and of the three-body problem, the orbit may leave the image, and the test requires at least three returns in agreement with a common period; a failure is then reported as inconclusive.  For the H\'enon--Heiles system and the double pendulum the image is the whole accessible region, so the edge cuts no island.  The integrator of the three-body problem may return only part of the requested iterates, and the test uses those available.%

\paragraph{Outcomes.}
Comparing $p$ with the number $N$ of islands of the instance gives four outcomes.  If $N=p$ the instance is \emph{verified}.  If $N>p$ it contains several chains, a \emph{merger}, which we confirm by launching an orbit from an island that the first orbit did not visit and finding its own cycle.  Because the area criterion can hide a merger when the islands of one chain are smaller than those of the other, which then count as fragments and leave $N=p$, mergers are also searched with an orbit launched from every component of at least $5$ pixels: components visited by the same orbit are grouped, and an instance with two or more groups of at least $100$ pixels each, every one with a period of at most $40$, is counted as a merger.  Conversely, an island cut by the pixelization into two pieces above the threshold gives $N>p$ although a single orbit visits all the components; such an instance, of which there is one, after MAE pretraining, is not a merger and is counted as verified. If $N<p$ the network has \emph{missed} islands of the chain, that is, some positions of the cycle carry no label.  The positions of the cycle are the iterates congruent modulo $p$; if every one of them lies, at least $90\%$ of the time, within one pixel of a component of the instance, island or fragment, the instance is counted as complete.  Finally, if no regular return is found after three starting points the instance \emph{fails}.  These outcomes assume that the whole chain is in the image, which holds for the H\'enon--Heiles system and the double pendulum.  For the maps that show only a window of the phase space, the Duffing oscillator and the windows of the three-body problem, an instance that touches the edge of the window is still verified if $N=p$, since no island can then be hidden, and still a merger if $N>p$; but a period larger than the number of visible islands, or no return at all, may only mean that islands or the orbit lie beyond the image, and the instance is reported as \emph{inconclusive}.%

\paragraph{Leaks.}
Independently of the periods, every pixel that an orbit launched in an instance paints in another instance is recorded, without any dilation of the orbit, together with the distance from its starting pixel to the boundary of its island.  A chain split between two instances would show up as an orbit started deep inside an island that regularly reaches another instance, whereas a start within one or two pixels of the boundary can fall on the neighbouring instance simply because the delineation is imprecise.%

\begin{figure*}
  \centering
  \includegraphics[width=\textwidth]{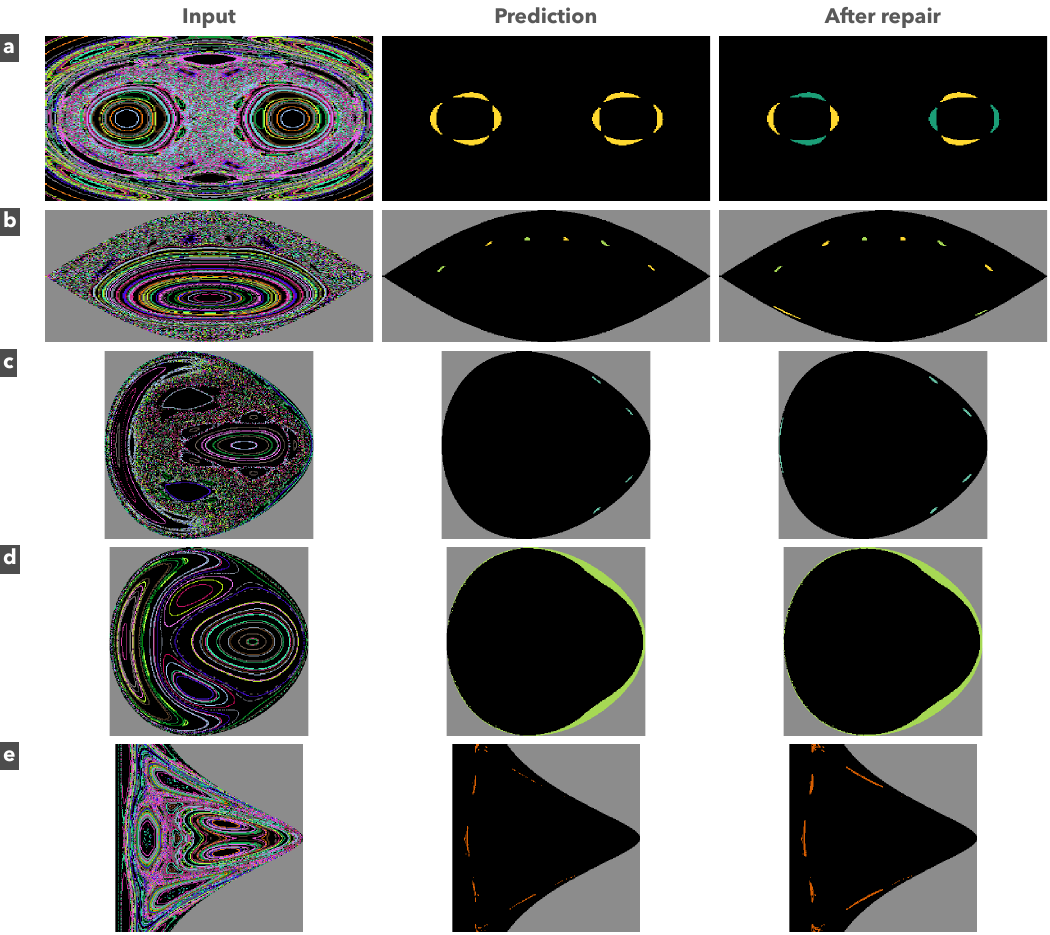}
  \caption{%
    One example of each kind of instance of the map-completion network that the check of Appendix~\ref{app:ood_verification} does not verify, two for missed islands (all of them are listed in Table~\ref{tab:ood_failures}), before and after the orbit-based repair of Appendix~\ref{app:ood_repair}.  Each row shows the input, the network prediction and the repaired prediction over the same window as in Fig.~\ref{fig:ood_deployment} (for the double pendulum and the H\'enon--Heiles system, the bounding box of the accessible region), with grey marking the excluded region; the predictions show only the instance of the row, in its colour in Fig.~\ref{fig:ood_deployment}.
    \textbf{(a)}~Merger, the only one: two period-four chains of the Duffing oscillator in one instance (one of them marked by white ellipses in Fig.~\ref{fig:ood_deployment}(b)), split by the repair.
    \textbf{(b)}~Missed island, the example discussed in Sec.~\ref{sec:deployment}: in the double pendulum at $E=-1$ the fourth island of a period-four chain, near the lower edge of the accessible region, is not labelled; the repair adds it.
    \textbf{(c)}~Missed island narrower than a pixel, chosen to show this case: the fifth island of a period-five chain of the H\'enon--Heiles system at $E=1/8$ lies along the edge of the accessible region; the repair adds it as a trail of single pixels.
    \textbf{(d)}~Failure, the one of the two in which the band is visible: a regular period-one band along the edge of the H\'enon--Heiles accessible region at $E=1/12$, labelled in pieces; the repair closes some of the gaps, but the band stays fragmented.
    \textbf{(e)}~Inconclusive, one of four: an instance of the three-body problem in the window to the right of the Earth, cut by the edge of the window, with seven visible islands and period eight.}
  \label{fig:ood_failures}
\end{figure*}

\begin{table}[t]
  \caption{
    The $18$ instances of the map-completion network that the check does not verify, with their colour in Fig.~\ref{fig:ood_deployment}, the number $N$ of islands, the period $p$ (``--'' when there is no regular return), the outcome of the check, and the outcome after the repair.  Colours are reused when a map has more than eight instances; H--H stands for H\'enon--Heiles, the three-body instances are all in the window to the right of the Earth, and ``inconcl.'' stands for inconclusive.}
  \label{tab:ood_failures}
  \begin{ruledtabular}
  \footnotesize
  \begin{tabular}{llrrll}
    Map & Colour & $N$ & $p$ & Check & Repaired \\
    \hline
    Duffing & yellow & 8 & 4 & merger & split \\
    Duffing & teal & 9 & 10 & missed & verified \\
    Duffing & orange & 9 & 10 & missed\footnotemark[1] & verified \\
    Duffing & pink & 4 & 8 & missed & verified \\
    Duffing & blue & 11 & 36 & missed & unchanged\footnotemark[2] \\
    Pendulum, $E=-1$ & green & 3 & 4 & missed & verified \\
    Pendulum, $E=-1$ & yellow & 3 & 4 & missed & verified \\
    Pendulum, $E=-0.94$ & beige & 2 & 3 & missed & verified \\
    H--H, $E=1/8$ & teal & 4 & 5 & missed & verified \\
    Three-body & purple & 5 & 7 & missed & verified \\
    Three-body & magenta & 5 & 11 & missed & verified \\
    Three-body & olive & 4 & 8 & missed & verified \\
    H--H, $E=1/12$ & green & 1 & -- & fail & fail \\
    H--H, $E=1/8$ & orange & 1 & -- & fail & fail \\
    Duffing & orange & 4 & -- & inconcl. & inconcl. \\
    Three-body & beige & 2 & 5 & inconcl. & inconcl. \\
    Three-body & grey & 1 & -- & inconcl. & inconcl. \\
    Three-body & dk.\ orange & 7 & 8 & inconcl. & inconcl. \\
  \end{tabular}
  \end{ruledtabular}
  \footnotetext[1]{The missing positions of the cycle are covered by small pieces of the same instance (complete with fragments).}
  \footnotetext[2]{The repair is rejected by the test on the number of islands.}
\end{table}

\paragraph{Instances that fail the check.}
Table~\ref{tab:ood_failures} lists the $18$ instances of the map-completion network that the check does not verify, and Fig.~\ref{fig:ood_failures} shows one of each kind.  The merger [Fig.~\ref{fig:ood_failures}(a)] consists of two period-four chains of the Duffing oscillator, images of one another under the symmetry $(x,\dot{x})\to(-x,-\dot{x})$ and interleaved around one ring; the orbits launched in its islands fall into two groups, one per chain.  In the eleven instances with missed islands [Fig.~\ref{fig:ood_failures}(b),(c)], the orbit never lands in another instance at the unlabelled positions of its cycle, only on pixels predicted as background or on fragments of the same instance.  The islands are therefore lost in the foreground prediction of the network, not in the clustering, which assigns every foreground pixel to an instance (Appendix~\ref{app:instance_inference}).  The missing islands are thin: in the double pendulum at $E=-1$ they are narrow holes in the chaotic sea near the lower edge of the accessible region, where no orbit of the input was launched and chaotic points partly fill the hole; in the H\'enon--Heiles system at $E=1/8$ the missing island lies along the edge of the accessible region and is narrower than a pixel.  The two failures [Fig.~\ref{fig:ood_failures}(d)] are regular period-one bands along the edge of the accessible region of the H\'enon--Heiles system, at $E=1/12$ and $E=1/8$.  Orbits launched in them run along the whole edge and are regular, with MEGNO indicators of $1.8$ and $2.1$ after a time $t=3000$, against $82$ and still growing for an orbit of the chaotic sea, but the bands become as thin as about one pixel and the network labels them only in pieces, so that the orbit lands on the other pieces and the return test fails.  The four inconclusive instances [Fig.~\ref{fig:ood_failures}(e)] touch the edge of the window of the Duffing oscillator or of the three-body problem.%

\subsection{Orbit-based repair}
\label{app:ood_repair}

The orbits of the check can also correct the instance map, in the manner of the grouping step of the labelling (Appendix~\ref{app:chains}).  The instances of the network are kept, and each is split into its connected components.  From the deepest pixel of every component of at least $5$ pixels we launch an orbit of $3\times10^{3}$ iterates.  Its period $n$ is the first return to the component, a return being counted when any pixel within one pixel of the orbit point belongs to it, and is accepted if the orbit makes no other return up to $3n$ and returns at no fewer than $80\%$ of the multiples of $n$.  This replaces the check at $2n$ and $3n$ of Appendix~\ref{app:chains}, so that an orbit that occasionally lands on a fragment of its own region is not rejected.  Components of an instance reached by an accepted orbit, within one pixel, are grouped, and an instance whose components fall into two or more groups is split; a group of fewer than $20$ pixels stays with the largest.  The pixels on which an accepted orbit lands and that the network predicts as background join the instance of the orbit.  We do not fill the regions that an orbit encloses: the islands that the network misses are typically elongated and about one pixel wide [Fig.~\ref{fig:ood_failures}(b),(c)], so that the orbit itself covers them; on the maps of Fig.~\ref{fig:ood_deployment}, filling would add $5$ pixels. Pixels of other instances are never relabelled, and components without an accepted orbit keep their label.  When every iterate of the accepted orbits of a group lies inside the image, the repair of the group is kept only if the repaired group has as many islands, counted as in Appendix~\ref{app:ood_verification}, as the period of its orbit; otherwise the group is left as predicted.  When an orbit leaves the image, the number of islands cannot be checked and the repair is applied.%

On every sixteenth map of the validation set, $63$ maps painted with $k=150$ orbits of $T_{\mathrm{w}}=1000$ iterates and decoded as in Sec.~\ref{sec:results}, the repair raises PQ from $0.650$ to $0.666$: $21$ more instances are matched, against one more false positive.  The gain comes from the splits ($22$ instances in $10$ maps); on the other maps the pixels added along the orbits rarely turn an unmatched instance into a match.  The test on the number of islands rejects $12$ repairs; without it the false positives grow by $5$ and PQ reaches $0.664$.  On $62$ further maps, offset by eight from the first set and not used to design the repair, PQ rises from $0.581$ to $0.594$ ($0.592$ without the test), and is the same for any island threshold between $0.1$ and $0.5$: the synthetic maps have few islands narrower than a pixel, and the threshold matters only for those.  Orbits of $300$ iterates give the same PQ.

On the maps of Fig.~\ref{fig:ood_deployment} the repair splits the Duffing merger into its two period-four chains and recovers $22$ missing islands in ten instances, among them the missing islands of the two chains of the double pendulum at $E=-1$ and the fifth island of a period-five chain of the H\'enon--Heiles system at $E=1/8$, which lies along the edge of the accessible region and is narrower than a pixel.  The test on the number of islands rejects one repair, of a Duffing instance of eleven pieces whose orbit has period $36$, which is left as predicted (Table~\ref{tab:ood_failures}, last column).  The two thin bands of the H\'enon--Heiles system remain failures of the check: at $E=1/12$ the repair closes some of the gaps along the edge, but the band stays fragmented, and at $E=1/8$ the orbit launched in the labelled piece does not return to it.
After the repair the check of Appendix~\ref{app:ood_verification} is no longer independent, since the repaired instances are built from the same orbits.%

\subsection{Comparison of the pretraining strategies}
\label{app:ood_arms}

\begin{figure*}
  \centering
  \includegraphics[width=0.85\textwidth]{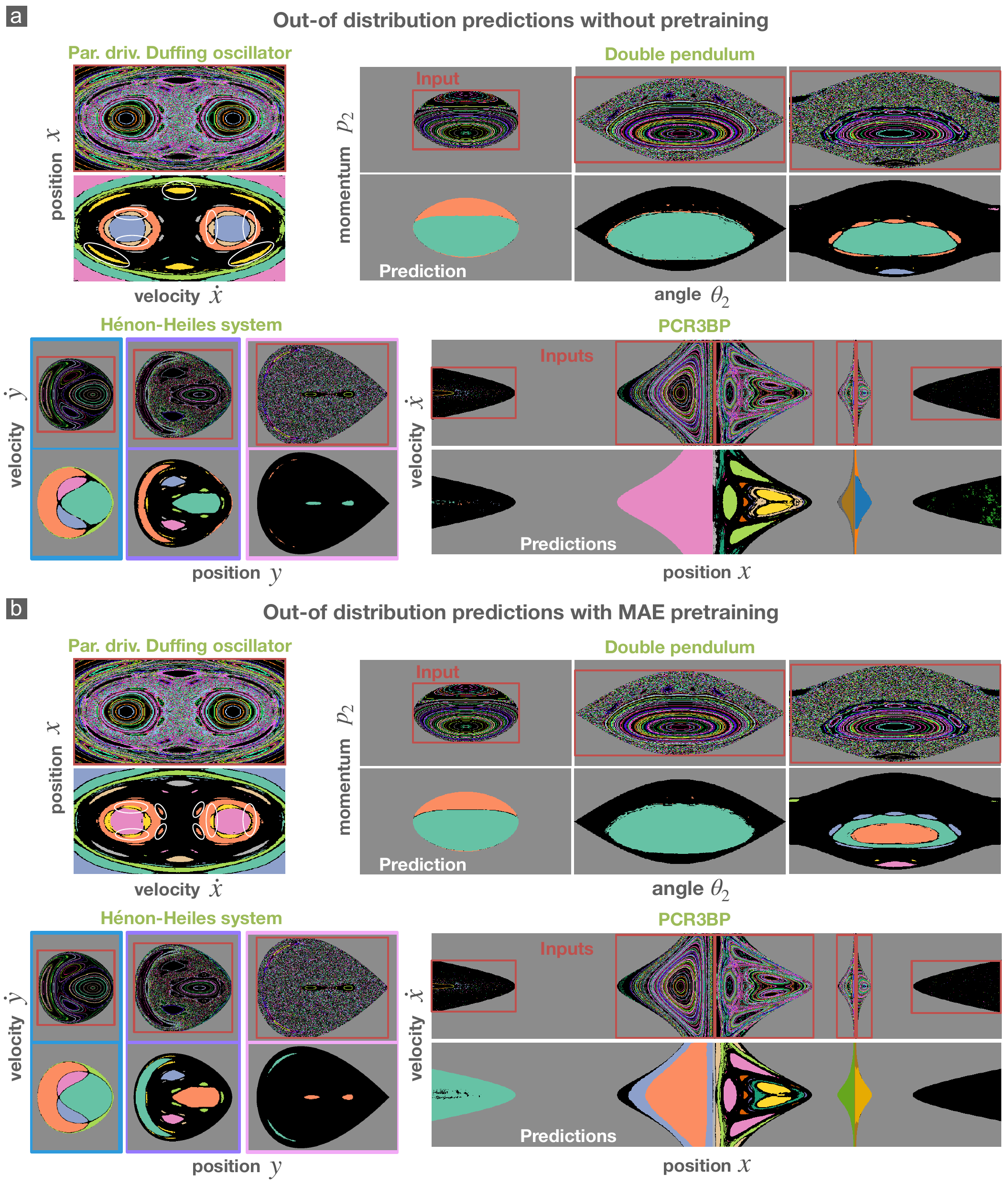}
  \caption{
    Predictions of the networks without pretraining \textbf{(a)} and with MAE pretraining
    only \textbf{(b)} for the same examples as in Fig.~\ref{fig:ood_deployment}.}
  \label{fig:ood_appendix}
\end{figure*}

\begin{table*}[t]
  \caption{
    Verification of the predictions on the maps of Fig.~\ref{fig:ood_deployment} for the three pretraining arms.  The upper block counts the instances by outcome of the first-return check; ``inconclusive'' counts the instances cut by the edge of the window (Duffing and three-body problem) in which the check cannot decide, because they show no regular return or a period larger than the number of visible islands.  The lower block counts the orbits launched in the islands (one per island) that paint at least one pixel in another instance, and, among them, those started more than two pixels from the boundary of their island.  The last column applies the check to the map-completion predictions after the orbit-based repair of Appendix~\ref{app:ood_repair}, which uses the same orbits, so that it is not an independent verification.}
  \label{tab:ood_verification}
  \begin{ruledtabular}
  \begin{tabular}{lcccc}
    & No pretraining & MAE & Map completion & + repair \\
    \hline
    Instances decoded                  & 55 & 53 & 61 & 62 \\
    Instances tested                   & 53 & 53 & 60 & 61 \\
    Verified ($N=p$)                   & 29 & 33 & \textbf{42} & 54 \\
    Merger                             & 3 & 1 & 1 & 0 \\
    Missed islands ($N<p$)             & 7 & 10 & 11 & 1 \\
    \quad of which complete with fragments & 0 & 0 & 1 & 0 \\
    No regular return (fail)           & 8 & 3 & 2 & 2 \\
    Inconclusive (cut by the window)   & 6 & 6 & 4 & 4 \\
    \hline
    Orbits launched                    & 268 & 220 & 213 & -- \\
    \quad leaving their instance       & 119 & 48 & \textbf{32} & -- \\
    \quad starting $>2$ px from the boundary & 11 & 6 & \textbf{1} & -- \\
  \end{tabular}
  \end{ruledtabular}
\end{table*}

Figure~\ref{fig:ood_appendix} repeats the deployment of Fig.~\ref{fig:ood_deployment} with the networks fine-tuned without pretraining and after MAE pretraining only, the two other arms of Fig.~\ref{fig:pretraining_ablation}, and Table~\ref{tab:ood_verification} applies the check of Appendix~\ref{app:ood_verification} to all three networks.  The network with map-completion pretraining has the largest fraction of verified instances and the fewest leaks, and the only leak that starts more than two pixels from the boundary of its island loses two of the 294 pixels of the orbit.  Without pretraining the leaks start well inside the islands, and the number of failures grows.  The numbers depend on the choices made above; raising the area fraction from a fifth to $0.5$ and $0.7$ lowers the number of verified instances from $42$ to $40$ and $36$ for map completion, from $33$ to $30$ and $30$ after MAE pretraining and from $29$ to $27$ and $26$ without pretraining, and does not change the order of the three networks.

\end{document}